\documentclass{aa}

\usepackage{txfonts}

\ifx\pdfoutput\thisisundefined
   \newcommand{\dofig}{P}
\else
   \newcommand{\dofig}{D}
\fi

\if\dofig F
   \newcommand{\putfig}[2]{\vspace{#2}}
\else
   \if\dofig D
      \usepackage[pdftex]{graphicx}
      \newcommand{\putfig}[2]{\includegraphics[height=#2]{#1.pdf}}
   \else
      \usepackage[dvips]{graphicx}
      \newcommand{\putfig}[2]{\includegraphics[height=#2]{#1.eps}}
   \fi
\fi

\newcommand{\dr}{\frac{\pi}{180\degr}}
\newcommand{\rd}{\frac{180\degr}{\pi}}
\newcommand{\Rt}{R_{\theta}}
\newcommand{\Ap}{A_{\phi}}

\newcommand{\Ma}{\hskip 1em}
\newcommand{\Mb}{\hskip 3em}

\newcommand{\SP}{\hphantom{-}}

\newcommand{\sub}[1]{_\mathrm{#1}}

\newcommand{\sign}{\hbox{sign\,}}

\newcommand{\keyw}[1]{\hbox{{\tt #1}}}

\newcommand{\keyi}[2]{\hbox{{\tt #1\hspace{1pt}}{$#2$}\/}}
\newcommand{\CRPIX}[1]{\keyi{CRPIX}{#1}}
\newcommand{\CDELT}[1]{\keyi{CDELT}{#1}}
\newcommand{\CROTA}[1]{\keyi{CROTA}{#1}}
\newcommand{\CRVAL}[1]{\keyi{CRVAL}{#1}}
\newcommand{\CTYPE}[1]{\keyi{CTYPE}{#1}}
\newcommand{\CUNIT}[1]{\keyi{CUNIT}{#1}}
\newcommand{\LONPOLE}[1]{\keyi{LONPOLE}{#1}}
\newcommand{\LATPOLE}[1]{\keyi{LATPOLE}{#1}}
\newcommand{\RADESYS}[1]{\keyi{RADESYS}{#1}}
\newcommand{\EQUINOX}[1]{\keyi{EQUINOX}{#1}}
\newcommand{\WCSNAME}[1]{\keyi{WCSNAME}{#1}}

\newcommand{\keyii}[3]{\hbox{{\tt #1\hspace{1pt}{$#2$}\_{$#3$}}\/}}
\newcommand{\PC}[2]{\keyii{PC}{#1}{#2}}
\newcommand{\CD}[2]{\keyii{CD}{#1}{#2}}
\newcommand{\PV}[2]{\keyii{PV}{#1}{#2}}

\newcommand{\PVi}[1]{\hbox{{\tt PV\hspace{1pt}{$i$}\_{#1}{$a$}}\/}}

\newcommand{\keyv}[1]{\hbox{{\tt #1}}}

\begin{document}

\title{Representations of celestial coordinates in FITS}

\author{Mark R. Calabretta\inst{(1)}
\and    Eric W. Greisen\inst{(2)}}

\institute{Australia Telescope National Facility,
           P.O. Box 76,
           Epping, NSW 1710, Australia
\and       National Radio Astronomy Observatory,
           P.O. Box O,
           Socorro, NM,
           USA 87801-0387}

\offprints{M. Calabretta, \\
           \email{mcalabre@atnf.csiro.au}}

\date{Draft dated 17 Jul 2002}

\abstract{
   In Paper~I, Greisen \& Calabretta (\cite{kn:Paper1}) describe a generalized
   method for assigning physical coordinates to FITS image pixels.  This paper
   implements this method for all spherical map projections likely to be of
   interest in astronomy.  The new methods encompass existing informal FITS
   spherical coordinate conventions and translations from them are described.
   Detailed examples of header interpretation and construction are given.
}

\maketitle


\section{Introduction}
\label{sec:intro}

This paper is the second in a series which establishes conventions by which
world coordinates may be associated with FITS (Hanisch et al., \cite{kn:NOST})
image, random groups, and table data.  Paper~I (Greisen \& Calabretta,
\cite{kn:Paper1}) lays the groundwork by developing general constructs and
related FITS header keywords and the rules for their usage in recording
coordinate information.  In Paper~III, Greisen et al.\ (\cite{kn:Paper3})
apply these methods to spectral coordinates.  Paper~IV (Calabretta et al.,
\cite{kn:Paper4}) extends the formalism to deal with general distortions of
the coordinate grid.  This paper, Paper~II, addresses the specific problem of
describing celestial coordinates in a two-dimensional projection of the sky.
As such it generalizes the informal but widely used conventions established by
Greisen (\cite{kn:G1}, \cite{kn:G2}) for the Astronomical Image Processing
System, hereinafter referred to as the {\em AIPS convention}.

Paper~I describes the computation of world coordinates as a multi-step
process.  Pixel coordinates are linearly transformed to intermediate world
coordinates that in the final step are transformed into the required world
coordinates.

\begin{figure}
   \centerline{\putfig{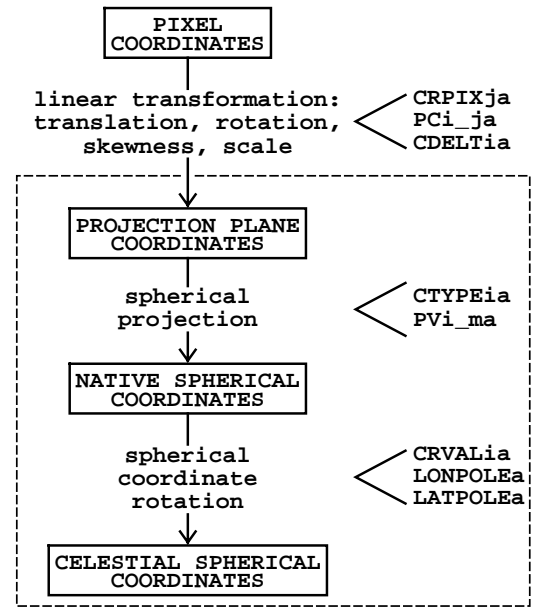}{8.0cm}}
   \caption[]{Conversion of pixel coordinates to celestial coordinates.  The
   {\sc intermediate world coordinates} of Paper~I, Fig.~1 are here
   interpreted as {\sc projection plane coordinates}, i.e.\ Cartesian
   coordinates in the plane of projection, and the multiple steps required to
   produce them have been condensed into one.  This paper is concerned in
   particular with the steps enclosed in the dotted box.}
   \label{fig:Process}
\end{figure}

In this paper we associate particular elements of the intermediate world
coordinates with Cartesian coordinates in the plane of the spherical
projection.  Figure~\ref{fig:Process}, by analogy with Fig.~1 of Paper~I,
focuses on the transformation as it applies to these projection plane
coordinates.  The final step is here divided into two sub-steps, a spherical
projection defined in terms of a convenient coordinate system which we refer
to as {\em native spherical coordinates}, followed by a spherical rotation of
these native coordinates to the required celestial coordinate system.

The original FITS paper by Wells et al.\ (\cite{kn:WGH}) introduced the
\CRPIX{ja}\footnote{The single-character alternate version code ``{\it a}'' on
the various FITS keywords was introduced in Paper~I\@.  It has values blank
and \keyv{A} through \keyv{Z}\@.} keyword to define the pixel coordinates
$(r_1,r_2,r_3,\ldots)$ of a {\em coordinate reference point}.  Paper~I retains
this but replaces the coordinate rotation keywords \CROTA{i} with a linear
transformation matrix.  Thus, the transformation of {\em pixel coordinates}
$(p_1,p_2,p_3,\ldots)$ to {\em intermediate world coordinates}
$(x_1,x_2,x_3,\ldots)$ becomes
\begin{eqnarray}
   x_i & = &           s_i  \sum_{j=1}^{N}      m_{ij}  (p_j - r_j) \, ,
             \label{eq:px} \\
       & = & \hphantom{s_i} \sum_{j=1}^{N} (s_i m_{ij}) (p_j - r_j) \, ,
             \nonumber
\end{eqnarray}
where $N$ is the number of axes given by the \keyw{NAXIS} keyword.  As
suggested by the two forms of the equation, the scales $s_i$, and matrix
elements $m_{ij}$ may be represented either separately or in combination.  In
the first form $s_i$ is given by \CDELT{ia} and $m_{ij}$ by \PC{i}{ja}; in the
second, the product $s_i m_{ij}$ is given by \CD{i}{ja}.  The two forms may
not coexist in one coordinate representation.

Equation~(\ref{eq:px}) establishes that the reference point is the origin of
intermediate world coordinates.  We require that the linear transformation be
constructed so that the plane of projection is defined by two axes of the
$x_i$ coordinate space.  We will refer to intermediate world coordinates in
this plane as {\em projection plane coordinates}, $(x,y)$, thus with
reference point at $(x,y) = (0,0)$.  Note that this does not necessarily
correspond to any plane defined by the $p_j$ axes since the linear
transformation matrix may introduce rotation and/or skew.

Wells et al.\ (\cite{kn:WGH}) established that all angles in FITS were to be
measured in degrees and this has been entrenched by the AIPS convention and
confirmed in the IAU-endorsed FITS standard (Hanisch et al., \cite{kn:NOST}).
Paper~I introduced the \CUNIT{ia} keyword to define the units of \CRVAL{ia}
and \CDELT{ia}.  Accordingly, we require \CUNIT{ia} = \keyv{`deg'} for the
celestial \CRVAL{ia} and \CDELT{ia}, whether given explicitly or not.
Consequently, the $(x,y)$ coordinates in the plane of projection are measured
in degrees.  For consistency, we use degree measure for native and celestial
spherical coordinates and for all other angular measures in this paper.

For linear coordinate systems Wells et al.\ (\cite{kn:WGH}) prescribed that
world coordinates should be computed simply by adding the relative world
coordinates, $x_i$, to the coordinate value at the reference point given by
\CRVAL{ia}.  Paper~I extends this by providing that particular values of
\CTYPE{ia} may be established by convention to denote non-linear
transformations involving predefined functions of the $x_i$ parameterized by
the \CRVAL{ia} keyword values and possibly other parameters.

In Sects.~\ref{sec:basics}, \ref{sec:celsys} and \ref{sec:projections} of this
paper we will define the functions for the transformation from $(x,y)$
coordinates in the plane of projection to celestial spherical coordinates for
all spherical map projections likely to be of use in astronomy.

The FITS header keywords discussed within the main body of this paper apply to
the primary image header and image extension headers.  Image fragments within
binary tables extensions defined by Cotton et al.\ (\cite{kn:CTP}) have
additional nomenclature requirements, a solution for which was proposed in
Paper~I\@.  Coordinate descriptions may also be associated with the random
groups data format defined by Greisen \& Harten (\cite{kn:GH}).  These issues
will be expanded upon in Sect.~\ref{sec:bintab}.

Section~\ref{sec:previous} considers the translation of older AIPS convention
FITS headers to the new system and provisions that may be made to support
older FITS-reading programs.  Section~\ref{sec:discussion} discusses the
concepts presented here, including worked examples of header interpretation
and construction.


\section{Basic concepts}
\label{sec:basics}

\subsection{Spherical projection}
\label{sec:sphproj}

As indicated in Fig.~\ref{fig:Process}, the first step in transforming $(x,y)$
coordinates in the plane of projection to celestial coordinates is to convert
them to native longitude and latitude, $(\phi,\theta)$.  The equations for the
transformation depend on the particular projection and this will be specified
via the \CTYPE{ia} keyword.  Paper~I defined ``4-3'' form for such purposes;
the rightmost three-characters are used as an {\em algorithm code} that in
this paper will specify the projection.  For example, the stereographic
projection will be coded as `\keyv{STG}'.  Some projections require additional
parameters that will be specified by the FITS keywords \PV{i}{ma} for
$m = 0,1,2,\ldots$, also introduced in Paper~I\@.  These parameters may be
associated with the longitude and/or latitude coordinate as specified for each
projection.  However, definition of the three-letter codes for the projections
and the equations, their inverses and the parameters which define them, form a
large part of this work and will be discussed in Sect.~\ref{sec:projections}.
The leftmost four characters of \CTYPE{ia} are used to identify the celestial
coordinate system and will be discussed in Sect.~\ref{sec:celsys}.

\subsection{Reference point of the projection}
\label{sec:refpoint}

The last step in the chain of transformations shown in Fig.~\ref{fig:Process}
is the spherical rotation from native coordinates, $(\phi,\theta)$, to
celestial\footnote{Usage here of the conventional symbols for right ascension
and declination for celestial coordinates is meant only as a mnemonic.  It
does not preclude other celestial systems.} coordinates $(\alpha,\delta)$.
Since a spherical rotation is completely specified by three Euler angles it
remains only to define them.

In principle, specifying the celestial coordinates of any particular native
coordinate pair provides two of the Euler angles (either directly or
indirectly).  In the AIPS convention, the \CRVAL{ia} keyword values for the
celestial axes\footnote{We will refer to these simply as ``{\em the}
\CRVAL{ia}'', and likewise for the other keyword values.} specify the celestial
coordinates of the reference point and this in turn is associated with a
particular point on the projection.  For zenithal projections that point is
the sphere's point of tangency to the plane of projection and this is the pole
of the native coordinate system.  Thus the AIPS convention links a celestial
coordinate pair to a native coordinate pair via the reference point.  Note
that this association via the reference point is purely conventional; it has
benefits which are discussed in Sect.~\ref{sec:projections} but in principle
any other point could have been chosen.

Section~\ref{sec:projections} presents the projection equations for the
transformation of $(x,y)$ to $(\phi,\theta)$.  The native coordinates of the
reference point would therefore be those obtained for $(x,y) = (0,0)$.
However, it may happen that this point lies outside the boundary of the
projection, for example as for the \keyv{ZPN} projection of
Sect.~\ref{sec:ZPN}.  Therefore, while this work follows the AIPS approach it
must of necessity generalize it.

Accordingly we specify only that a fiducial celestial coordinate pair
$(\alpha_0,\delta_0)$ given by the \CRVAL{ia} will be associated with a
fiducial native coordinate pair $(\phi_0,\theta_0)$ defined explicitly for
each projection.  For example, zenithal projections all have
$(\phi_0,\theta_0) = (0,90\degr)$, while cylindricals have $(\phi_0,\theta_0)
= (0,0)$.  The AIPS convention has been honored here as far as practicable by
constructing the projection equations so that $(\phi_0,\theta_0)$ transforms
to the reference point, $(x,y) = (0,0)$.  Thus, apart from the one exception
noted, the fiducial celestial and native coordinates are the celestial and
native coordinates of the reference point and we will not normally draw a
distinction.

It is important to understand why $(\phi_0,\theta_0)$ differs for different
projection types.  There are two main reasons; the first makes it difficult
for it to be the same, while the second makes it desirable that it differs.
Of the former, some projections such as Mercator's, diverge at the native
pole, therefore they cannot have the reference point there because that would
imply infinite values for \CRPIX{ja}.  On the other hand, the gnomonic
projection diverges at the equator so it can't have the reference point there
for the same reason.  Possibly $(\phi_0,\theta_0)$ chosen at some mid-latitude
could satisfy all projections, but that leads us to the second reason.

Different projection types are best suited to different purposes.  For
example, zenithal projections are best for mapping the region in the vicinity
of a point, often a pole; cylindrical projections are appropriate for the
neighborhood of a great circle, usually an equator; and the conics are
suitable for small circles such as parallels of latitude.  Thus, it would be
awkward if a cylindrical used to map, say, a few degrees on either side of the
galactic plane, had its reference point, and thus \CRPIX{ja} and \CRVAL{ia},
at the native pole, way outside the map boundary.  In formulating the
projection equations themselves the native coordinate system is chosen to
simplify the geometry as much as possible.  For the zenithals the natural
formulation has $(x,y) = (0,0)$ at the native pole, whereas for the
cylindricals the equations are simplest if $(x,y) = (0,0)$ at a point on the
equator.

As discussed above, a third Euler angle must be specified and this will be
given by the native longitude of the celestial pole, $\phi\sub{p}$, specified
by the new FITS keyword

\begin{center}
\begin{tabular}{l}
\noalign{\vspace{-5pt}}
   \LONPOLE{a} \hspace{1em} (floating-valued). \\
\noalign{\vspace{-5pt}}
\end{tabular}
\end{center}

\noindent
The default value of \LONPOLE{a} will be $0$ for $\delta_0\ge\theta_0$ or
$180\degr$ for $\delta_0<\theta_0$.  This is the condition for the celestial
latitude to increase in the same direction as the native latitude at the
reference point.  Thus, for example, in zenithal projections the default is
always $180\degr$ (unless $\delta_0 = 90\degr$) since $\theta_0 = 90\degr$.
In cylindrical projections, where $\theta_0 = 0$, the default value for
\LONPOLE{a} is $0$ for $\delta_0 \ge 0$, but it is $180\degr$ for
$\delta_0 < 0$.

\subsection{Spherical coordinate rotation}

Since $(\phi_0,\theta_0)$ differs for different projections it is apparent
that the relationship between $(\alpha_0,\delta_0)$ and the required Euler
angles also differs.

For zenithal projections, $(\phi_0,\theta_0) = (0,90\degr)$ so the \CRVAL{ia}
specify the celestial coordinates of the native pole,
i.e.\ $(\alpha_0,\delta_0) = (\alpha\sub{p},\delta\sub{p})$.  There is a
simple relationship between the Euler angles for consecutive rotations about
the Z-, X-, and Z-axes and $\alpha\sub{p}$, $\delta\sub{p}$ and $\phi\sub{p}$;
the ZXZ Euler angles are $(\alpha\sub{p}+90\degr, 90\degr-\delta\sub{p},
\phi\sub{p}-90\degr)$.  Given this close correspondence it is convenient to
write the Euler angle transformation formul\ae\ directly in terms of
$\alpha\sub{p}$, $\delta\sub{p}$ and $\phi\sub{p}$:
\begin{equation}
   \begin{array}{rcl}
      \alpha & = & \alpha\sub{p} + \arg\,(\sin\theta\cos\delta\sub{p} -
                       \cos\theta\sin\delta\sub{p}\cos(\phi-\phi\sub{p}), \\
             &   & \hphantom{\alpha\sub{p} + \arg\,(}
                       -\cos\theta\sin(\phi-\phi\sub{p})) \, , \\
      \delta & = & \sin^{-1}(\sin\theta \sin\delta\sub{p} + \cos\theta
                   \cos\delta\sub{p}\cos(\phi-\phi\sub{p}) ) \, .
   \end{array}
   \label{eq:nat2std}
\end{equation}
where $\arg\,()$ is an inverse tangent function that returns the correct
quadrant, i.e.\ if $(x,y) = (r\cos\beta,r\sin\beta)$ with $r > 0$ then
$\arg\,(x,y) = \beta$.  Note that, if $\delta\sub{p} = 90\degr$,
Eqs.~(\ref{eq:nat2std}) become
\begin{equation}
   \begin{array}{rcl}
      \alpha & = & \alpha\sub{p} + \phi - \phi\sub{p} - 180\degr \, , \\
      \delta & = & \theta \, ,
   \end{array}
   \label{eq:nat2std90}
\end{equation}
which may be used to define a simple change in the origin of longitude.
Likewise for $\delta\sub{p} = -90\degr$
\begin{equation}
   \begin{array}{rcl}
      \alpha & = & \alpha\sub{p} - \phi + \phi\sub{p} \, , \\
      \delta & = & -\theta \, .
   \end{array}
   \label{eq:nat2stdm90}
\end{equation}
The inverse equations are
\begin{equation}
   \begin{array}{rcl}
      \phi & = & \phi\sub{p} + \arg\,(\sin\delta\cos\delta\sub{p} -
                     \cos\delta\sin\delta\sub{p}\cos(\alpha-\alpha\sub{p}), \\
           &   & \hphantom{\phi\sub{p} + \arg\,(}
                     -\cos\delta\sin(\alpha-\alpha\sub{p})) \, , \\
    \theta & = & \sin^{-1}(\sin\delta \sin\delta\sub{p} + \cos\delta
                 \cos\delta\sub{p}\cos(\alpha-\alpha\sub{p})) \, ,
   \end{array}
   \label{eq:std2nat}
\end{equation}
Useful relations derived from Eqs.~(\ref{eq:nat2std}) and (\ref{eq:std2nat})
are
\begin{eqnarray}
   \cos\delta \cos(\alpha - \alpha\sub{p})
      & = & \sin\theta\cos\delta\sub{p}
            - \cos\theta \sin\delta\sub{p} \cos(\phi-\phi\sub{p}) \, ,
            \nonumber \\
   \cos\delta \sin(\alpha - \alpha\sub{p})
      & = & -\cos\theta \sin(\phi - \phi\sub{p}) \, ,
   \label{eq:nat2std2}
\end{eqnarray}
\begin{eqnarray}
   \cos\theta \cos(\phi - \phi\sub{p})
      & = & \sin\delta \cos\delta\sub{p}
            -\cos\delta \sin\delta \sub{p} \cos(\alpha - \alpha\sub{p}) \, ,
            \nonumber \\
   \cos\theta \sin(\phi - \phi\sub{p})
      & = & -\cos\delta \sin(\alpha - \alpha\sub{p}) \, .
   \label{eq:std2nat2}
\end{eqnarray}
A matrix method of handling the spherical coordinate rotation is described in
Appendix~\ref{apx:methods}.

\subsection{Non-polar $(\phi_0,\theta_0)$}
\label{sec:nonpolar}

Projections such as the cylindricals and conics for which
$(\phi_0,\theta_0) \neq (0,90\degr)$ are handled by providing formulae to
compute $(\alpha\sub{p},\delta\sub{p})$ from $(\alpha_0,\delta_0)$ whence the
above equations may be used.

Given that $(\alpha_0,\delta_0)$ are the celestial coordinates of the point
with native coordinates $(\phi_0,\theta_0)$, Eqs.~(\ref{eq:nat2std2}) and
(\ref{eq:std2nat2}) may be inverted to obtain
\begin{eqnarray}
   \delta\sub{p} & = & \arg\,(\cos\theta_0 \cos(\phi\sub{p}-\phi_0) \, ,
                              \sin\theta_0) \; \pm \nonumber \\
                 &   & \cos^{-1} \left(
                          \frac{\sin\delta_0}
                               {\sqrt{1 -
                                \cos^2\theta_0\sin^2(\phi\sub{p}-\phi_0)}}
                       \right) \, ,
   \label{eq:org2pola}
\end{eqnarray}
\begin{eqnarray}
   \sin(\alpha_0-\alpha\sub{p}) & = & \sin(\phi\sub{p}-\phi_0)\cos\theta_0 /
                                          \cos\delta_0 \, ,
   \label{eq:org2polb} \\
   \cos(\alpha_0-\alpha\sub{p}) & = & \frac{\sin\theta_0 -
                                      \sin\delta\sub{p}\sin\delta_0}
                                      {\cos\delta\sub{p}\cos\delta_0} \, ,
   \label{eq:org2polc}
\end{eqnarray}
whence Eqs.~(\ref{eq:nat2std}) may be used to determine the celestial
coordinates.  Note that Eq.~(\ref{eq:org2pola}) contains an ambiguity in the
sign of the inverse cosine and that all three indicate that some combinations
of $\phi_0$, $\theta_0$, $\alpha_0$, $\delta_0$, and $\phi\sub{p}$ are not
allowed.  For these projections, we must therefore adopt additional
conventions:
\begin{enumerate}
\item Equations~(\ref{eq:org2polb}) and (\ref{eq:org2polc}) indicate that
   $\alpha\sub{p}$ is undefined when $\delta_0 = \pm90\degr$.  This simply
   represents the longitude singularity at the pole and forces us to define
   $\alpha\sub{p} = \alpha_0$ in this case.
\item If $\delta\sub{p} = \pm90\degr$ then the longitude at the native pole is
   $\alpha\sub{p} = \alpha_0 + \phi\sub{p} - \phi_0 - 180\degr$ for
   $\delta\sub{p} = 90\degr$ and
   $\alpha\sub{p} = \alpha_0 - \phi\sub{p} + \phi_0$ for
   $\delta\sub{p} = -90\degr$.
\item Some combinations of $\phi_0$, $\theta_0$, $\delta_0$, and $\phi\sub{p}$
   produce an invalid argument for the $\cos^{-1}()$ in
   Eq.~(\ref{eq:org2pola}).  This is indicative of an inconsistency for which
   there is no solution for $\delta\sub{p}$.  Otherwise
   Eq.~(\ref{eq:org2pola}) produces two solutions for $\delta\sub{p}$.  Valid
   solutions are ones that lie in the range $-90\degr$ to $+90\degr$, and it
   is possible in some cases that neither solution is valid.

   Note, however, that if $\phi_0 = 0$, as is usual, then when \LONPOLE{a}
   ($\equiv \phi\sub{p}$) takes its default value of $0$ or $180\degr$
   (depending on $\theta_0$) then {\em any} combination of $\delta_0$ and
   $\theta_0$ produces a valid argument to the $\cos^{-1}()$ in
   Eq.~(\ref{eq:org2pola}), and at least one of the solutions is valid.
\item Where Eq.~(\ref{eq:org2pola}) has two valid solutions the one closest to
   the value of the new FITS keyword

   \begin{center}
   \begin{tabular}{l}
   \noalign{\vspace{-5pt}}
      \LATPOLE{a} \hspace{1em} (floating-valued). \\
   \noalign{\vspace{-5pt}}
   \end{tabular}
   \end{center}

   \noindent
   is chosen.  It is acceptable to set \LATPOLE{a} to a number greater than
   $+90\degr$ to choose the northerly solution (the default if \LATPOLE{a} is
   omitted), or a number less than $-90\degr$ to select the southern solution.
\item Equation~(\ref{eq:org2pola}) often only has one valid solution (because
   the other lies outside the range $-90\degr$ to $+90\degr$).  In this case
   \LATPOLE{a} is ignored.
\item For the special case where $\theta_0 = 0$, $\delta_0 = 0$, and
   $\phi\sub{p} - \phi_0 =\pm90\degr$ then $\delta\sub{p}$ is not
   determined and \LATPOLE{a} specifies it completely. \LATPOLE{a} has no
   default value in this case.
\end{enumerate}

\noindent
These rules governing the application of Eqs.~(\ref{eq:org2pola}),
(\ref{eq:org2polb}) and (\ref{eq:org2polc}) are certainly the most complex of
this formalism.  FITS writers are well advised to check the values of
$\phi_0$, $\theta_0$, $\alpha_0$, $\delta_0$, and $\phi\sub{p}$ against
them to ensure their validity.

\subsection{User-specified $(\phi_0,\theta_0)$}
\label{sec:userspec}

In Sect.~\ref{sec:refpoint} we formally decoupled $(\alpha_0,\delta_0)$ from
the reference point and associated it with $(\phi_0,\theta_0)$.  One
implication of this is that it should be possible to allow $(\phi_0,\theta_0)$
to be user-specifiable.  This may be useful in some circumstances, mainly to
allow \CRVAL{ia} to match a point of interest rather than some predefined
point which may lie well outside the image and be of no particular interest.
We therefore reserve keywords \PVi{1} and \PVi{2} attached to {\em longitude}
coordinate $i$ to specify $\phi_0$ and $\theta_0$ respectively.

By itself, this prescription discards the AIPS convention and lacks utility
because it breaks the connection between \CRVAL{ia} and any point whose pixel
coordinates are given in the FITS header.  New keywords could be invented to
define these pixel coordinates but this would introduce additional complexity
and still not satisfy the AIPS convention.  The solution adopted here is to
provide an option to force $(x,y) = (0,0)$ at $(\phi_0,\theta_0)$ by
introducing an implied offset $(x_0,y_0)$ which is computed for
$(\phi_0,\theta_0)$ from the relevant projection equations given in
Sect.~\ref{sec:projections}.  This is to be applied to the $(x,y)$ coordinates
when converting to or from pixel coordinates.  The operation is controlled by
the value of \PVi{0} attached to {\em longitude} coordinate $i$; the offset
is to be applied only when this differs from its default value of zero.

This construct should be considered advanced usage, of which
Figs.~\ref{fig:ZEAex} and \ref{fig:AITex} provide an example.  Normally we
expect that \PVi{1} and \PVi{2} will either be omitted or set to the
projection-specific defaults given in Sect.~\ref{sec:projections}.

\subsection{Encapsulation}
\label{sec:encapsulation}

So that all required transformation parameters can be contained completely
within the recognized world coordinate system (WCS) header cards, the values
of \LONPOLE{a} and \LATPOLE{a} may be recorded as \PVi{3} and \PVi{4},
attached to {\em longitude} coordinate~{\it i}, and these take precedence
where a conflict arises.

We recommend that FITS writers include the \PVi{1}, \PVi{2}, \PVi{3}, and
\PVi{4} cards in the header, even if only to denote the correct use of the
default values.

Note carefully that these are associated with the {\em longitude} coordinate,
whereas the projection parameters defined later are all associated with the
{\em latitude} coordinate.

\subsection{Change of coordinate system}

A change of coordinate system may be effected in a straightforward way if the
transformation from the original system, $(\alpha,\delta)$, to the new system,
$(\alpha',\delta')$, and its inverse are known.  The new coordinates of the
$(\phi_0,\theta_0)$, namely $(\alpha'_0,\delta'_0)$, are obtained simply by
transforming $(\alpha_0,\delta_0)$.  To obtain $\phi'\sub{p}$, first transform
the coordinates of the pole of the new system to the original celestial system
and then transform the result to native coordinates via Eq.~(\ref{eq:std2nat})
to obtain $(\phi'\sub{p},\theta'\sub{p})$.  As a check, compute
$\delta'\sub{p}$ via Eq.~(\ref{eq:org2pola}) and verify that
$\theta'\sub{p} = \delta'\sub{p}$.

\subsection{Comparison with linear coordinate systems}
\label{sec:compare}

It must be stressed that the coordinate transformation described here differs
from the linear transformation defined by Wells et al.\ (\cite{kn:WGH}) even
for some simple projections where at first glance they may appear to be the
same.  Consider the plate carr\'{e}e projection defined in Sect.~\ref{sec:CAR}
with $\phi = x$, $\theta = y$ and illustrated in Fig.~\ref{fig:CAR}.
Figure~\ref{fig:Process} shows that while the transformation from $(x,y)$ to
$(\phi,\theta)$ may be linear (in fact identical), there still remains the
non-linear transformation from $(\phi,\theta)$ to $(\alpha,\delta)$.  Hence
the linear coordinate description defined by the unqualified \CTYPE{ia} pair
of `\keyv{RA}', `\keyv{DEC}' which uses the Wells et al.\ prescription will
generally differ from that of `\keyv{RA---CAR}', `\keyv{DEC---CAR}' with the
same \CRVAL{ia}, etc.\  If \LONPOLE{a} assumes its default value then they
will agree to first order at points near the reference point but gradually
diverge at points away from it.

\begin{figure}
   \centerline{\putfig{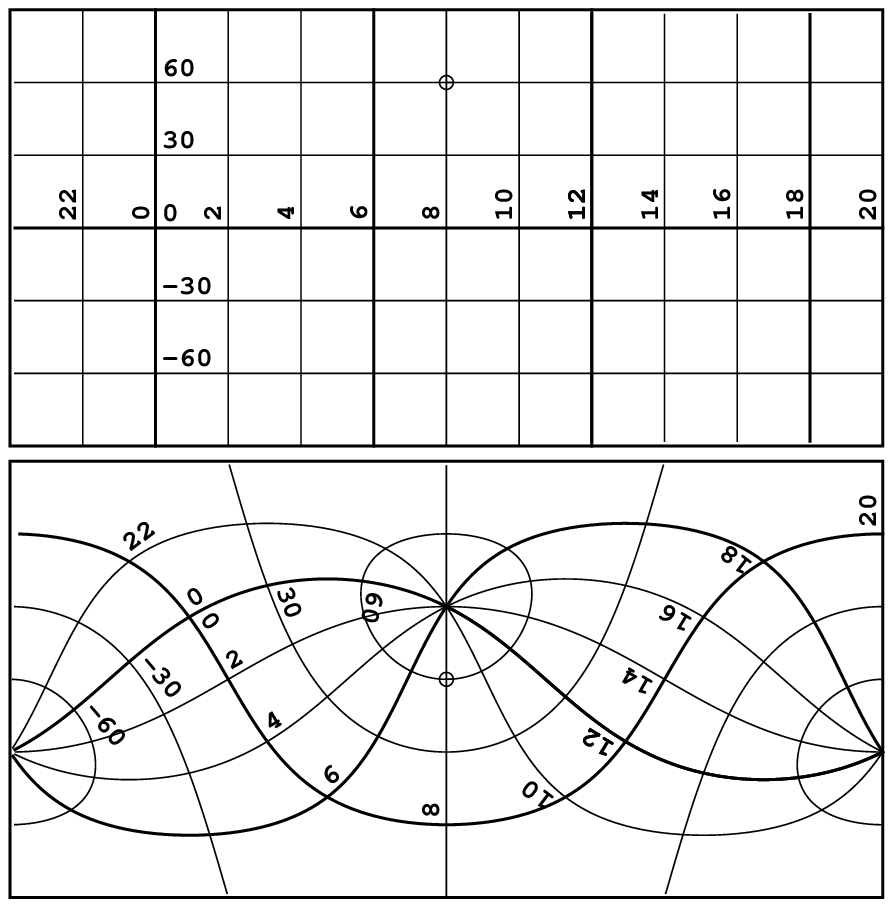}{260pt}}
   \caption[]{A linear equatorial coordinate system (top) defined via the
      methods of Wells et al.\ (\cite{kn:WGH}), and the corresponding oblique
      system constructed using the methods of this paper.  The reference
      coordinate $(\alpha_0,\delta_0)$ for each is at right ascension
      $8^{\rm hr}$, declination $+60\degr$ (marked).  The two sets of FITS
      header cards differ only in their \CTYPE{ia} keyword values.  The
      non-oblique graticule could be obtained in the current system by setting
      $(\alpha_0,\delta_0,\phi\sub{p})=(120\degr,0,0)$.}
   \label{fig:Grids}
\end{figure}

Figure~\ref{fig:Grids} illustrates this point for a plate carr\'{e}e
projection with reference coordinates of $8^{\rm hr}$ right ascension and
$+60\degr$ declination and with $\phi\sub{p} = 0$.  It is evident that since
the plate carr\'{e}e has $(\phi_0,\theta_0) = (0,0)$, a non-oblique graticule
may only be obtained by setting $\delta_0 = 0$ with $\phi\sub{p} = 0$.  It
should also be noted that where a larger map is to be composed of tiled
submaps the coordinate description of a submap should only differ in the value
of its reference pixel coordinate.


\section{Celestial coordinate systems}
\label{sec:celsys}

As mentioned in Sect.~\ref{sec:sphproj}, Paper~I defined ``4-3'' form for the
\CTYPE{ia} keyword value; i.e., the first four characters specify the
coordinate type, the fifth character is a ``-'', and the remaining three
characters specify an algorithm code for computing the world coordinate
value.  Thus while the right half specifies the transformation to be applied
in computing the spherical coordinate pair, $(\alpha,\delta)$, the left half
simply identifies the celestial system with which $(\alpha,\delta)$ are to be
associated.  In this sense \CTYPE{ia} contains an active part which drives the
transformation and a passive part which describes the results.

Consistent with past practice, an equatorial coordinate pair is denoted by
`\keyv{RA--}' and `\keyv{DEC-}', and other celestial systems are of the form
`\keyv{$x$LON}' and `\keyv{$x$LAT}' for longitude and latitude pairs, where $x
= $ {\tt G} for galactic\footnote{``New'' galactic coordinates are assumed
here, Blaauw et al.\ (\cite{kn:BGPW}).  Users of the older system or future
systems should adopt a different value of $x$ and document its meaning.},
{\tt E} for ecliptic, {\tt H} for helioecliptic\footnote{Ecliptic and
helioecliptic systems each have their equator on the ecliptic.  However, the
reference point for ecliptic longitude is the vernal equinox while that for
helioecliptic longitude is the sun vector.}, and {\tt S} for supergalactic
coordinates.  Since representation of planetary, lunar and solar coordinate
systems could exceed the 26 possibilities afforded by the single character
$x$, we also here allow `\keyv{$yz$LN}' and `\keyv{$yz$LT}' pairs.  Additional
values of $x$ and $yz$ will undoubtedly be defined.  A basic requirement,
however, is that the coordinate system be right-handed and that the pole be at
latitude $+90\degr$.  A coordinate pair such as azimuth and zenith distance
would have to be represented as a negative azimuth and elevation with implied
conversion.  In some situations negation of the longitude coordinate may be
implemented via a sign reversal of the appropriate \CDELT{ia}.  It remains the
responsibility of the authors of new coordinate system types to define them
properly and to gain general recognition for them from the FITS community.
However, FITS interpreters should be able to associate coordinate pairs even
if the particular coordinate system is not recognized.

Paper~I clarified that, while the subscript of \CRPIX{ja} and the $j$
subscript of \PC{i}{ja} and \CD{i}{ja} refer to pixel coordinate elements, the
$i$ subscript of \PC{i}{ja} and \CD{i}{ja} and the subscripts of \CDELT{ia},
\CTYPE{ia} and \CRVAL{ia} refer to world coordinate elements.  However we now
have three different sets of world coordinates, $(x,y)$, $(\phi,\theta)$, and
$(\alpha,\delta)$.  This leads us to associate $x$, $\phi$, and $\alpha$ on
the one hand, and $y$, $\theta$, and $\delta$ on the other.  This simply
means, for example, that if {\tt CTYPE3 = 'GLON-AIT'}, then the third element
of the intermediate world coordinate calculated via Eq.~(\ref{eq:px})
corresponds to what we have been calling the $x$-coordinate in the plane of
projection, the association being between $\alpha$ and $x$.  In this way pairs
of \CTYPE{ia} with complementary left halves and matching right halves define
which elements of the intermediate world coordinate vector form the plane of
projection.


\subsection{Equatorial and ecliptic coordinates}
\label{sec:equatorial}

Several systems of equatorial coordinates (right ascension and declination)
are in common use.  Apart from the International Celestial Reference System
(ICRS, IAU~\cite{kn:ICRS}), the axes of which are by definition fixed with
respect to the celestial sphere, each system is parameterized by time.  In
particular, mean equatorial coordinates (Hohenkerk et al., \cite{kn:Hoh})
are defined in terms of the epoch (i.e.\ instant of time) of the mean equator
and equinox (i.e.\ pole and origin of right ascension).  The same applies for
ecliptic coordinate systems.  Several additional keywords are therefore
required to specify these systems fully.  We introduce the new keyword

\begin{center}
\begin{tabular}{l}
\noalign{\vspace{-5pt}}
   \RADESYS{a} \hspace{1em} (character-valued) \\
\noalign{\vspace{-5pt}}
\end{tabular}
\end{center}

\noindent
to specify the particular system.  {Recognized values are given in
Table~\ref{ta:radesys}.  Apart from \keyv{FK4-NO-E} these keywords are
applicable to ecliptic as well as equatorial coordinates.

Wells et al.\ (\cite{kn:WGH}) introduced the keyword \keyw{EPOCH} to mean the
epoch of the mean equator and equinox.  However we here replace it with

\begin{center}
\begin{tabular}{l}
\noalign{\vspace{-5pt}}
   \EQUINOX{a} \hspace{1em} (floating-valued), \\
\noalign{\vspace{-5pt}}
\end{tabular}
\end{center}

\noindent
since the word ``epoch'' is often used in astrometry to refer to the time of
observation.  The new keyword\footnote{\keyw{EQUINOX} has since been
adopted by Hanisch et al.\ (\cite{kn:NOST}) which deprecates \keyw{EPOCH}\@.}
takes preference over \keyw{EPOCH} if both are given.  Note that \EQUINOX{a}
applies to ecliptic as well as to equatorial coordinates.

\begin{table}
   \caption[]{Allowed values of \RADESYS{a}.}
   \begin{tabular}{ll}
      \hline
      \noalign{\smallskip}
      \RADESYS{a}       & Definition \\
      \noalign{\smallskip}
      \hline
      \noalign{\smallskip}
      `\keyv{ICRS}'     & International Celestial Reference System \\
      `\keyv{FK5}'      & mean place, \\
                        & \hskip 0.4cm new (IAU 1984) system \\
      `\keyv{FK4}'      & mean place, \\
                        & \hskip 0.4cm old (Bessell-Newcomb) system \\
      `\keyv{FK4-NO-E}' & mean place, \\
                        & \hskip 0.4cm old system but without e-terms \\
      `\keyv{GAPPT}'    & geocentric apparent place, \\
                        & \hskip 0.4cm IAU 1984 system \\
      \noalign{\smallskip}
      \hline
   \end{tabular}
   \label{ta:radesys}
\end{table}

For \RADESYS{a} values of \keyv{FK4} and \keyv{FK4-NO-E}, any stated equinox
is Besselian and, if neither \EQUINOX{a} nor \keyw{EPOCH} are given, a default
of 1950.0 is to be taken.  For \keyv{FK5}, any stated equinox is Julian and,
if neither keyword is given, it defaults to 2000.0.

If the \EQUINOX{a} keyword is given it should always be accompanied by
\RADESYS{a}.  However, if it should happen to appear by itself then
\RADESYS{a} defaults to \keyv{FK4} if \EQUINOX{a} $< 1984.0$, or to \keyv{FK5}
if \EQUINOX{a} $\ge\ 1984.0$.  Note that these defaults, while probably true
of older files using the \keyw{EPOCH} keyword, are not required of them.

\RADESYS{a} defaults to ICRS if both it and \EQUINOX{a} are absent.

Geocentric apparent equatorial and ecliptic coordinates (\keyv{GAPPT}) require
the epoch of the equator and equinox of date.  This will be taken as the time
of observation rather than \EQUINOX{a}.  The time of observation may also be
required for other astrometric purposes in addition to the usual astrophysical
uses, for example, to specify when the mean place was correct in accounting
for proper motion, including ``fictitious'' proper motions in the conversion
between the \keyv{FK4} and \keyv{FK5} systems.  The old \keyw{DATE-OBS}
keyword may be used for this purpose.  However, to provide a more convenient
specification we here introduce the new keyword

\begin{center}
\begin{tabular}{l}
\noalign{\vspace{-5pt}}
   \keyw{MJD-OBS} \hspace{1em} (floating-valued), \\
\noalign{\vspace{-5pt}}
\end{tabular}
\end{center}

\noindent
that provides \keyw{DATE-OBS} as a Modified Julian Date
(${\rm JD} - 2400000.5$) but is identical to it in all other respects.
\keyw{MJD-OBS} does not have a version code since there can only be one time
of observation.  Following the year-2000 conventions for \keyw{DATE} keywords
(Bunclark \& Rots \cite{kn:BR}), this time refers by default to the start of
the observation unless another interpretation is clearly explained in the
comment field.  In the present case the distinction is not important.  We
leave it to future agreements to clarify systems of time measurement and other
matters related to time.

The combination of \CTYPE{ia}, \RADESYS{a}, and \EQUINOX{a} define the
coordinate system of the \CRVAL{ia} and of the celestial coordinates that
result from the sequence of transformations summarized by
Fig.~\ref{fig:Process}.  However, \keyv{FK4} coordinates are not strictly
spherical since they include a contribution from the elliptic terms of
aberration, the so-called ``e-terms'' which amount to $\leq 343$ milliarcsec.
Strictly speaking, therefore, a map obtained from, say, a radio synthesis
telescope, should be regarded as \keyv{FK4-NO-E} unless it has been
appropriately resampled or a distortion correction provided (Paper IV).  In
common usage, however, \CRVAL{ia} for such maps is usually given in \keyv{FK4}
coordinates.  In doing so, the e-terms are effectively corrected to first
order only.


\section{Alternate FITS image representations}
\label{sec:bintab}


\subsection{Random groups visibility data}

The random-groups extension to FITS (Greisen \& Harten \cite{kn:GH}) has been
used to transmit interferometer fringe visibility sample data.  It has been
customary among users of this format to convey the suggested projection type
as the last four characters of the random parameter types (\keyi{PTYPE}{n}) of
the Fourier plane coordinates $(u,v,w)$.  Any rotation of these coordinates
was carried, however, by \CROTA{i} associated with the one-pixel array axis
used to convey the field declination.  Within the new convention users of
random groups for visibility data should be prepared to read, carry forward,
and use as needed, the new keywords \PC{i}{ja}, \CD{i}{ja}, \LONPOLE{a},
\LATPOLE{a}, \PV{i}{ja}, \keyw{MJD-OBS}, \EQUINOX{a}, and \RADESYS{a}.  In
particular, any rotation of the $(u,v)$ coordinate system should be
represented via \PC{i}{ja} or \CD{i}{ja} (but not both) attached to the two
degenerate axes used to convey the celestial coordinates of the field center.


\subsection{Pixel list and image array column elements}

Paper~I defined the coordinate keywords used to describe a multi-dimensional
image array in a single element of a FITS binary table and a tabulated list of
pixel coordinates in a FITS ASCII or binary table.  In this section we extend
this to the keywords specific to celestial coordinates.


\subsubsection{Keyword naming convention}

\begin{table}
   \caption[]{Alternate coordinate keywords; the data type of the alternate
      keyword matches that of the primary keyword}.
   \protect\begin{tabular}{llll}
      \hline
      \noalign{\smallskip}
      Keyword         & Primary        & BINTABLE        & Pixel \\
      Description     & Array          & Array           & List  \\
      \noalign{\smallskip}
      \hline
      \noalign{\smallskip}
      Coord Rotation  & \LONPOLE{a}    & \keyi{LONP}{na} & \keyi{LONP}{na} \\
      Coord Rotation  & \LATPOLE{a}    & \keyi{LATP}{na} & \keyi{LATP}{na} \\
      Coord Epoch     & \EQUINOX{a}    & \keyi{EQUI}{na} & \keyi{EQUI}{na} \\
      Date of Obs     & \keyw{MJD-OBS} & \keyi{MJDOB}{n} & \keyi{MJDOB}{n} \\
      Reference Frame & \RADESYS{a}    & \keyi{RADE}{na} & \keyi{RADE}{na} \\
      \noalign{\smallskip}
      \hline
   \end{tabular}
   \label{ta:bintdef}
\end{table}

Table~\ref{ta:bintdef} lists the corresponding set of coordinate system
keywords for use with each type of FITS image representation.

The data type of the alternate keyword matches that of the primary keyword and
the allowed values are the same.  The following notes apply to the naming
conventions used in Table~\ref{ta:bintdef}:

\begin{itemize}
\item
   $a$ is a 1-character coordinate version code and may be blank (for the
   primary version of the coordinate description) or any single uppercase
   character from \keyv{A} through \keyv{Z}\@.
\item
   $n$ is an integer table column number without any leading zeros (1 -- 999).
\end{itemize}

When using the \keyv{BINTABLE} image array format (Cotton et al.,
\cite{kn:CTP}), if the table only contains a single image column or if there
are multiple image columns but they all have the same value for any of the
keywords in Table~\ref{ta:bintdef} then the simpler form of the keyword name,
as used for primary arrays, may be used.  For example, if all the images in
the table have the same epoch then one may use a single \EQUINOX{a} keyword
rather than multiple \keyi{EQUI}{na} keywords.  The other keywords, however,
must always be specified using the more complex keyword name with the column
number suffix and the axis number prefix.

In principle, more than one pixel list image can be stored in a single FITS
table by defining more than one pair of $p_1$ and $p_2$ pixel coordinate
columns.  Under the convention defined here, however, all the images must
share the same values for the  keywords in Table~\ref{ta:bintdef}.

Example binary table and pixel list headers are given in
Sect.~\ref{sec:interp2}


\section{Spherical map projections}
\label{sec:projections}

\begin{figure*}
   \if\dofig F
      \vspace{187pt}
   \else
      \centerline{\putfig{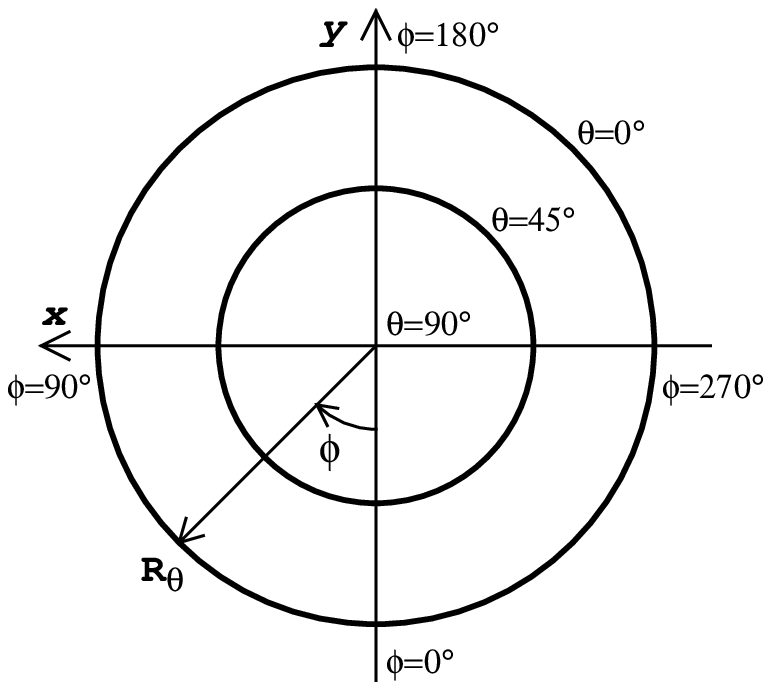}{187pt} \hfil
                  \putfig{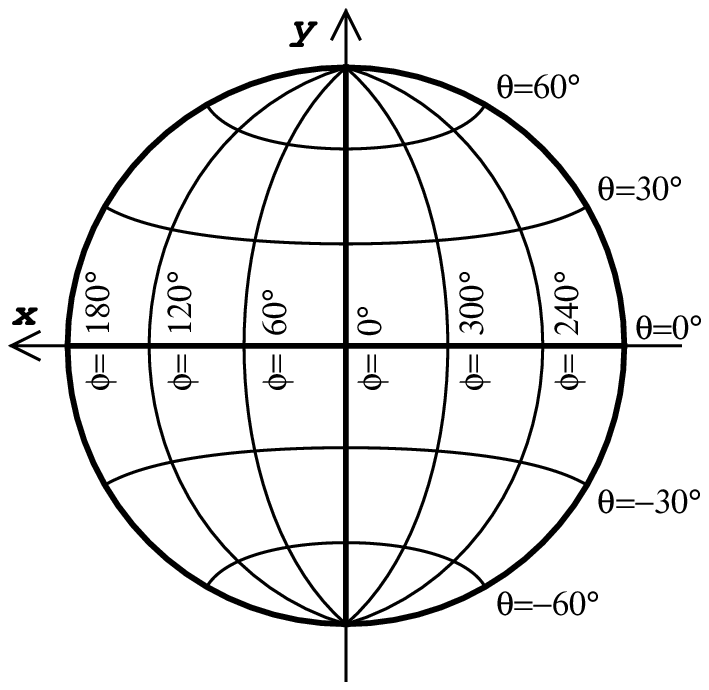}{187pt}}
   \fi
   \caption[]{(Left) native coordinate system with its pole at the reference
      point, i.e.\ $(\phi_0,\theta_0) = (0,90\degr)$, and (right) with the
      intersection of the equator and prime meridian at the reference point
      i.e.\ $(\phi_0,\theta_0) = (0,0)$.}
\label{fig:Native}
\end{figure*}

In this section we present the transformation equations for all spherical map
projections likely to be of use in astronomy.  Many of these such as the
gnomonic, orthographic, zenithal equidistant, Sanson-Flamsteed, Hammer-Aitoff
and COBE quadrilateralized spherical cube are in common use.  Others with
special properties such as the stereographic, Mercator, and the various equal
area projections could not be excluded.  A selection of the conic and
polyconic projections, much favored by cartographers for their minimal
distortion properties, has also been included.  However, we have omitted
numerous other projections which we considered of mathematical interest only.
Evenden (\cite{kn:Eve}) presents maps of the Earth for 73 different
projections, although without mathematical definition, including most of those
described here.  These are particularly useful in judging the distortion
introduced by the various projections.  Snyder (\cite{kn:Sny}) provides
fascinating background material on the subject; historical footnotes in this
paper, mainly highlighting astronomical connections, are generally taken from
this source.  It should be evident from the wide variety of projections
described here that new projections could readily be accommodated, the main
difficulty being in obtaining general recognition for them from the FITS
community.

Cartographers have often given different names to special cases of a class of
projection.  This applies particularly to oblique projections which, as we
have seen in Sect.~\ref{sec:basics}, the current formalism handles in a
general way.  While we have tended to avoid such special cases, the gnomonic,
stereographic, and orthographic projections, being specializations of the
zenithal perspective projection, are included for conformance with the AIPS
convention.  It is also true that zenithal and cylindrical projections may be
thought of as special cases of conic projections (see
Sect.~\ref{sec:conics}).  However, the limiting forms of the conic equations
tend to become intractable and infinite-valued projection parameters may be
involved.  Even when the conic equations don't have singularities in these
limits it is still likely to be less efficient to use them than the simpler
special-case equations.  Moreover, we felt that it would be unwise to disguise
the true nature of simple projections by implementing them as special cases of
more general ones.  In the same vein, the cylindrical equal area projection,
being a specialization of the cylindrical perspective projection, stands on
its own right, as does the Sanson-Flamsteed projection which is a limiting
case of Bonne's projection.  A list of aliases is provided in
Appendix~\ref{apx:aliases}, Table~\ref{ta:aliases}.

The choice of a projection often depends on particular special properties that
it may have.  Certain {\em equal area} projections (also known as
{\em authalic}, {\em equiareal}, {\em equivalent}, {\em homalographic},
{\em homolographic}, or {\em homeotheric}) have the property that equal areas
on the sphere are projected as equal areas in the plane of projection.  This
is obviously a useful property when surface density must be preserved.
{\em Conformality} is a property which applies to points in the plane of
projection which are locally distortion-free.  Practically speaking, this
means that the projected meridian and parallel through the point intersect at
right angles and are equiscaled.  A projection is said to be {\em conformal}
or {\em orthomorphic} if it has this property at all points.  Such a
projection cannot be equal area.  It must be stressed that conformality is a
local property, finite regions in conformal projections may be very severely
distorted.  A number of projections have other special properties and these
will be noted for each.

Maps of the Earth are conventionally displayed with terrestrial latitude
increasing upwards and longitude to the right, i.e.\ north up and east to the
{\em right}, as befits a sphere seen from the outside.  On the other hand,
since the celestial sphere is seen from the inside, north is conventionally up
and east to the {\em left}.  The AIPS convention arranged that celestial
coordinates at points near the reference point should be calculable to first
order via the original linear prescription of Wells et al.\ (\cite{kn:WGH}),
i.e.\ $(\alpha,\delta)\approx(\alpha_0,\delta_0)+(x,y)$.  Consequently, the
\CDELT{ia} keyword value associated with the right ascension was
{\em negative} while that for the declination was {\em positive}.  The
handedness of the $(x,y)$ coordinates as calculated by the AIPS convention
equivalent of Eq.~(\ref{eq:px}) is therefore opposite to that of the
$(p_1,p_2)$ pixel coordinates.

In accordance with the image display convention of Paper~I we think of the
$p_1$-pixel coordinate increasing to the right with $p_2$ increasing upwards,
i.e.\ a right-handed system.  This means that the $(x,y)$ coordinates must be
{\em left-handed} as shown in Fig.~\ref{fig:Native}.  Note, however, that the
approximation $(\alpha,\delta)\approx(\alpha_0,\delta_0)+(x,y)$ cannot hold
unless 1) $(\phi_0,\theta_0)$, and hence $(\alpha_0,\delta_0)$, do actually
map to the reference point (Sect.~\ref{sec:refpoint}), 2) $\phi\sub{p}$
assumes its default value (Sect.~\ref{sec:compare}), and 3) the projection is
scaled true at the reference point (some are not as discussed in
Sect.~\ref{sec:choice}).  Figure~\ref{fig:Native} also illustrates the
orientation of the native coordinate system with respect to the $(x,y)$
coordinate system for the two main cases.

Cartographers, for example Kellaway (\cite{kn:Kel}), think of spherical
projections as being a projection of the surface of a sphere onto a plane,
this being the {\em forward} direction; the deprojection from plane back to
sphere is thus the {\em inverse} or {\em reverse} direction.  However, this is
at variance with common usage in FITS where the transformation from pixel
coordinates to world coordinates is considered the forward direction.  We take
the cartographic view in this section as being the natural one and trust that
any potential ambiguity may readily be resolved by context.

\begin{figure*}
   \centerline{\putfig{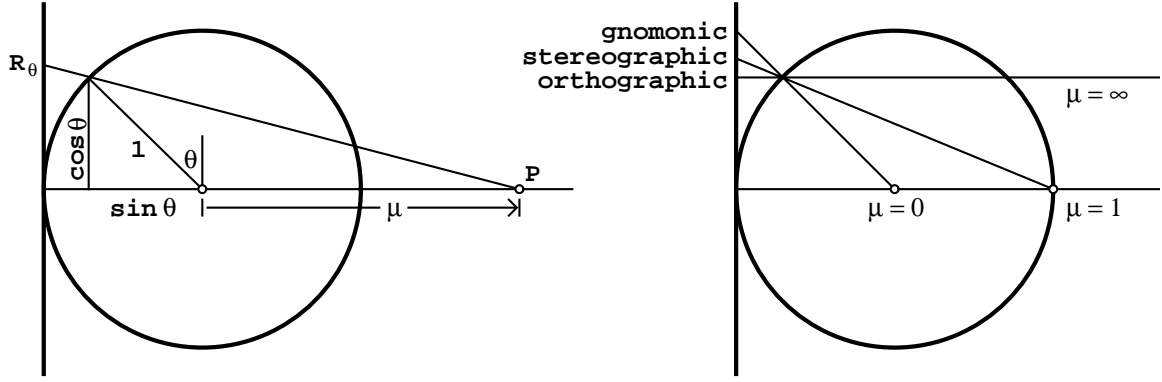}{145pt}}
   \caption[]{(Left) geometry of the zenithal perspective projections, the
      point of projection at $\mathbf{P}$ is $\mu$ spherical radii from the
      center of the sphere; (right) the three important special cases.  This
      diagram is at 3:2 scale compared to the graticules of
      Figs.~\ref{fig:TAN}, \ref{fig:STG} and \ref{fig:SIN}.}
   \label{fig:Zenithal}
\end{figure*}

The requirement stated in Sect.~\ref{sec:intro} that $(x,y)$ coordinates in
the plane of projection be measured in ``degrees'' begs clarification.
Spherical projections are usually defined mathematically in terms of a scale
factor, $r_0$, known as the ``radius of the generating sphere''.  However, in
this work $r_0$ is set explicitly to $180\degr / \pi$ in order to maintain
backwards compatibility with the AIPS convention.  This effectively sets the
circumference of the generating sphere to $360\degr$ so that arc length is
measured naturally in degrees (rather than radians as for a unit sphere).
However, this true angular measure on the generating sphere becomes distorted
when the sphere is projected onto the plane of projection.  So while the
``degree'' units of $r_0$ are notionally carried over by conventional
dimensional analysis to the $(x,y)$ they no longer represent a true angle
except near the reference point (for most projections)}.

In addition to the $(x,y)$ coordinates, the native spherical coordinates,
$(\phi,\theta)$, celestial coordinates, $(\alpha,\delta)$, and all other
angles in this paper are measured in degrees.  In the equations given below,
the arguments to all trigonometric functions are in degrees and all inverse
trigonometric functions return their result in degrees.  Whenever a conversion
between radians and degrees is required it is shown explicitly.  All of the
graticules presented in this section have been drawn to the same scale in
$(x,y)$ coordinates in order to represent accurately the exaggeration and
foreshortening found in some projections.  It will also be apparent that since
FITS image planes are rectangular and the boundaries of many projections are
curved, there may sometimes be cases when the FITS image must contain pixels
that are outside the boundary of the projection.  These pixels should be
blanked correctly and geometry routines should return a sensible error code to
indicate that their celestial coordinates are undefined.


\subsection{Zenithal (azimuthal) projections}
\label{sec:zenithals}

{\em Zenithal} or {\em azimuthal} projections all map the sphere directly onto
a plane.  The native coordinate system is chosen to have the polar axis
orthogonal to the plane of projection at the reference point as shown on the
left side of Fig.~\ref{fig:Native}.  Meridians of native longitude are
projected as uniformly spaced rays emanating from the reference point and the
parallels of native latitude are mapped as concentric circles centered on the
same point.  Since all zenithal projections are constructed with the pole of
the native coordinate system at the reference point we set
\begin{equation}
   (\phi_0, \theta_0)\sub{zenithal} = (0,90\degr) \, .
\end{equation}

Zenithal projections are completely specified by defining the radius as a
function of native latitude, $\Rt$.  Rectangular Cartesian coordinates,
$(x,y)$, in the plane of projection as defined by Eq.~(\ref{eq:px}), are then
given by
\begin{eqnarray}
      x & = & \SP\Rt\sin\phi \, , \label{eq:azx} \\
      y & = &   -\Rt\cos\phi \, , \label{eq:azy}
\end{eqnarray}
which may be inverted as
\begin{eqnarray}
   \phi & = & \arg\,(-y, x)   \, , \label{eq:azphi} \\
   \Rt  & = & \sqrt{x^2 + y^2} \, . \label{eq:azrt}
\end{eqnarray}


\subsubsection{\keyv{AZP}: zenithal perspective}
\label{sec:AZP}

Zenithal (azimuthal) perspective projections are generated from a point and
carried through the sphere to the plane of projection as illustrated in
Fig.~\ref{fig:Zenithal}.  We need consider only the case where the plane of
projection is tangent to the sphere at its pole; the projection is simply
rescaled if the plane intersects some other parallel of native latitude.  If
the source of the projection is at a distance $\mu$ spherical radii from the
center of the sphere with $\mu$ increasing in the direction away from the
plane of projection, then it is straightforward to show that
\begin{equation}
   \Rt = \rd \frac{(\mu+1) \cos\theta}{\mu+\sin\theta} \, ,
   \label{eq:AZPRt}
\end{equation}
with $\mu \neq -1$ being the only restriction.  When $\Rt$ is given
Eq.~(\ref{eq:AZPRt}) has two solutions for $\theta$, one for each side of the
sphere.  The following form of the inverse equation always gives the planeward
solution for any $\mu$
\begin{equation}
   \theta = \arg\,(\rho, 1) - \sin^{-1} \left(
            \frac{\rho\mu}{\sqrt{\rho^2 + 1}} \right) \, ,
\end{equation}
where
\begin{equation}
   \rho = \dr \frac{\Rt}{\mu + 1} \, .
\end{equation}
For $|\mu| \neq 1$ the sphere is divided by a native parallel at latitude
$\theta\sub{x}$ into two unequal segments that are projected in superposition:
\begin{equation}
   \theta\sub{x} = \left\{ \begin{array}{ll}
                              \sin^{-1}(-1/\mu) & \mbox{\ldots $|\mu| > 1$} \\
                              \sin^{-1}(-\mu)   & \mbox{\ldots $|\mu| < 1$}
                           \end{array} \right. \, . \label{eq:AZPTx}
\end{equation}
For $|\mu| > 1$, the projection is bounded and both segments are projected in
the correct orientation, whereas for $|\mu| \leq 1$ the projection is
unbounded and the anti-planewards segment is inverted.

A {\em near-sided} perspective projection may be obtained with $\mu < -1$.
This correctly represents the image of a sphere, such as a planet, when viewed
from a distance $|\mu|$ times the planetary radius.  The coordinates of the
reference point may be expressed in planetary longitude and latitude,
$(\lambda,\beta)$.  Also, the signs of the relevant \CDELT{ia} may be chosen
so that longitude increases as appropriate for a sphere seen from the outside
rather than from within.

\begin{figure*}
   \centerline{\putfig{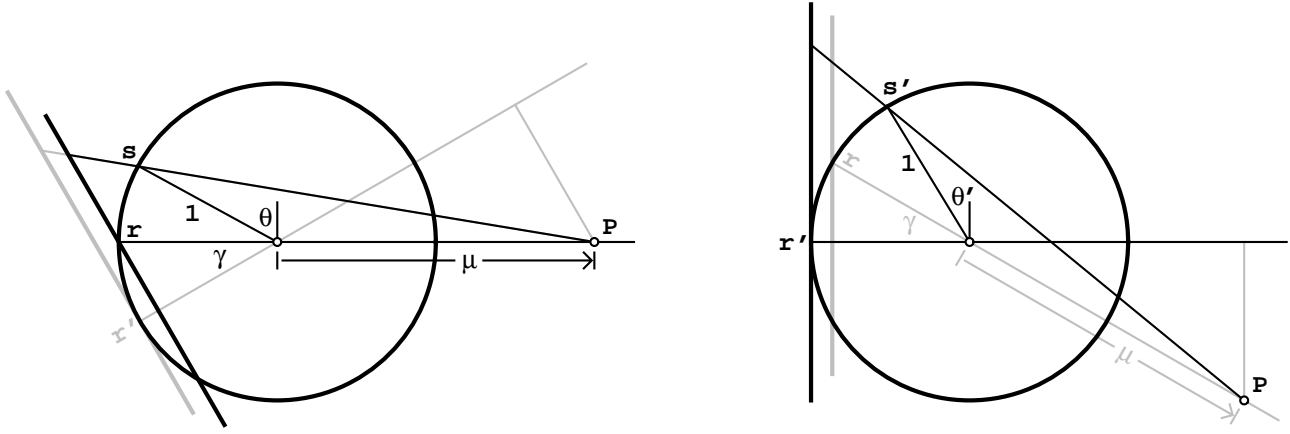}{165pt}}
   \caption[]{Alternate geometries of slant zenithal perspective projections
      with $\mu = 2$ and $\gamma = 30\degr$: (left) tilted plane of projection
      (\keyv{AZP}), (right) displaced point of projection $\mathbf{P}$
      (\keyv{SZP}).  Grey lines in each diagram indicate the other point of
      view.  They differ geometrically only by a scale factor, the effect of
      translating the plane of projection.  Each projection has its native
      pole at the reference point, $\mathbf{r}$ and $\mathbf{r'}$, but these
      are geometrically different points.  Thus the native latitudes $\theta$
      and $\theta'$ of the geometrically equivalent points $\mathbf{s}$ and
      $\mathbf{s'}$ differ.  Consequently the two sets of projection equations
      have a different form.  This diagram is at 3:2 scale compared to the
      graticules of Figs.~\ref{fig:AZP} and \ref{fig:SZP}.}
   \label{fig:SlantZen}
\end{figure*}

It is particularly with regard to planetary mapping that we must generalize
\keyv{AZP} to the case where the plane of projection is tilted with respect to
the axis of the generating sphere, as shown on the left side of
Fig.~\ref{fig:SlantZen}.  It can be shown (Sect.~\ref{sec:construct1}) that
this geometry is appropriate for spacecraft imaging with non-zero look-angle,
$\gamma$, the angle between the camera's optical axis and the line to the
center of the planet.

Such {\em slant} zenithal perspective projections are not radially symmetric
and their projection equations must be expressed directly in terms of $x$ and
$y$:
\begin{eqnarray}
   x & = & \SP R \sin\phi \, ,            \label{eq:AZPx} \\
   y & = &   - R \sec\gamma \cos\phi \, , \label{eq:AZPy}
\end{eqnarray}
where
\begin{equation}
   R = \rd \frac{(\mu+1) \cos\theta}
                {(\mu+\sin\theta) + \cos\theta \cos\phi \tan\gamma} \,
\end{equation}
is a function of $\phi$ as well as $\theta$.  Equations~(\ref{eq:AZPx}) and
(\ref{eq:AZPy}) reduce to the non-slant case for $\gamma = 0$.  The inverse
equations are
\begin{eqnarray}
   \phi   & = & \arg\,(-y \cos\gamma, x)   \, , \\
   \theta & = & \left\{ \begin{array}{ll}
                           \psi - \omega \\
                           \psi + \omega + 180\degr
                        \end{array} \right. \, ,
\end{eqnarray}
where
\begin{eqnarray}
     \psi & = & \arg\,(\rho, 1) \, , \\
   \omega & = & \sin^{-1} \left( \frac{\rho\mu}{\sqrt{\rho^2 + 1}}
                          \right) \, , \\
     \rho & = & \frac{R}{\rd (\mu + 1) + y \sin\gamma} \, , \\
        R & = & \sqrt{x^2 + y^2\cos^2\gamma}           \, .
\end{eqnarray}
For $|\mu| < 1$ only one of the solutions for $\theta$ will be valid,
i.e.\ lie in the range $[-90\degr, 90\degr]$ after normalization.  Otherwise
there will be two valid solutions; the one closest to $90\degr$ should be
chosen.

\begin{figure}
   \centerline{\putfig{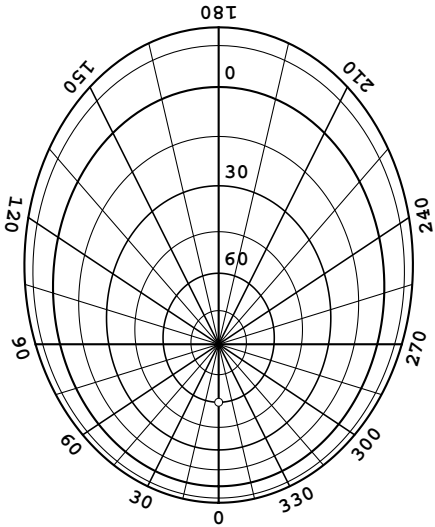}{210pt}}
   \caption[]{Slant zenithal perspective (\keyv{AZP}) projection with
      $\mu = 2$ and $\gamma = 30\degr$ which corresponds to the left-hand side
      of Fig.~\ref{fig:SlantZen}.  The reference point of the corresponding
      \keyv{SZP} projection is marked at $(\phi,\theta) = (0,60\degr)$.}
   \label{fig:AZP}
\end{figure}

With $\gamma \neq 0$ the projection is not scaled true at the reference point.
In fact the $x$ scale is correct but the $y$ scale is magnified by
$\sec\gamma$, thus stretching parallels of latitude near the pole into ellipses
(see Fig.~\ref{fig:AZP}).  This also shows the native meridians projected as
rays emanating from the pole.  For constant $\theta$, each parallel of native
latitude defines a cone with apex at the point of projection.  This cone
intersects the tilted plane of projection in a conic section.
Equations~(\ref{eq:AZPx}) and (\ref{eq:AZPy}) reduce to the parametric
equations of an ellipse, parabola, or hyperbola; the quantity
\begin{equation}
   C = (\mu + \sin\theta)^2 - \tan^2\gamma \cos^2\theta
\end{equation}
determines which:
\begin{equation}
   \begin{array}{llll}
      C & > & 0 & {\rm ...ellipse},   \\
      C & = & 0 & {\rm ...parabola},  \\
      C & < & 0 & {\rm ...hyperbola}.
   \end{array}
\end{equation}
If $|\mu \cos\gamma| \le 1$ then the open conic sections are possible and
$C = 0$ when
\begin{equation}
   \theta = \pm \gamma - \sin^{-1}(\mu \cos\gamma) \, .
\end{equation}
For $C > 0$ the eccentricity of the ellipse is a function of $\theta$, as is
the offset of its center in $y$.

Definition of the perimeter of the projection is more complicated for the
slant projection than the orthogonal case.  As before, for $|\mu| > 1$ the
sphere is divided into two unequal segments that are projected in
superposition.  The boundary between these two segments is what would be seen
as the limb of the planet in spacecraft photography.  It corresponds to native
latitude
\begin{equation}
   \theta\sub{x} =  \sin^{-1}(-1/\mu) \, ,
\end{equation}
which is projected as an ellipse, parabola, or hyperbola for
$|\mu \cos\gamma|$ greater than, equal to, or less than $1$ respectively.

In general, for $|\mu \cos\gamma| > 1$, the projection is bounded, otherwise
it is unbounded.  However, the latitude of divergence is now a function of
$\phi$:
\begin{equation}
   \theta_\infty = \left\{ \begin{array}{ll}
                               \psi - \omega \\
                               \psi + \omega + 180\degr
                            \end{array} \right. \, , \label{eq:AZPTinf}
\end{equation}
where
\begin{eqnarray}
   \psi   & = &   - \tan^{-1}( \, \tan\gamma \cos\phi ) \, , \\
   \omega & = & \SP \sin^{-1} \left(
                                \frac{\mu}{\sqrt{1 + \tan^2\gamma \cos^2\phi}}
                              \right) \, .
\end{eqnarray}
Zero, one, or both of the values of $\theta_\infty$ given by
Eq.~(\ref{eq:AZPTinf}) may be valid, i.e.\ lie in the range
$[-90\degr, 90\degr]$ after normalization.

The FITS keywords \PVi{1} and \PVi{2}, attached to {\em latitude}
coordinate~{\it i}, will be used to specify, respectively, $\mu$ in spherical
radii and $\gamma$ in degrees, both with default value 0.


\subsubsection{\keyv{SZP}: slant zenithal perspective}
\label{sec:SZP}

While the generalization of the \keyv{AZP} projection to tilted planes of
projection is useful for certain applications it does have a number of
drawbacks, in particular, unequal scaling at the reference point.

\begin{figure}
   \centerline{\putfig{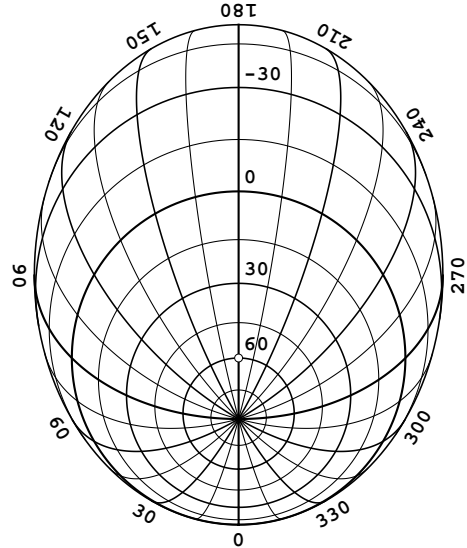}{210pt}}
   \caption[]{Slant zenithal perspective (\keyv{SZP}) projection with
      $\mu = 2$ and $(\phi\sub{c},\theta\sub{c}) = (180\degr,60\degr)$, which
      corresponds to the right-hand side of Fig.~\ref{fig:SlantZen}.  The
      reference point of the corresponding \keyv{AZP} projection is marked at
      $(\phi,\theta) = (180\degr,60\degr)$.}
   \label{fig:SZP}
\end{figure}

Figure~\ref{fig:SlantZen} shows that moving the point of projection,
$\mathbf{P}$, off the axis of the generating sphere is equivalent, to within a
scale factor, to tilting the plane of projection.  However this approach has
the advantage that the plane of projection remains tangent to the sphere.
Thus the projection is conformal at the native pole as can be seen by the
circle around the native pole in Fig.~\ref{fig:SZP}.  It is also quite
straightforward to formulate the projection equations with $\mathbf{P}$ offset
in $x$ as well as $y$.

It is interesting to note that this slant zenithal perspective (\keyv{SZP})
projection also handles the case that corresponds to $\gamma = 90\degr$ in
\keyv{AZP}\@.  \keyv{AZP} fails in this extreme since $\mathbf{P}$ falls in
the plane of projection -- effectively a scale factor of zero is applied to
\keyv{AZP} over the corresponding \keyv{SZP} case.  One of the more important
aspects of \keyv{SZP} is the application of its limiting case with
$\mu = \infty$ in aperture synthesis radio astronomy as discussed in
Sect.~\ref{sec:SIN}.  One minor disadvantage is that the native meridians are
projected as curved conic sections rather than straight lines.

If the Cartesian coordinates of $\mathbf{P}$ measured in units of the
spherical radius are $(x\sub{p},y\sub{p},z\sub{p})$, then
\begin{eqnarray}
  x & = & \SP \rd \frac{z\sub{p} \cos\theta \sin\phi -
                        x\sub{p} (1 - \sin\theta)}
                       {z\sub{p} - (1 - \sin\theta)} \, , \label{eq:SZPx} \\
  y & = &   - \rd \frac{z\sub{p} \cos\theta \cos\phi +
                        y\sub{p} (1 - \sin\theta)}
                       {z\sub{p} - (1 - \sin\theta)} \, . \label{eq:SZPy}
\end{eqnarray}
To invert these equations, compute $\theta$ first via
\begin{eqnarray}
  \theta & = & \sin^{-1} \left(
                            \frac{-b \pm \sqrt{b^2 - ac}}{a}
                         \right) \, , \label{eq:SZPtheta}
\end{eqnarray}
where
\begin{eqnarray}
   a  & = & X'^2 + Y'^2 + 1             \, , \\
   b  & = & X'(X-X') + Y'(Y-Y')         \, , \\
   c  & = & (X - X')^2 + (Y - Y')^2 - 1 \, , \\
   (X, Y)  & = & \dr (x, y)             \, , \\
   (X',Y') & = & (X - x\sub{p}, Y - y\sub{p}) / z\sub{p} \, .
                 \label{eq:SZPXYd}
\end{eqnarray}
Choose $\theta$ closer to $90\degr$ if Eq.~(\ref{eq:SZPtheta}) has two valid
solutions; then $\phi$ is given by
\begin{equation}
   \phi = \arg\,(-(Y - Y'(1 - \sin\theta)), \, X - X'(1 - \sin\theta)) \, .
   \label{eq:SZPphi}
\end{equation}

The Cartesian coordinates of $\mathbf{P}$ are simply related to the parameters
used in \keyv{AZP}\@.  If $\mu$ is the distance of $\mathbf{P}$ from the
center of the sphere $\mathbf{O}$, and the line through $\mathbf{P}$ and
$\mathbf{O}$ intersects the sphere at $(\phi\sub{c},\theta\sub{c})$ on the
planewards side (point $\mathbf{r}$ in Fig.~\ref{fig:SlantZen}, right), then
$\gamma = 90\degr - \theta\sub{c}$ and
\begin{eqnarray}
   x\sub{p} & = &   - \mu \cos\theta\sub{c} \sin\phi\sub{c} \, , \\
   y\sub{p} & = & \SP \mu \cos\theta\sub{c} \cos\phi\sub{c} \, , \\
   z\sub{p} & = & \SP \mu \sin\theta\sub{c} + 1 \, ,
\end{eqnarray}
where $\mu$ is positive if $\mathbf{P}$ lies on the opposite side of
$\mathbf{O}$ from $\mathbf{r}$ and negative otherwise.  For a non-degenerate
projection we require $z\sub{p} \neq 0$ and this is the only constraint on the
projection parameters.

For $|\mu| > 1$ the sphere is divided into two unequal segments that are
projected in superposition.  The limb is defined by computing the native
latitude $\theta\sub{x}$ as a function of $\phi$
\begin{equation}
   \theta\sub{x} = \left\{ \begin{array}{ll}
                              \psi - \omega \\
                              \psi + \omega + 180
                           \end{array} \right. \, , \label{eq:SZPTx}
\end{equation}
where
\begin{eqnarray}
   \psi   & = & \arg\,(\rho, \sigma) \, , \\
   \omega & = & \sin^{-1} \left( \frac{1}{\sqrt{\rho^2 + \sigma^2}}
                          \right)    \, , \\
   (\rho, \sigma) & = & (z\sub{p} - 1, \,
                         x\sub{p} \sin\phi - y\sub{p} \cos\phi) \, , \\
                  & = & (\mu \sin\theta\sub{c}, \,
                         -\mu \cos\theta\sub{c} \cos(\phi - \phi\sub{c}) \, .
\end{eqnarray}
Zero, one, or both of the values of $\theta\sub{x}$ given by
Eq.~(\ref{eq:SZPTx}) may be valid, i.e.\ lie in the range
$[-90\degr, 90\degr]$ after normalization.  A second boundary constraint
applies if $|1 - z\sub{p}| \le 1$ in which case the projection diverges at
native latitude:
\begin{equation}
   \theta_\infty = \sin^{-1}(1 - z\sub{p}) \, .
\end{equation}

The FITS keywords \PVi{1}, \PVi{2}, and \PVi{3}, attached to {\em latitude}
coordinate~{\it i}, will be used to specify, respectively, $\mu$ in
spherical radii with default value 0, $\phi\sub{c}$ in degrees with default
value 0, and $\theta\sub{c}$ in degrees with default value $90\degr$.


\subsubsection{\keyv{TAN}: gnomonic}
\label{sec:TAN}

\begin{figure}
   \centerline{\putfig{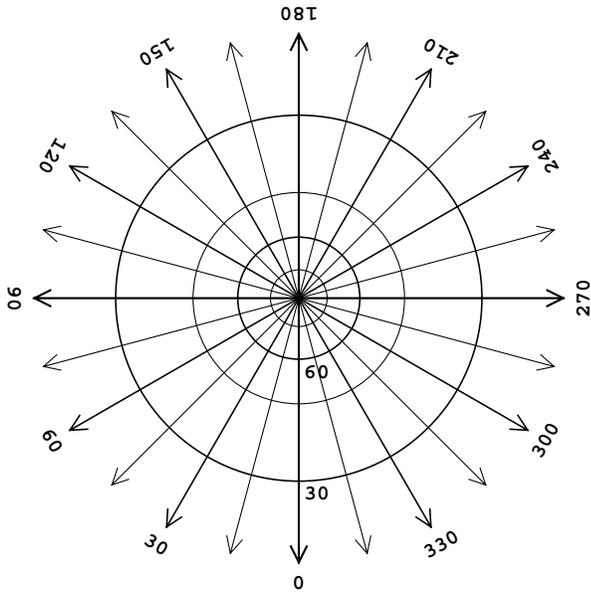}{230pt}}
   \caption[]{Gnomonic (\keyv{TAN}) projection; diverges at $\theta = 0$.}
   \label{fig:TAN}
\end{figure}

The zenithal perspective projection with $\mu = 0$, the gnomonic
projection\footnote{The gnomonic projection is the oldest known, dating to
Thales of Miletus (ca.\,624-547\,{\sc b.c.}).  The stereographic and
orthographic date to Hipparchus (ca.\,190-after 126\,{\sc b.c.}).}, is widely
used in optical astronomy and was given its own code within the AIPS
convention, namely \keyv{TAN}\footnote{Referring to the dependence of $\Rt$ on
the angular separation between the tangent point and field point, i.e.\ the
native {\em co}-latitude.}.  For $\mu = 0$, Eq.~(\ref{eq:AZPRt}) reduces to
\begin{equation}
   \Rt = \rd \cot \theta \, , \label{eq:TANRt}
\end{equation}
with inverse
\begin{eqnarray}
   \theta & = & \tan^{-1} \left( \frac{180\degr}{\pi\Rt} \right) \, .
   \label{eq:TANtheta}
\end{eqnarray}
The gnomonic projection is illustrated in Fig.~\ref{fig:TAN}.  Since the
projection is from the center of the sphere, all great circles are projected
as straight lines.  Thus, the shortest distance between two points on the
sphere is represented as a straight line interval, which, however, is not
uniformly divided.  The gnomonic projection diverges at $\theta = 0$, but one
may use a gnomonic projection onto the six faces of a cube to display the
whole sky.  See Sect.~\ref{sec:TSC} for details.


\subsubsection{\keyv{STG}: stereographic}
\label{sec:STG}

\begin{figure}
   \centerline{\putfig{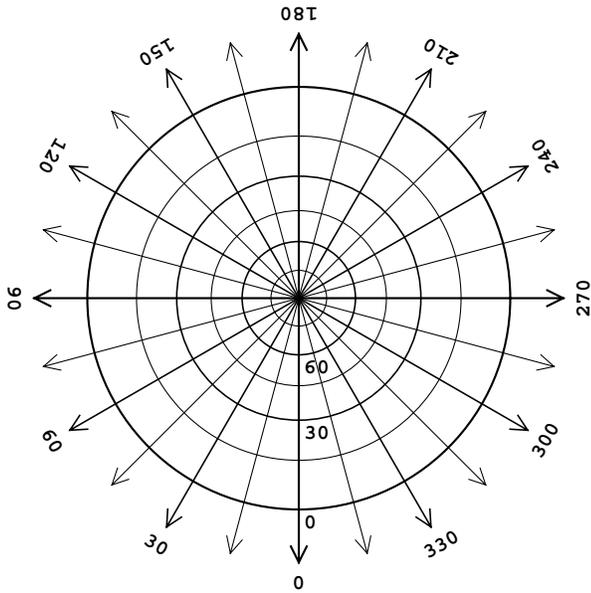}{230pt}}
   \caption[]{Stereographic (\keyv{STG}) projection; diverges at
      $\theta = -90\degr$.}
   \label{fig:STG}
\end{figure}

The stereographic projection, the second important special case of the
zenithal perspective projection defined by the AIPS convention, has $\mu = 1$,
for which Eq.~(\ref{eq:AZPRt}) becomes
\begin{eqnarray}
      \Rt & = & \rd \frac{2\cos\theta}{1 + \sin\theta} \, , \\
          & = & {\frac{360\degr}{\pi}}
                   \tan \left( \frac{90\degr - \theta}{2} \right)
                   \nonumber \, ,
\end{eqnarray}
with inverse
\begin{equation}
   \theta = 90\degr - 2 \tan^{-1} \left( \frac{\pi\Rt}{360\degr} \right) \, .
\end{equation}
The stereographic projection illustrated in Fig.~\ref{fig:STG} is the
conformal (orthomorphic) zenithal projection\footnote{First noted by
astronomer Edmond~Halley (1656-1742).}, everywhere satisfying the isoscaling
requirement
\begin{equation}
   \frac{\partial R_\theta}{\partial \theta} =
      \frac{-\pi R_\theta}{180\degr \cos\theta} \, .
\end{equation}
It also has the amazing property that it maps all circles on the sphere to
circles in the plane of projection, although concentric circles on the sphere
are not necessarily concentric in the plane of projection.  This property made
it the projection of choice for Arab astronomers in constructing astrolabes.


\subsubsection{\keyv{SIN}: slant orthographic}
\label{sec:SIN}

\begin{figure}
   \if\dofig F
      \vspace{215pt}
   \else
      \centerline{\putfig{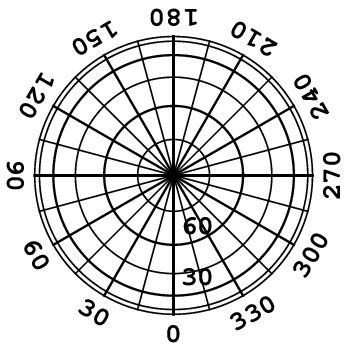}{100pt}}
      \centerline{\putfig{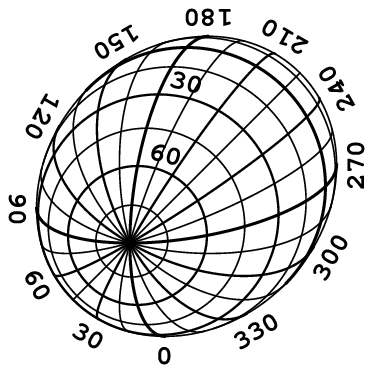}{110pt} \hss
                  \putfig{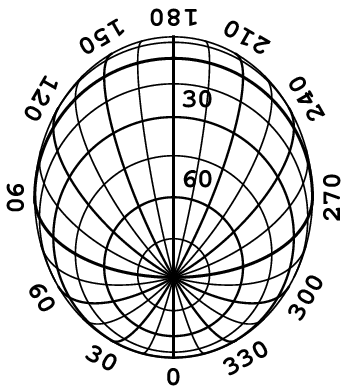}{115pt}}
   \fi
   \caption[]{Slant orthographic (\keyv{SIN}) projection: (top) the
      orthographic projection, $(\xi,\eta) = (0,0)$, north and south sides
      begin to overlap at $\theta = 0$; (bottom left)
      $(\phi_c,\theta_c) = (225\degr,60\degr)$,
      i.e.\ $(\xi,\eta) = (-1/\sqrt{6}, 1/\sqrt{6})$;
      (bottom right) projection appropriate for an east-west array observing
      at $\delta_0 = 60\degr$, $(\phi_c,\theta_c) = (180\degr,60\degr)$,
      $(\xi,\eta) = (0, 1/\sqrt{3})$.}
   \label{fig:SIN}
\end{figure}

The zenithal perspective projection with $\mu = \infty$, the orthographic
projection, is illustrated in the upper portion of Fig.~\ref{fig:SIN} (at
consistent scale).  It represents the visual appearance of a sphere, e.g.\ a
planet, when seen from a great distance.

\begin{figure}
   \centerline{\putfig{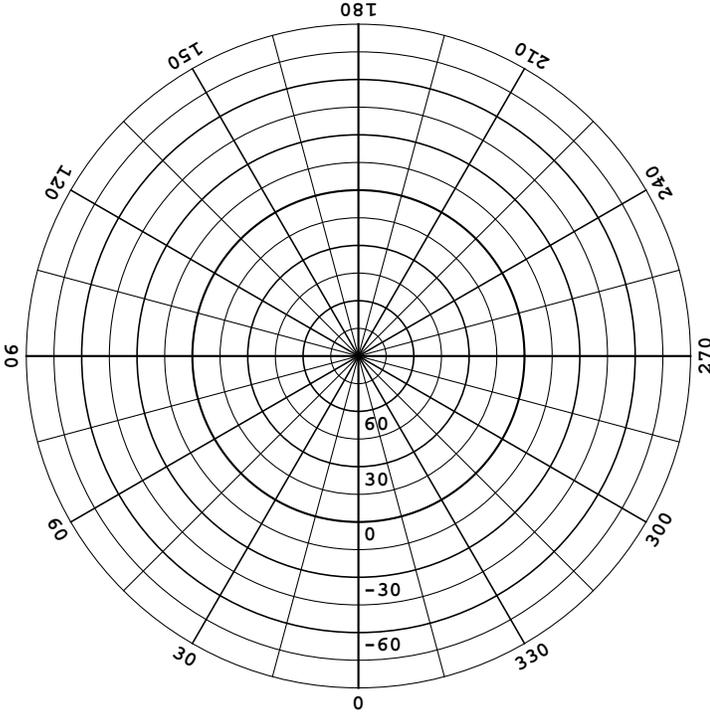}{270pt}}
   \caption[]{Zenithal equidistant (\keyv{ARC}) projection; no limits.}
   \label{fig:ARC}
\end{figure}

The orthographic projection is widely used in aperture synthesis radio
astronomy and was given its own code within the AIPS convention, namely
\keyv{SIN}\footnote{Similar etymology to \keyv{TAN}\@.}.  Use of this
projection code obviates the need to specify an infinite value as a parameter
of \keyv{AZP}\@.  In this case, Eq.~(\ref{eq:AZPRt}) becomes
\begin{equation}
      \Rt = \rd \cos\theta \, , \\
\end{equation}
with inverse
\begin{equation}
   \theta = \cos^{-1} \left( \dr \Rt \right) \, .
\end{equation}

In fact, use of the orthographic projection in radio interferometry is an
approximation, applicable only for small field sizes.  However, an exact
solution exists where the interferometer baselines are co-planar.  It reduces
to what Greisen (\cite{kn:G1}) called the \keyv{NCP} projection for the
particular case of an East-West interferometer (Brouw \cite{kn:Br}).  The
projection equations (derived in Appendix~\ref{apx:SYN}) are
\begin{eqnarray}
   x  & = & \SP \rd \left[\, \cos\theta \sin\phi +
                    \xi  \, (1 - \sin\theta) \right] \, , \label{eq:SYNx} \\
   y  & = &   - \rd \left[\, \cos\theta \cos\phi -
                    \eta \, (1 - \sin\theta) \right] \, . \label{eq:SYNy}
\end{eqnarray}
These are the equations of the ``slant orthographic'' projection, equivalent
to Eqs.~(\ref{eq:SZPx}) and (\ref{eq:SZPy}) of the \keyv{SZP} projection in
the limit $\mu = \infty$, with
\begin{eqnarray}
   \xi  & = & \SP \cot\theta\sub{c} \sin\phi\sub{c} \, , \\
   \eta & = &   - \cot\theta\sub{c} \cos\phi\sub{c} \, .
\end{eqnarray}

It can be shown that the slant orthographic projection is equivalent to an
orthographic projection centered at $(x,y) = \rd (\xi, \eta)$ which has been
stretched in the $\phi\sub{c}$ direction by a factor of
${\rm cosec}\,\theta\sub{c}$.  The projection equations may be inverted using
Eqs.~(\ref{eq:SZPtheta}) and (\ref{eq:SZPphi}) except that
Eq.~(\ref{eq:SZPXYd}) is replaced with
\begin{equation}
   (X', Y') = (\xi, \eta) \, .
\end{equation}
The outer boundary of the \keyv{SIN} projection is given by
Eq.~(\ref{eq:SZPTx}) in the limit $\mu = \infty$:
\begin{equation}
   \theta\sub{x} = - \tan^{-1} \left( \xi\sin\phi - \eta\cos\phi \right) \, .
\end{equation}

Two example graticules are illustrated in the lower portion of
Fig.~\ref{fig:SIN}.  We here extend the original \keyv{SIN} projection of the
AIPS convention to encompass the slant orthographic projection, with the
dimensionless $\xi$ and $\eta$ given by keywords \PVi{1} and \PVi{2},
respectively, attached to {\em latitude} coordinate~{\it i}, both with default
value 0.


\subsubsection{\keyv{ARC}: zenithal equidistant}
\label{sec:ARC}

Some non-perspective zenithal projections are also of interest in astronomy.
The zenithal equidistant projection first appeared in Greisen (\cite{kn:G1})
as \keyv{ARC}\@.  It is widely used as the approximate projection of Schmidt
telescopes.  As illustrated in Fig.~\ref{fig:ARC}, the native meridians are
uniformly divided to give equispaced parallels.  Thus
\begin{eqnarray}
   \Rt & = & 90\degr - \theta \, , \label{eq:ARCRt}
\end{eqnarray}
which is trivially invertible.  This projection was also known in antiquity.


\subsubsection{\keyv{ZPN}: zenithal polynomial}
\label{sec:ZPN}

\begin{figure}
   \centerline{\putfig{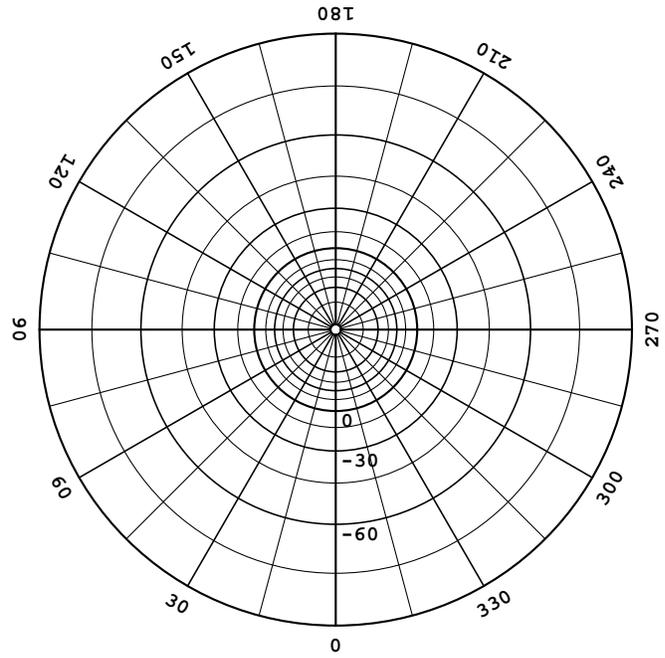}{250pt}}
   \caption[]{Zenithal polynomial projection (\keyv{ZPN}) with parameters,
      0.050, 0.975, -0.807, 0.337, -0.065, 0.010, 0.003, -0.001; limits depend
      upon the parameters.}
   \label{fig:ZPN}
\end{figure}

The zenithal polynomial projection, \keyv{ZPN}, generalizes the \keyv{ARC}
projection by adding polynomial terms up to a large degree in the zenith
distance.  We define it as
\begin{equation}
   \Rt = \rd \sum_{m=0}^{29} P_m \left(\dr (90\degr - \theta) \right)^i \, .
\end{equation}
Note the dimensionless units of $P_m$ imparted by $\pi/180\degr$.

Since its inverse cannot be expressed analytically, \keyv{ZPN} should only be
used when the geometry of the observations require it.  In particular, it
should never be used as an $n^{\rm th}$-degree expansion of one of the
standard zenithal projections.

If $P_0$ is non-zero the native pole is mapped to an open circle centered on
the reference point as illustrated in Fig.~\ref{fig:ZPN}.  In other words,
$(\phi_0,\theta_0) = (0,90\degr)$ is not at $(x,y) = (0,0)$, which in fact lies
outside the boundary of the projection.  However, we do not dismiss
$P_0 \neq 0$ as a possibility since it is not inconsistent with the formalism
presented in Sect.~\ref{sec:refpoint} and could conceivably be useful for
images which do not contain the reference point.  Needless to say, care should
be exercised in constructing and interpreting such systems particularly in
that $(\alpha_0,\delta_0)$ (i.e.\ the \CRVAL{ia}) do not specify the celestial
coordinates of the reference point (the \CRPIX{ja}).

$P_m$ (dimensionless) is given by the keywords \PVi{0}, \PVi{1}, $\ldots$,
\PVi{29}, attached to {\em latitude} coordinate~{\it i}, all of which have
default values of zero.


\subsubsection{\keyv{ZEA}: zenithal equal-area}
\label{sec:ZEA}

\begin{figure}
   \centerline{\putfig{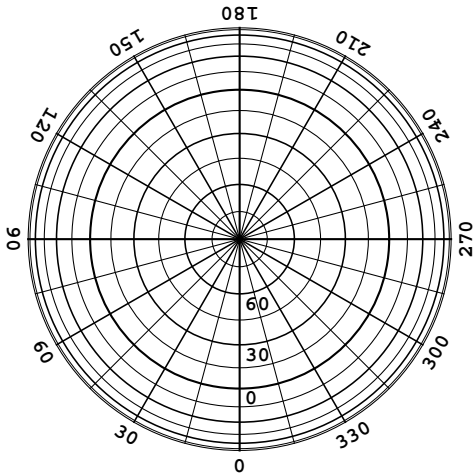}{180pt}}
   \caption[]{Zenithal equal area projection (\keyv{ZEA}); no limits.}
   \label{fig:ZEA}
\end{figure}

Lambert's zenithal equal-area projection illustrated in Fig.~\ref{fig:ZEA} is
constructed by defining $\Rt$ so that the area enclosed by the native parallel
at latitude $\theta$ in the plane of projection is equal to the area of the
corresponding spherical cap.  It may be generated using
\begin{eqnarray}
     \Rt & = & \rd \sqrt{2 (1 - \sin\theta)} \label{eq:ZEARt} \\
         & = & {\frac{360\degr}{\pi}}
                  \sin \left( \frac{90\degr - \theta}{2} \right) \, ,
                  \nonumber
\end{eqnarray}
with inverse
\begin{equation}
  \theta = 90\degr - 2 \sin^{-1} \left(\frac{\pi\Rt}{360\degr}\right) \, .
\end{equation}

\subsubsection{\keyv{AIR}: Airy projection}
\label{sec:AIR}

The Airy projection\footnote{Devised in 1861 by astronomer royal George
Biddell Airy, 1801-1892.} minimizes the error for the region within latitude
$\theta\sub{b}$ (Evenden \cite{kn:Eve}).  It is defined by
\begin{equation}
   \Rt = -2\rd \left(
                  \frac{\ln(\cos\xi)}{\tan\xi} +
                  \frac{\ln(\cos\xi\sub{b})}{\tan^2\xi\sub{b}}\tan\xi
               \right) \, ,  \label{eq:AIRt}
\end{equation}
where
\begin{eqnarray}
          \xi & = & \frac{90\degr - \theta}{2}   \, , \nonumber \\
   \xi\sub{b} & = & \frac{90\degr - \theta\sub{b}}{2} \, . \nonumber
\end{eqnarray}
When $\theta\sub{b}$ approaches $90\degr$, the second term of
Eq.~(\ref{eq:AIRt}) approaches its asymptotic value of $-\frac{1}{2}$.  For
all $\theta\sub{b}$, the projection is unbounded at the native south pole.
Inversion of Eq.~(\ref{eq:AIRt}), a transcendental equation in $\theta$, must
be done via iterative methods.

\begin{figure}
   \centerline{\putfig{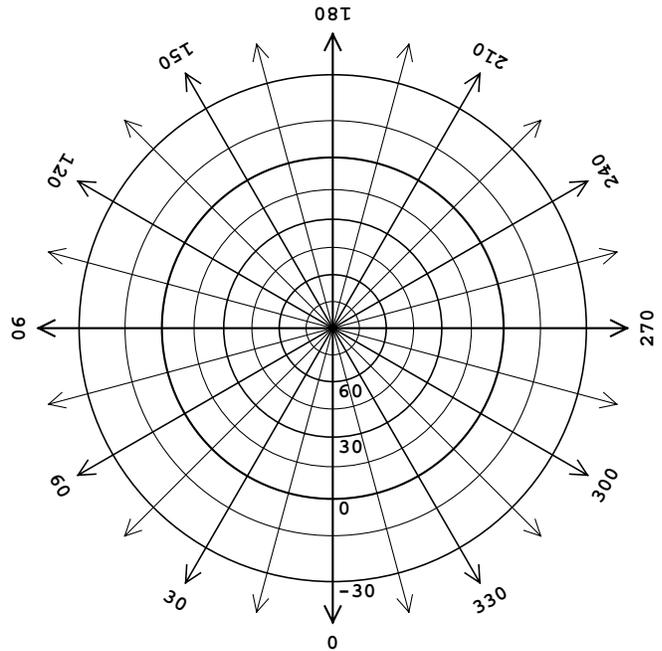}{230pt}}
   \caption[]{Airy projection (\keyv{AIR}) with $\theta\sub{b} = 45\degr$;
      diverges at $\theta = -90\degr$.}
   \label{fig:AIR}
\end{figure}

The FITS keyword \PVi{1}, attached to {\em latitude} coordinate~{\it i}, will
be used to specify $\theta\sub{b}$ in degrees with a default of $90\degr$.
This projection is illustrated in Fig.~\ref{fig:AIR}.


\subsection{Cylindrical projections}
\label{sec:cylindricals}

Cylindrical projections are so named because the surface of projection is a
cylinder.  The native coordinate system is chosen to have its polar axis
coincident with the axis of the cylinder.  Meridians and parallels are mapped
onto a rectangular graticule so that cylindrical projections are described by
formul\ae\ which return $x$ and $y$ directly.  Since all cylindrical
projections are constructed with the native coordinate system origin at the
reference point, we set
\begin{equation}
   (\phi_0, \theta_0)\sub{cylindrical} = (0,0) \, .
\end{equation}
Furthermore, all cylindrical projections have
\begin{equation}
  x \propto \phi \, .
\end{equation}
Cylindrical projections are often chosen to map the regions adjacent to a
great circle, usually the equator, with minimal distortion.


\subsubsection{\keyv{CYP}: cylindrical perspective}
\label{sec:CYP}

\begin{figure}
   \centerline{\putfig{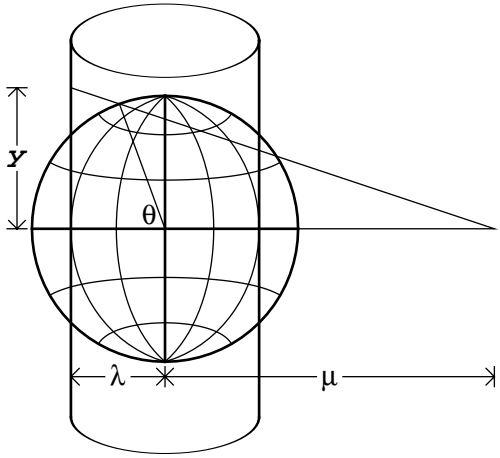}{6.4cm}}
   \caption[]{Geometry of cylindrical projections.}
   \label{fig:Cylindrl}
\end{figure}

Figure~\ref{fig:Cylindrl} illustrates the geometry for the construction of
cylindrical perspective projections.  The sphere is projected onto a cylinder
of radius $\lambda$ spherical radii from points in the equatorial plane of the
native system at a distance $\mu$ spherical radii measured from the center of
the sphere in the direction opposite the projected surface.  The cylinder
intersects the sphere at latitudes $\theta\sub{x} = \cos^{-1}\lambda$.  It is
straightforward to show that
\begin{eqnarray}
   x & = & \lambda \phi \, , \\
   y & = & \rd \left( \frac{\mu+\lambda}{\mu+\cos\theta}\right)
               \sin\theta \, .
\end{eqnarray}
This may be inverted as
\begin{eqnarray}
     \phi & = & \frac{x}{\lambda} \, , \\
   \theta & = & \arg\,(1,\eta) + \sin^{-1}\left(
                   \frac{\eta\mu}{\sqrt{\eta^2+1}} \right) \, ,
\end{eqnarray}
where
\begin{equation}
     \eta = \dr \frac{y}{\mu+\lambda} \, .
\end{equation}
Note that all values of $\mu$ are allowable except $\mu = -\lambda$.  For FITS
purposes, we define the keywords \PVi{1} to convey $\mu$ and \PVi{2} for
$\lambda$, both measured in spherical radii, both with default value 1, and
both attached to {\em latitude} coordinate~{\it i}.

\begin{figure}
   \centerline{\putfig{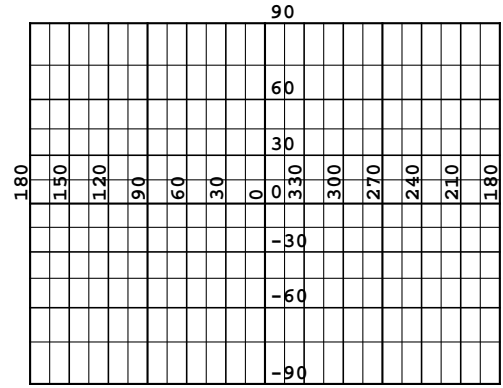}{155pt}}
   \caption[]{Gall's stereographic projection, \keyv{CYP} with $\mu = 1$,
      $\theta\sub{x} = 45\degr$; no limits.}
   \label{fig:CYP}
\end{figure}

The case with $\mu = \infty$ is covered by the class of cylindrical equal area
projections.  No other special-cases need be defined since cylindrical
perspective projections have not previously been used in FITS\@.  Aliases for
a number of special cases are listed in Appendix~\ref{apx:aliases},
Table~\ref{ta:aliases}.  Probably the most important of these is Gall's
stereographic projection, which minimizes distortions in the equatorial
regions.  It has $\mu = 1, \lambda = \sqrt{2}/2$, giving
\begin{eqnarray*}
   x & = & \phi / \sqrt{2} \, , \\
   y & = & \rd \left( 1 + \frac{\sqrt{2}}{2} \right)
             \tan \left( \frac{\theta}{2} \right) \, .
\end{eqnarray*}
It is illustrated in Fig.~\ref{fig:CYP}.

\begin{figure}
   \centerline{\putfig{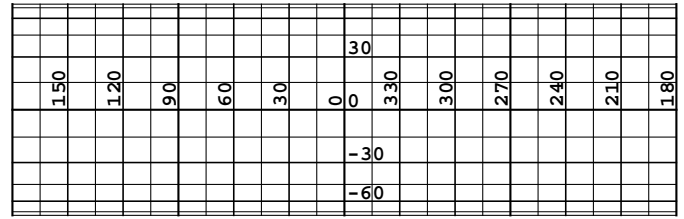}{90pt}}
   \caption[]{Lambert's equal area projection, \keyv{CEA} with $\lambda = 1$;
      no limits.}
   \label{fig:CEA}
\end{figure}


\subsubsection{\keyv{CEA}: cylindrical equal area}
\label{sec:CEA}

The cylindrical equal area projection is the special case of the cylindrical
perspective projection with $\mu = \infty$.  It is conformal at latitudes
$\pm\theta\sub{c}$ where $\lambda = \cos^2\theta\sub{c}$.  The formul\ae\
are
\begin{eqnarray}
        x & = & \phi \, , \\
        y & = & \rd \frac{\sin\theta}{\lambda} \, ,
\end{eqnarray}
which reverse to
\begin{eqnarray}
     \phi & = & x \, , \\
   \theta & = & \sin^{-1} \left( \dr \lambda y \right) \, .
\end{eqnarray}
Note that the scaling parameter $\lambda$ is applied to the $y$-coordinate
rather than the $x$-coordinate as in \keyv{CYP}\@.  The keyword \PVi{1}
attached to {\em latitude} coordinate~{\it i} is used to specify the
dimensionless $\lambda$ with default value 1.

Lambert's\footnote{The mathematician, astronomer and physicist Johann Heinrich
Lambert (1728-1777) was the first to make significant use of calculus in
constructing map projections.  He formulated and gave his name to a number of
important projections as listed in Table~\ref{ta:aliases}.} equal area
projection, the case with $\lambda = 1$, is illustrated in Fig.~\ref{fig:CEA}.
It shows the extreme compression of the parallels of latitude at the poles
typical of all cylindrical equal area projections.


\subsubsection{\keyv{CAR}: plate carr\'{e}e}
\label{sec:CAR}

\begin{figure}
   \centerline{\putfig{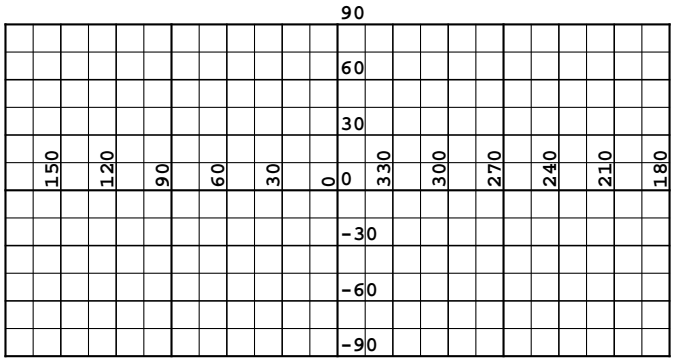}{140pt}}
   \caption[]{The plate carr\'{e}e projection (\keyv{CAR}); no limits.}
   \label{fig:CAR}
\end{figure}

The equator and all meridians are correctly scaled in the plate carr\'{e}e
projection\footnote{Although colloquially referred to as ``Cartesian'',
Claudius~Ptolemy (ca.\,90-ca.\,170\,{\sc a.d.}), influential cartographer and
author of the Ptolemaic model of the solar system, credits Marinus of Tyre
with its invention in about {\sc a.d.}\,100, thus predating Descartes by some
1500 years.}, whose main virtue is that of simplicity.  Its formul\ae\ are
\begin{eqnarray}
   x & = & \phi   \, , \label{eq:CARx} \\
   y & = & \theta \, . \label{eq:CARy}
\end{eqnarray}
The projection is illustrated in Fig.~\ref{fig:CAR}.

\subsubsection{\keyv{MER}: Mercator}
\label{sec:MER}

\begin{figure}
   \centerline{\putfig{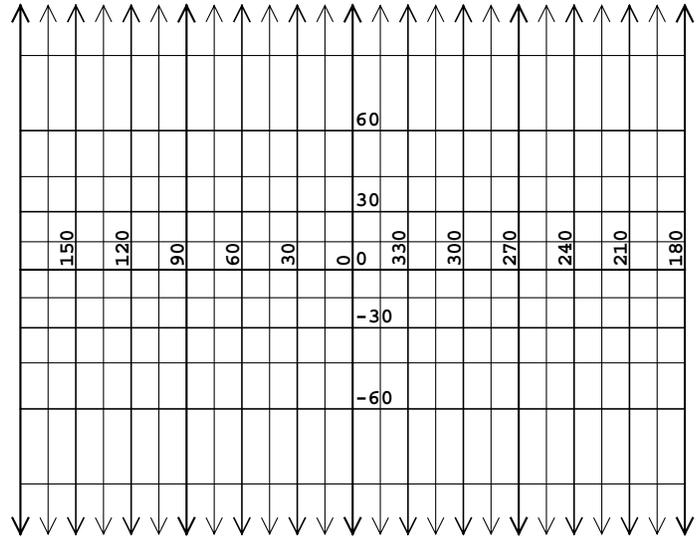}{210pt}}
   \caption[]{Mercator's projection (\keyv{MER}); diverges at
      $\theta = \pm90\degr$.}
   \label{fig:MER}
\end{figure}

Since the meridians and parallels of all cylindrical projections intersect at
right angles the requirement for conformality reduces to that of equiscaling
at each point.  This is expressed by the differential equation
\begin{equation}
   \frac{\partial y}{\partial\theta} =
       \frac{-1}{\cos\theta} \frac{\partial x}{\partial\phi} \, ,
\end{equation}
the solution\footnote{Gerardus Mercator (1512-1594), a prominent Flemish
map-maker, effectively solved this equation by numerical integration.
Presented in 1569 it thus predates Newton's {\em theory of fluxions} by nearly
a century.} of which gives us Mercator's projection:
\begin{eqnarray}
   x & = & \phi \, , \\
   y & = & \rd \ln \tan \left( \frac{90\degr+\theta}{2} \right) \, ,
\end{eqnarray}
with inverse
\begin{equation}
   \theta = 2 \tan^{-1} \left( {\rm e}^{y\pi/180\degr} \right) - 90\degr\, .
\end{equation}
This projection, illustrated in Fig.~\ref{fig:MER}, has been widely used in
navigation since it has the property that lines of constant bearing (known as
{\em rhumb lines} or {\em loxodromes}) are projected as straight lines.  This
is a direct result of its conformality and the fact that its meridians do not
converge.

Refer to Sect.~\ref{sec:AITGLSMER} for a discussion of the usage of
\keyv{MER} in AIPS\@.


\subsection{Pseudocylindrical and related projections}
\label{sec:pseudocylindricals}

{\em Pseudocylindricals} are like cylindrical projections except that the
parallels of latitude are projected at diminishing lengths towards the polar
regions in order to reduce lateral distortion there.  Consequently the
meridians are curved.  Pseudocylindrical projections lend themselves to the
construction of {\em interrupted} projections in terrestrial cartography.
However, this technique is unlikely to be of use in celestial mapping and is
not considered here.  Like ordinary cylindrical projections, the
pseudocylindricals are constructed with the native coordinate system origin at
the reference point.  Accordingly we set
\begin{equation}
   (\phi_0, \theta_0)\sub{pseudocylindrical} = (0, 0) \, .
\end{equation}
The Hammer-Aitoff projection is a modified zenithal projection, not a
pseudocylindrical, but is presented with this group on account of its
superficial resemblance to them.


\subsubsection{\keyv{SFL}: Sanson-Flamsteed}
\label{sec:SFL}

\begin{figure}
   \centerline{\putfig{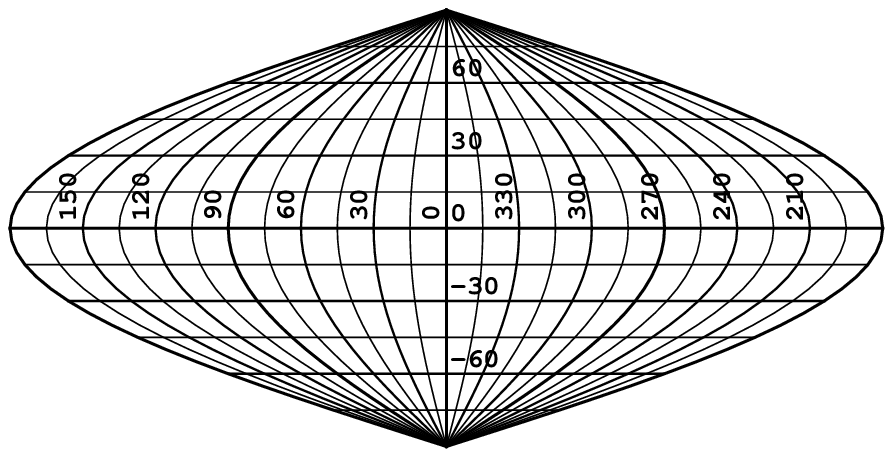}{135pt}}
   \caption[]{Sanson-Flamsteed projection (\keyv{SFL}); no limits.}
   \label{fig:SFL}

   \vskip 30pt

   \centerline{\putfig{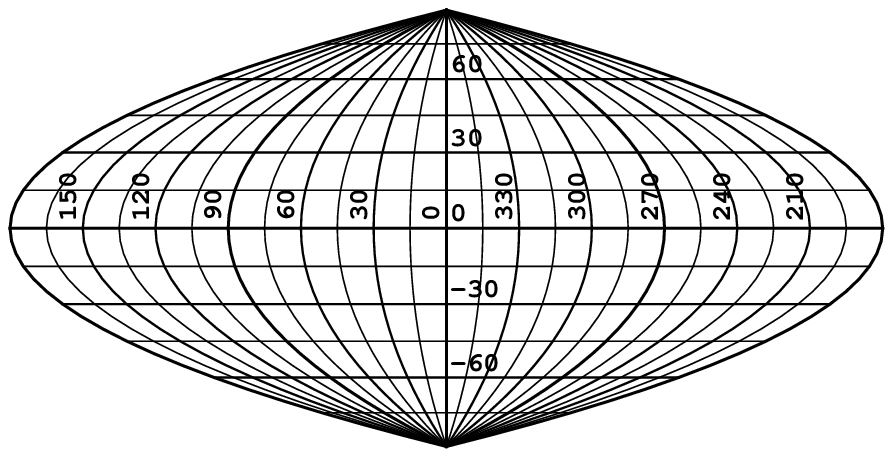}{135pt}}
   \caption[]{Parabolic projection (\keyv{PAR}); no limits.}
   \label{fig:PAR}

   \vskip 30pt

   \centerline{\putfig{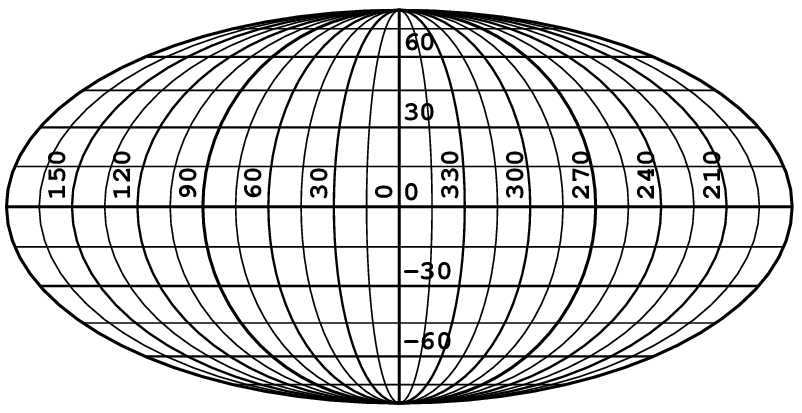}{120pt}}
   \caption[]{Mollweide's projection (\keyv{MOL}); no limits.}
   \label{fig:MOL}

   \vskip 30pt

   \centerline{\putfig{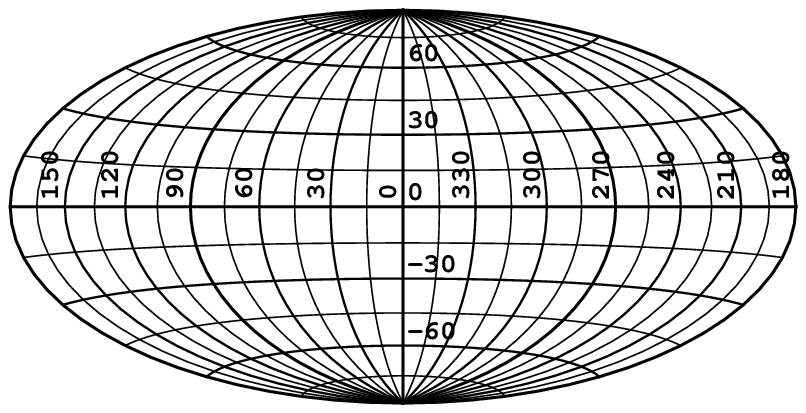}{120pt}}
   \caption[]{Hammer-Aitoff projection (\keyv{AIT}); no limits.}
   \label{fig:AIT}
\end{figure}

Bonne's projection (Sect.~\ref{sec:BON}) reduces to the pseudocylindrical
Sanson-Flamsteed\footnote{Nicolas Sanson d'Abbeville (1600-1667) of France and
John Flamsteed (1646-1719), the first astronomer royal of England, popularized
this projection, which was in existence at least as early as 1570.} projection
when $\theta_1 = 0$.  Parallels are equispaced and projected at their true
length which makes it an equal area projection.  The formul\ae\ are
\begin{eqnarray}
  x & = & \phi \cos\theta \, , \\
  y & = & \theta \, ,
\end{eqnarray}
which reverse into
\begin{eqnarray}
    \phi & = & \frac{x}{\cos y} \, , \\
  \theta & = & y \, .
\end{eqnarray}
This projection is illustrated in Fig.~\ref{fig:SFL}.  Refer to
Sect.~\ref{sec:AITGLSMER} for a discussion relating \keyv{SFL} to the
\keyv{GLS} projection in AIPS\@.


\subsubsection{\keyv{PAR}: parabolic}
\label{sec:PAR}

The parabolic or Craster pseudocylindrical projection is illustrated in
Fig.~\ref{fig:PAR}.  The meridians are projected as parabolic arcs which
intersect the poles and correctly divide the equator, and the parallels of
latitude are spaced so as to make it an equal area projection.  The
formul\ae\ are
\begin{eqnarray}
   x & = & \phi \left( 2\cos \frac{2\theta}{3} - 1 \right) \, , \\
   y & = & 180\degr \sin\frac{\theta}{3} \, ,
\end{eqnarray}
with inverse
\begin{eqnarray}
   \theta & = & 3 \sin^{-1} \left( \frac{y}{180\degr} \right) \, , \\
     \phi & = & \rd \frac{x}{1 - 4 (y/180\degr)^2} \, .
\end{eqnarray}


\subsubsection{\keyv{MOL}: Mollweide's}
\label{sec:MOL}

In Mollweide's pseudocylindrical projection\footnote{Presented in 1805 by
astronomer and mathematician Karl Brandan Mollweide (1774-1825).}, the
meridians are projected as ellipses that correctly divide the equator and
the parallels are spaced so as to make the projection equal area.  The
formul\ae\ are
\begin{eqnarray}
   x & = & \frac{2\sqrt{2}}{\pi} \phi \cos\gamma \, , \\
   y & = & \sqrt{2}\rd \sin\gamma \, ,
\end{eqnarray}
where $\gamma$ is defined as the solution of the transcendental equation
\begin{equation}
   \sin\theta = \frac{\gamma}{90\degr} + \frac{\sin 2\gamma}{\pi} \, .
\end{equation}
The inverse equations may be written directly without reference to $\gamma$,
\begin{eqnarray}
     \phi & = & \pi x / \left(2\sqrt{2-(\dr y)^2} \,\right) \, , \\
   \theta & = & \sin^{-1}\left(\frac{1}{90\degr}\sin^{-1} \left(
                 \dr\frac{y}{\sqrt{2}}\right) + \right.\nonumber \\
          &   & \hskip 1cm \left.\mbox{} \frac{y}{180\degr}
                   \sqrt{2 - (\dr y)^2} \,\,\right) \, .
\end{eqnarray}
Mollweide's projection is illustrated in Fig.~\ref{fig:MOL}.


\begin{figure*}
   \if\dofig F
      \vspace{330pt}
   \else
      \centerline{\putfig{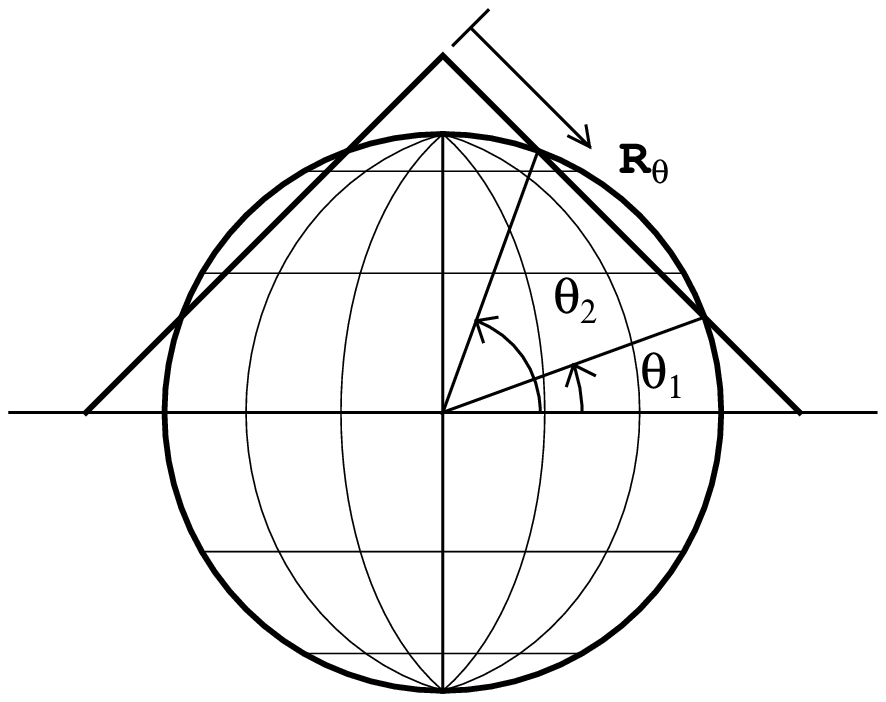}{160pt} \hfil
                  \putfig{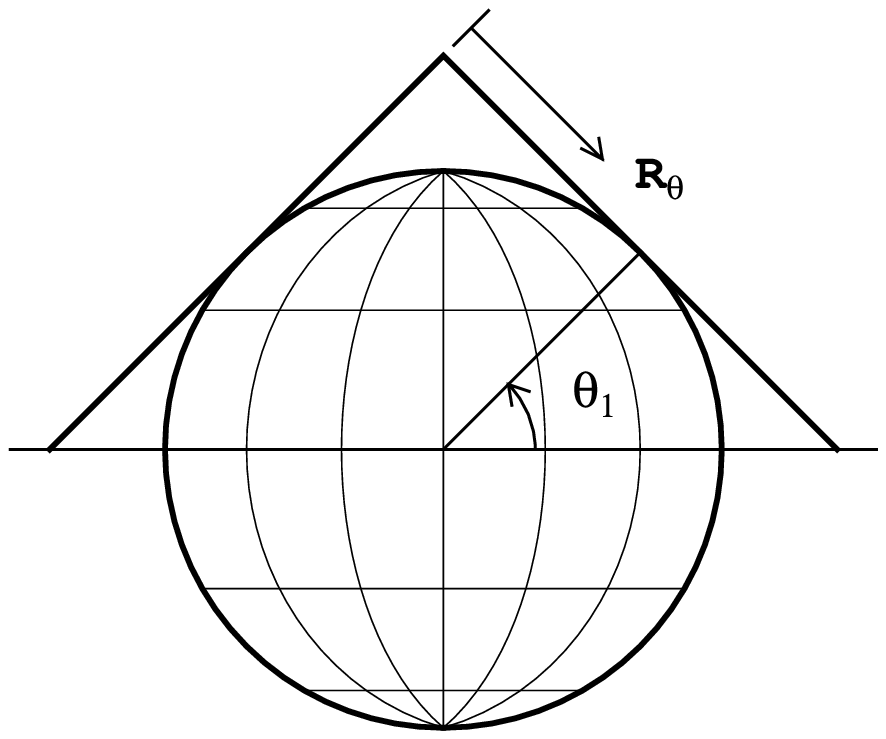}{160pt}}
      \vskip -20pt
      \centerline{\putfig{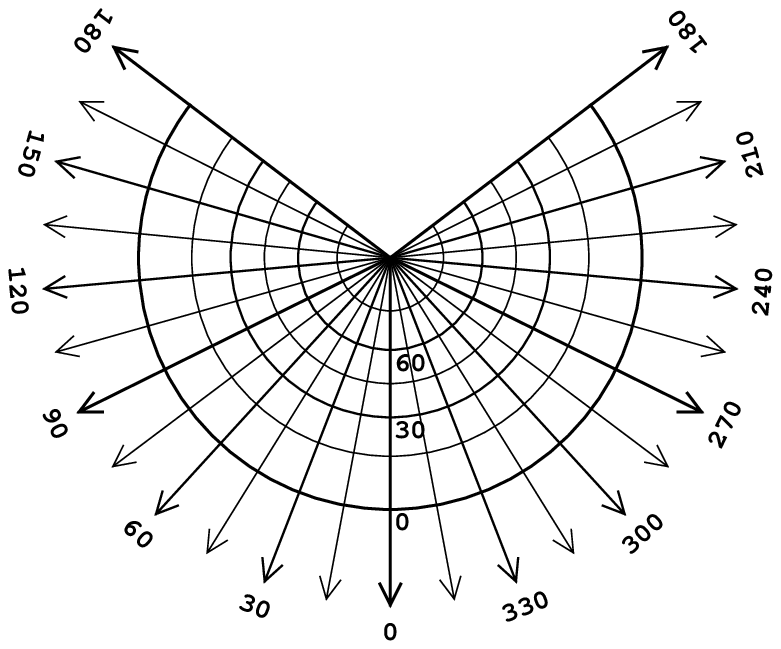}{190pt} \hfil
                  \putfig{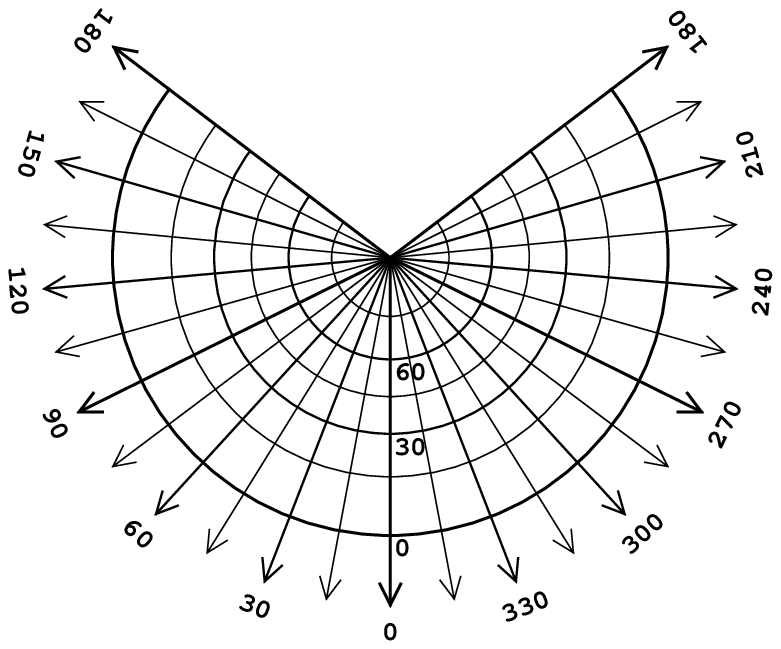}{190pt}}
   \fi
   \caption[]{Construction of conic perspective projections (\keyv{COP}) and
      the resulting graticules; (left) two-standard projection with
      $\theta_1 = 20\degr$, $\theta_2 = 70\degr$; (right) one-standard
      projection with $\theta_1 = \theta_2 = 45\degr$.  Both projections have
      $\theta\sub{a} = 45\degr$ and this accounts for their similarity.  Both
      diverge at $\theta = \theta\sub{a} \pm 90\degr$.}
   \label{fig:COP}
\end{figure*}

\subsubsection{\keyv{AIT}: Hammer-Aitoff}
\label{sec:AIT}

The Hammer-Aitoff\footnote{David Aitoff (1854-1933) developed his projection
from the zenithal equidistant projection in 1889 and in 1892 Ernst Hammer
(1858-1925) applied his idea more usefully to the zenithal equal area
projection.  See Jones (\cite{kn:Jon}).} projection illustrated in
Fig.~\ref{fig:AIT} is developed from the equatorial case of the zenithal equal
area projection by doubling the equatorial scale and longitude coverage.  The
whole sphere is mapped thereby while preserving the equal area property.
Note, however, that the equator is not evenly divided.

This projection reduces distortion in the polar regions compared to
pseudocylindricals by making the meridians and parallels more nearly
orthogonal.  Together with its equal area property this makes it one of most
commonly used all-sky projections.

The formul\ae\ for the projection and its inverse are derived in Greisen
(\cite{kn:G2}) and Calabretta (\cite{kn:C1}) among others.  They are
\begin{eqnarray}
        x & = & 2 \gamma \cos\theta \sin\frac{\phi}{2} \, , \\
        y & = & \gamma \sin \theta \, ,
\end{eqnarray}
where
\begin{equation}
   \gamma = \rd \sqrt{\frac{2}{1 + \cos\theta\cos(\phi/2)}} \, .
\end{equation}
The reverse equations are
\begin{eqnarray}
     \phi & = & 2 \arg \left(2Z^2 - 1, \dr\frac{Z}{2} x\right) \, , \\
   \theta & = & \sin^{-1} (\dr yZ) \, ,
\end{eqnarray}
where
\begin{eqnarray}
   Z & = & \sqrt{1 - \left(\dr\frac{x}{4}\right)^2 -
                     \left(\dr\frac{y}{2}\right)^2} \, , \\
     & = & \sqrt{\frac{1}{2} \left( 1 + \cos\theta \cos\frac{\phi}{2}
                 \right)} \, .
\end{eqnarray}
Note that $\frac{1}{2} \le Z^2 \le 1$.  Refer to Sect.~\ref{sec:AITGLSMER}
for a discussion of the usage of \keyv{AIT} in AIPS\@.


\subsection{Conic projections}
\label{sec:conics}

In conic projections the sphere is thought to be projected onto the surface of
a cone which is then opened out.  The native coordinate system is chosen so
that the poles are coincident with the axis of the cone.  Native meridians are
then projected as uniformly spaced rays that intersect at a point (either
directly or by extrapolation), and parallels are projected as equiangular
arcs of concentric circles.

{\em Two-standard} conic projections are characterized by two latitudes,
$\theta_1$ and $\theta_2$, whose parallels are projected at their true length.
In the conic perspective projection these are the latitudes at which the cone
intersects the sphere.  {\em One-standard} conic projections have
$\theta_1 = \theta_2$ and the cone is tangent to the sphere as shown in
Fig.~\ref{fig:COP}.  Since conics are designed to minimize distortion in the
regions between the two standard parallels they are constructed so that the
point on the prime meridian mid-way between the two standard parallels maps
to the reference point so we set
\begin{equation}
   (\phi_0, \theta_0)\sub{conic} = (0, \theta\sub{a}) \, , \\
\end{equation}
where
\begin{equation}
   \theta\sub{a} = (\theta_1 + \theta_2) / 2 \, . \label{eq:theta_a}
\end{equation}
Being concentric, the parallels may be described by $\Rt$, the radius for
latitude $\theta$, and $\Ap$, the angle for longitude $\phi$.  An offset in
$y$ is also required to force $(x,y) = (0,0)$ at
$(\phi, \theta) = (0, \theta\sub{a})$.  All one- and two-standard conics have
\begin{eqnarray}
  \Ap & = & C \phi \, ,
\end{eqnarray}
where $C$, a constant known as the constant of the cone, is such that the
apical angle of the projected cone is $360\degr\,C$.  Since the standard
parallels are projected as concentric arcs at their true length we have
\begin{eqnarray}
   C = \frac{180\degr \cos\theta_1}{\pi R_{\theta_1}}
     = \frac{180\degr \cos\theta_2}{\pi R_{\theta_2}} \, .
   \label{eq:ConC}
\end{eqnarray}
Cartesian coordinates in the plane of projection are
\begin{eqnarray}
  x & = & \SP\Rt\sin(C\phi) \, , \\
  y & = &   -\Rt\cos(C\phi) + Y_0 \, ,
\end{eqnarray}
and these may be inverted as
\begin{eqnarray}
   \Rt & = & \sign\theta\sub{a} \sqrt{x^2+(Y_0-y)^2} \, , \label{eq:ConRt} \\
  \phi & = & \arg \left( \frac{Y_0-y}{\Rt} , \frac{x}{\Rt} \right) / C \, .
             \label{eq:ConPhi}
\end{eqnarray}
To complete the inversion the equation for $\theta$ as a function of $\Rt$ is
given for each projection.  The equations given here correctly invert southern
conics, i.e.\ those with $\theta\sub{a} < 0$.  An example is shown in
Fig.~\ref{fig:COE}.

The conics will be parameterized in FITS by $\theta\sub{a}$ (given by
Eq.~(\ref{eq:theta_a})) and $\eta$ where
\begin{equation}
   \eta = | \theta_1 - \theta_2 | / 2 \, .
\end{equation}
The keywords \PVi{1} and \PVi{2} attached to {\em latitude}
coordinate~{\it i} will be used to specify $\theta\sub{a}$ and $\eta$
respectively, both measured in degrees.  \PVi{1} has no default while \PVi{2}
defaults to 0.  It is recommended that both keywords always be given.  Where
$\theta_1$ and $\theta_2$ are required in the equations below they may be
computed via
\begin{eqnarray}
  \theta_1 & = & \theta\sub{a} - \eta \, , \label{eq:ConTheta1} \\
  \theta_2 & = & \theta\sub{a} + \eta \, . \label{eq:ConTheta2}
\end{eqnarray}
This sets $\theta_2$ to the larger of the two.  The order, however, is
unimportant.

As noted in Sect.~\ref{sec:projections}, the zenithal projections are special
cases of the conics with $\theta_1 = \theta_2 = 90\degr$.  Likewise, the
cylindrical projections are conics with $\theta_1 = -\theta_2$.  However, we
strongly advise against using conics in these cases for the reasons given
previously.  Nevertheless, the only formal requirement on $\theta_1$ and
$\theta_2$ in the equations presented below is that
$-90\degr \leq \theta_1,\theta_2 \leq 90\degr$.


\subsubsection{\keyv{COP}: conic perspective}
\label{sec:COP}

Development of Colles' conic perspective projection is shown in
Fig.~\ref{fig:COP}.  The projection formul\ae\ are
\begin{eqnarray}
    C & = & \sin \theta\sub{a} \, , \\
  \Rt & = & \rd \cos\eta\, \left[ \cot\theta\sub{a} -
                                  \tan(\theta-\theta\sub{a}) \right] \, , \\
  Y_0 & = & \rd \cos \eta \cot \theta\sub{a} \, .
\end{eqnarray}
The inverse may be computed with
\begin{equation}
   \theta = \theta\sub{a} + \tan^{-1}\left( \cot\theta\sub{a} -
                \dr\frac{\Rt}{\cos\eta} \right) \, .
\end{equation}


\subsubsection{\keyv{COE}: conic equal area}
\label{sec:COE}

\begin{figure}
   \centerline{\putfig{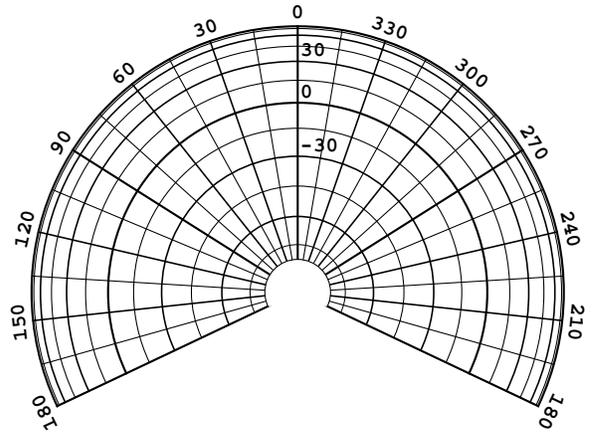}{170pt}}
   \caption[]{Conic equal area projection (\keyv{COE}) with
      $\theta_1 = -20\degr$, and $\theta_2 = -70\degr$, also illustrating the
      inversion of southern hemisphere conics; no limits.}
   \label{fig:COE}
\end{figure}

The standard parallels in Alber's conic equal area projection are projected as
concentric arcs at their true length and separated so that the area between
them is the same as the corresponding area on the sphere.  The other parallels
are then drawn as concentric arcs spaced so as to preserve the area.  The
projection formul\ae\ are
\begin{eqnarray}
    C & = & \gamma/2 \, , \label{eq:COEC} \\
  \Rt & = & \rd \frac{2}{\gamma} \sqrt{1 + \sin\theta_1\sin\theta_2
                    - \gamma\sin\theta} \, , \\
  Y_0 & = & \rd\frac{2}{\gamma} \sqrt{1 + \sin\theta_1\sin\theta_2
                - \gamma\sin((\theta_1+\theta_2)/2)} \, , \label{eq:COEy0}
\end{eqnarray}
where
\begin{equation}
   \gamma = \sin\theta_1 + \sin\theta_2 \, . \label{eq:COEgamma}
\end{equation}
The inverse may be computed with
\begin{eqnarray}
   \theta & = & \sin^{-1} \left(
                   \frac{1}{\gamma} +
                   \frac{\sin\theta_1\sin\theta_2}{\gamma} -
                   \gamma \left( \frac{\pi\Rt}{360\degr} \right)^2
                \right) \, . \label{eq:COEtheta}
\end{eqnarray}
This projection is illustrated in Fig.~\ref{fig:COE}.


\subsubsection{\keyv{COD}: conic equidistant}
\label{sec:COD}

\begin{figure}
   \centerline{\putfig{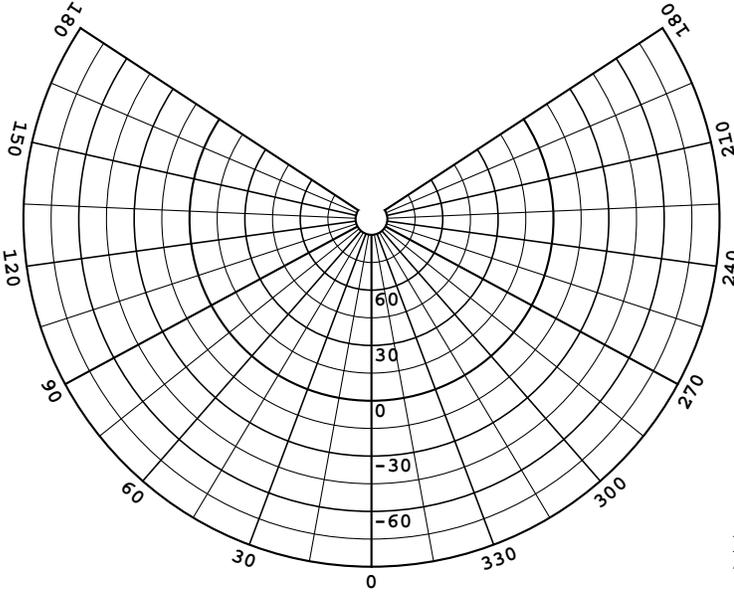}{230pt}}
   \caption[]{Conic equidistant projection (\keyv{COD}) with
      $\theta_1 = 20\degr$ and $\theta_2 = 70\degr$; no limits.}
   \label{fig:COD}
\end{figure}

In the conic equidistant projection the standard parallels are projected at
their true length and at their true separation.  The other parallels are then
drawn as concentric arcs spaced at their true distance from the standard
parallels.  The projection formul\ae\ are
\begin{eqnarray}
    C & = & \rd \frac{\sin\theta\sub{a}\sin\eta}{\eta} \, , \\
  \Rt & = & \theta\sub{a} - \theta + \eta \cot\eta\cot\theta\sub{a} \, , \\
  Y_0 & = & \eta \cot\eta\cot\theta\sub{a}  \, .
\end{eqnarray}
The inverse may be computed with
\begin{equation}
   \theta = \theta\sub{a} + \eta\cot\eta\cot\theta\sub{a} - \Rt \, .
\end{equation}
For $\theta_1 = \theta_2$ these expressions reduce to
\begin{eqnarray}
    C & = & \sin\theta\sub{a} \, , \\
  \Rt & = & \theta\sub{a} - \theta + \rd \cot\theta\sub{a} \, , \\
  Y_0 & = & \rd \cot\theta\sub{a}  \, ,
\end{eqnarray}
and inverse
\begin{equation}
   \theta = \theta\sub{a} + \rd\cot\theta\sub{a} - \Rt \,  .
\end{equation}
This projection is illustrated in Fig.~\ref{fig:COD}.


\subsubsection{\keyv{COO}: conic orthomorphic}
\label{sec:COO}

\begin{figure}
   \centerline{\putfig{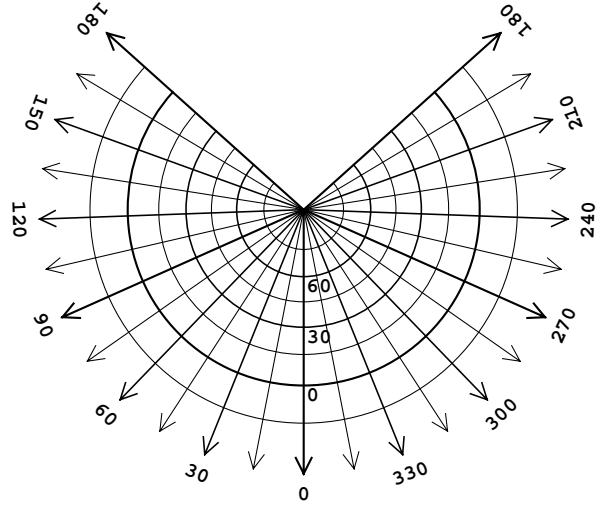}{195pt}}
   \caption[]{Conic orthomorphic projection (\keyv{COO}) with
      $\theta_1 = 20\degr$ and $\theta_2 = 70\degr$; diverges at
      $\theta = -90\degr$.}
   \label{fig:COO}
\end{figure}

The requirement for conformality of conic projections is
\begin{equation}
   \frac{\partial R_\theta}{\partial \theta} =
      \frac{-\pi R_\theta}{180\degr \cos\theta} C \, .
\end{equation}
Solution of this differential equation gives rise to the formul\ae\
for Lambert's conic orthomorphic projection:
\begin{eqnarray}
    C & = & \frac{\ln\left(\frac{\cos\theta_2}{\cos\theta_1}\right)}
                 {\ln\left[\frac{\tan\left(\frac{90\degr
                 -\theta_2}{2}\right)}{\tan\left(\frac{90\degr
                 -\theta_1}{2}\right)}\right]} \, , \\
  \Rt & = & \psi \left[ \tan\left(\frac{90\degr-\theta}{2}
                      \right) \right] ^C \, , \\
  Y_0 & = & \psi \left[ \tan\left(\frac{90\degr-\theta\sub{a}}{2}
                      \right) \right] ^C \, ,
\end{eqnarray}
where
\begin{eqnarray}
  \psi & = & \rd \frac{\cos\theta_1}{C\left[ \tan\left(
                       \frac{90\degr-\theta_1}{2}
                       \right) \right] ^C} \, , \\
       & = & \rd \frac{\cos\theta_2}{C\left[ \tan\left(
                       \frac{90\degr-\theta_2}{2}
                       \right) \right] ^C} \, .
\end{eqnarray}
The inverse may be computed with
\begin{equation}
  \theta = 90\degr - 2 \tan^{-1} \left( \left[
             \frac{\Rt}{\psi} \right]^\frac{1}{C} \right) \, .
\end{equation}
When $\theta_1 = \theta_2$ the expression for $C$ may be replaced with
$C = \sin\theta_1$.  This projection is illustrated in Fig.~\ref{fig:COO}.


\subsection{Polyconic and pseudoconic projections}
\label{sec:polyconics}

{\em Polyconics} are generalizations of the standard conic projections; the
parallels of latitude are projected as circular arcs which may or may not be
concentric, and meridians are curved rather than straight as in the standard
conics.  {\em Pseudoconics} are a sub-class with concentric parallels.  The
two polyconics presented here have parallels projected at their true length
and use the fact that for a cone tangent to the sphere at latitude $\theta_1$,
as shown in Fig.~\ref{fig:COP}, we have $R_{\theta_1} = \rd\cot\theta_1$.
Since both are constructed with the native coordinate system origin at the
reference point we set
\begin{equation}
   (\phi_0, \theta_0)\sub{polyconic} = (0, 0) \, .
\end{equation}


\subsubsection{\keyv{BON}: Bonne's equal area}
\label{sec:BON}

\begin{figure}
   \centerline{\putfig{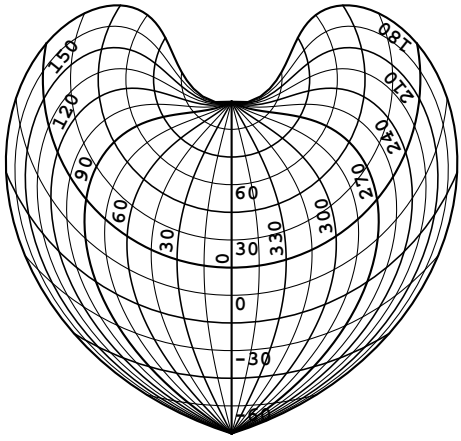}{175pt}}
   \caption[]{Bonne's projection (\keyv{BON}) with $\theta_1 = 45\degr$; no
      limits.}
   \label{fig:BON}
\end{figure}

In Bonne's pseudoconic projection\footnote{The ancestry of this important
projection may be traced back to Ptolemy.  Attribution for its invention is
uncertain but it was certainly used before the birth of Rigobert Bonne
(1727-1795) who did much to popularize it.} all parallels are projected as
concentric equidistant arcs of circles of true length and true spacing.  This
is sufficient to guarantee that it is an equal area projection.  It is
parameterized by the latitude $\theta_1$ for which
$R_{\theta_1} = \rd\cot\theta_1$.  The projection is conformal at this
latitude and along the central meridian.  The equations for Bonne's projection
become divergent for $\theta_1 = 0$ and this special case is handled as the
Sanson-Flamsteed projection.  The projection formul\ae\ are
\begin{eqnarray}
        x & = & \SP\Rt \sin\Ap \, , \\
        y & = &   -\Rt \cos\Ap + Y_0 \, ,
\end{eqnarray}
where
\begin{eqnarray}
      \Ap & = & \frac{180\degr}{\pi\Rt} \phi\cos\theta \, , \\
      \Rt & = & Y_0 - \theta \, , \\
      Y_0 & = & \rd \cot\theta_1 + \theta_1 \, .
\end{eqnarray}
The inverse formul\ae\ are then
\begin{eqnarray}
   \theta & = & Y_0 - \Rt \, , \\
     \phi & = & \dr \Ap \Rt\, / \cos\theta \, ,
\end{eqnarray}
where
\begin{eqnarray}
      \Rt & = & \sign\theta_1 \sqrt{x^2 + (Y_0-y)^2} \, , \\
      \Ap & = & \arg \left(\frac{Y_0-y}{\Rt}, \frac{x}{\Rt} \right) \, .
\end{eqnarray}
This projection is illustrated in Fig.~\ref{fig:BON}.  The keyword \PVi{1}
attached to {\em latitude} coordinate~{\it i} will be used to give $\theta_1$
in degrees with no default value.


\subsubsection{\keyv{PCO}: polyconic}
\label{sec:PCO}

\begin{figure}
   \centerline{\putfig{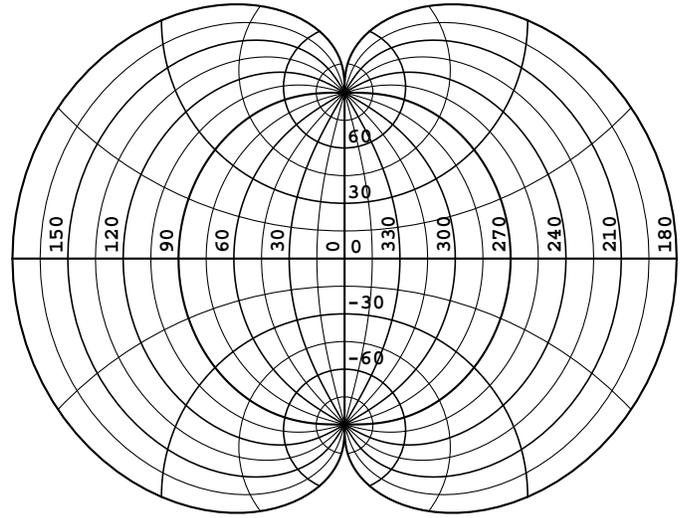}{200pt}}
   \caption[]{Polyconic projection (\keyv{PCO}); no limits.}
   \label{fig:PCO}
\end{figure}

Each parallel in Hassler's polyconic projection is projected as an arc of a
circle of radius $\rd\cot\theta$ at its true length, $360\degr\cos\theta$, and
correctly divided.  The scale along the central meridian is true and
consequently the parallels are not concentric.  The projection formul\ae\ are
\begin{eqnarray}
  x & = & \rd \cot\theta \sin(\phi\sin\theta) \, , \\
  y & = & \theta + \rd \cot\theta \left[1 - \cos(\phi\sin\theta)\right] \, .
\end{eqnarray}
Inversion requires iterative solution for $\theta$ of the equation
\begin{equation}
   x^2 - \frac{360\degr}{\pi}(y-\theta)\cot\theta + (y-\theta)^2 = 0 \, .
   \label{eq:PCOinvt}
\end{equation}
Once $\theta$ is known $\phi$ is given by
\begin{equation}
   \phi = \frac{1}{\sin\theta} \arg \left( \rd -
             (y-\theta)\tan\theta, x \tan\theta \right) \, .
\end{equation}
The polyconic projection is illustrated in Fig.~\ref{fig:PCO}.

\subsection{Quad-cube projections}
\label{sec:quadcubes}

Quadrilateralized spherical cube (quad-cube) projections belong to the class
of polyhedral projections\footnote{Polyhedral projections date from
renaissance times when the artist and mathematician Albrecht~D\"{u}rer
(1471-1528) described, although did not implement, the tetrahedral,
dodecahedral, and icosahedral cases.  Snyder (\cite{kn:Sny}) traces subsequent
development into the twentieth century, including one by R.~Buckminster Fuller
onto the non-Platonic ``cuboctahedron'' with constant scale along each edge.}
in which the sphere is projected onto the surface of an enclosing polyhedron,
typically one of the five Platonic polyhedra: the tetrahedron, hexahedron
(cube), octahedron, dodecahedron, and icosahedron.

The starting point for polyhedral projections is the gnomonic projection of
the sphere onto the faces of an enclosing polyhedron.  This may then be
modified to make the projection equal area or impart other special properties.
However, minimal distortion claims often made for polyhedral projections
should be balanced against the many interruptions of the flattened polyhedron;
the breaks should be counted as extreme distortions.  Their importance in
astronomy is not so much in visual representation but in solving the problem
of distributing $N$ points as uniformly as possible over the sphere.  This may
be of particular importance in optimizing computationally intensive
applications.  The general problem may be tackled by {\em pixelizations} such
as HEALPix (Hierarchical Equal Area isoLatitude PIXelization, Gorski
\cite{kn:GoH}), which define a distribution of points on the sphere but do
not relate these to points in a plane of projection and are therefore outside
the scope of this paper.

The quad-cubes have been used extensively in the COBE project and are
described by Chan \& O'Neill (\cite{kn:CO}) and O'Neill \& Laubscher
(\cite{kn:OL}).  The icosahedral case has also been studied by Tegmark
(\cite{kn:Teg}).  It is close to optimal, providing a 10\% improvement over
the cubic case.  However, we have not included it here since it relies on
image pixels being organized in an hexagonal close-packed arrangement rather
than the simple rectangular arrangement supported by FITS\@.

The six faces of quad-cube projections are numbered and laid out as
\begin{displaymath}
\begin{array}{cccccccc}
   \hskip 90pt
   &   &   &   & 0 \\
   & 4 & 3 & 2 & 1 & 4 & 3 & 2 \\
   &   &   &   & 5
\end{array}
\end{displaymath}
where faces 2, 3 and 4 may appear on one side or the other (or both).  The
layout used depends only on the FITS writer's choice of $\phi\sub{c}$ in
Table~\ref{ta:cubeface}.  It is also permissible to split faces between sides,
for example to put half of face 3 to the left and half to the right to create
a symmetric layout.  FITS readers should have no difficulty determining the
layout since the origin of $(x,y)$ coordinates is at the center of face 1 and
each face is $90\degr$ on a side.  The range of $x$ therefore determines the
layout.  Other arrangements are possible and Snyder (\cite{kn:Sny})
illustrates the ``saw-tooth'' layout of Reichard's 1803 map of the Earth.
While these could conceivably have benefit in celestial mapping we judged the
additional complication of representing them in FITS to be unwarranted.  The
layout used in the COBE project itself has faces 2, 3, and 4 to the left.

\begin{table}
   \caption[]{Assignment of parametric variables and central longitude and
      latitude by face number for quadrilateralized spherical cube
      projections.}
   \begin{displaymath}
      \begin{array}{clllrcrr}
         \hline
         \noalign{\smallskip}
            {\rm Face} & \SP\xi & \SP\eta & \SP\zeta &
                                & \phi\sub{c} &          & \theta\sub{c} \\
         \noalign{\smallskip}
         \hline
         \noalign{\smallskip}
                0     &  \SP m  &     -l  &  \SP n   &
                                &    0\degr   &          &    90\degr \\
                1     &  \SP m  &  \SP n  &  \SP l   &
                                &    0\degr   &          &     0\degr \\
                2     &     -l  &  \SP n  &  \SP m   &
                 \SP -270\degr  &   {\rm or}  &  90\degr &     0\degr \\
                3     &     -m  &  \SP n  &     -l   &
                     -180\degr  &   {\rm or}  & 180\degr &     0\degr \\
                4     &  \SP l  &  \SP n  &     -m   &
                      -90\degr  &   {\rm or}  & 270\degr &     0\degr \\
                5     &  \SP m  &  \SP l  &     -n   &
                                &    0\degr   &          &   -90\degr \\
         \noalign{\smallskip}
         \hline
      \end{array}
   \end{displaymath}
   \label{ta:cubeface}
\end{table}

The native coordinate system has its pole at the center of face 0 and origin
at the center of face 1 (see Fig.~\ref{fig:TSC}) which is the reference point
whence
\begin{equation}
   (\phi_0, \theta_0)\sub{quad-cube} = (0, 0) \, .
\end{equation}
The face number may be determined from the native longitude and latitude by
computing the direction cosines:
\begin{eqnarray}
   l & = & \cos \theta \cos\phi \, , \nonumber \\
   m & = & \cos \theta \sin\phi \, , \label{eq:dircos} \\
   n & = & \sin \theta \, . \nonumber
\end{eqnarray}
The face number is that which maximizes the value of $\zeta$ in
Table~\ref{ta:cubeface}.  That is, if $\zeta$ is the largest of $n$, $l$, $m$,
$-l$, $-m$, and $-n$, then the face number is 0 through 5, respectively.  Each
face may then be given a Cartesian coordinate system, $(\xi,\eta)$, with
origin in the center of each face as per Table~\ref{ta:cubeface}.  The
formul\ae\ for quad-cubes are often couched in terms of the variables
\begin{eqnarray}
  \chi & = & \xi  / \zeta \, , \\
  \psi & = & \eta / \zeta \, .
\end{eqnarray}

Being composed of six square faces the quad-cubes admit the possibility of
efficient data storage in FITS\@.  By stacking them on the third axis of a
three-dimensional data structure the storage required for an all-sky map may
be halved.  This axis will be denoted by a \CTYPE{ia} value of
`\keyv{CUBEFACE}'.  In this case the value of $(x,y)$ computed via
Eq.~(\ref{eq:px}) for the center of each face must be $(0,0)$ and the FITS
interpreter must increment this by $(\phi\sub{c},\theta\sub{c})$ using its
choice of layout.  Since the \keyv{CUBEFACE} axis type is purely a storage
mechanism the linear transformation of Eq.~(\ref{eq:px}) must preserve the
\keyv{CUBEFACE} axis pixel coordinates.


\subsubsection{\keyv{TSC}: tangential spherical cube}
\label{sec:TSC}

\begin{figure}
   \centerline{\putfig{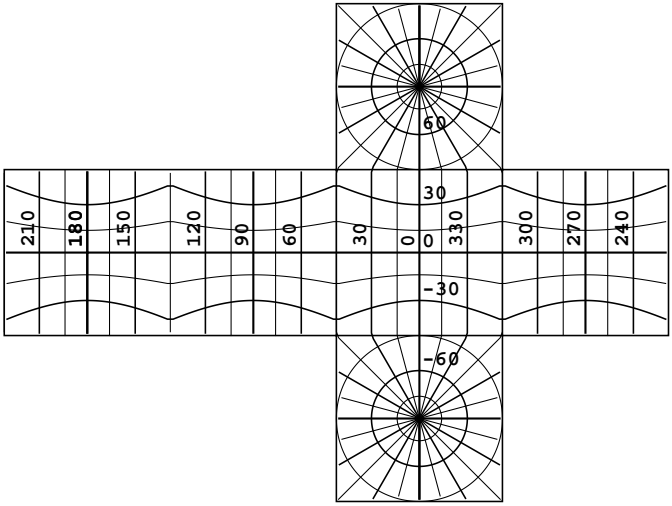}{200pt}}
   \caption[]{Tangential spherical cube projection (\keyv{TSC}); no limits.}
   \label{fig:TSC}
\end{figure}

While perspective quad-cube projections could be developed by projecting a
sphere onto an enclosing cube from any point of projection, inside or outside
the sphere, it is clear that only by projecting from the center of the sphere
will every face be treated equally.  Thus the tangential spherical cube
projection (\keyv{TSC}) consists of six faces each of which is a gnomonic
projection of a portion of the sphere.  As discussed in Sect.~\ref{sec:AZP},
gnomonic projections map great circles as straight lines but unfortunately
diverge very rapidly away from the poles and can only represent a portion of
the sphere without extreme distortion.  The \keyv{TSC} projection partly
alleviates this by projecting great circles as piecewise straight lines.  To
compute the forward projection first determine $\chi$ and $\psi$ as
described above, then
\begin{eqnarray}
   x & = & \phi\sub{c}   + 45\degr \chi \, , \\
   y & = & \theta\sub{c} + 45\degr \psi \, .
\end{eqnarray}
To invert these first determine to which face the $(x,y)$ coordinates refer,
then compute
\begin{eqnarray}
   \chi & = & (x - \phi\sub{c})   / 45\degr \, , \\
   \psi & = & (y - \theta\sub{c}) / 45\degr \, ,
\end{eqnarray}
then
\begin{eqnarray}
   \zeta  & = & 1/\sqrt{1 + \chi^2 + \psi^2} \, .
\end{eqnarray}
Once $\zeta$ is known $\xi$ and $\eta$ are obtained via
\begin{eqnarray}
   \xi    & = & \chi \zeta \, , \\
   \eta   & = & \psi \zeta \, .
\end{eqnarray}
The direction cosines $(l,m,n)$ may be identified with $(\xi,\eta,\zeta)$ with
with the aid of Table~\ref{ta:cubeface}, whence $(\phi,\theta)$ may readily be
computed.  The projection is illustrated in Fig.~\ref{fig:TSC} for the full
sphere.


\subsubsection{\keyv{CSC}: COBE quadrilateralized spherical cube}
\label{sec:CSC}

\begin{figure}
   \centerline{\putfig{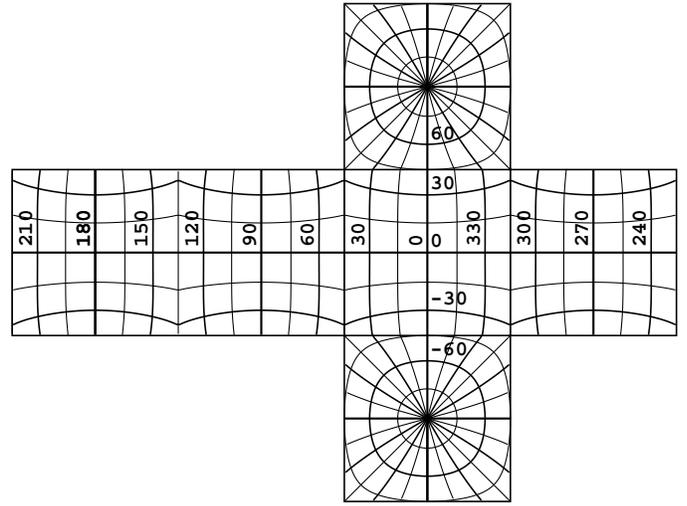}{200pt}}
   \caption[]{COBE quadrilateralized spherical cube projection (\keyv{CSC});
      no limits.}
   \label{fig:CSC}
\end{figure}

The COBE quadrilateralized spherical cube projection illustrated in
Fig.~\ref{fig:CSC} modifies the tangential spherical cube projection in such a
way as to make it approximately equal area.  The forward equations are
\begin{eqnarray}
   x & = & \phi\sub{c}   + 45\degr\, F (\chi, \psi) \, , \\
   y & = & \theta\sub{c} + 45\degr\, F (\psi, \chi) \, ,
\end{eqnarray}
where the function $F$ is given by
\begin{eqnarray}
   F(\chi,\psi) & = & \chi \gamma^* + \chi^3 (1 - \gamma^*) + \nonumber \\
   & & \chi \psi^2 (1 - \chi^2)
          \left[
             \, \Gamma + (M - \Gamma) \chi^2 + \vphantom{\sum_{i=0}^{\infty}}
          \right. \nonumber \\
   & & \hskip 12pt
          \left.
             (1 - \psi^2) \sum_{i=0}^{\infty}
             \sum_{j=0}^{\infty} C_{ij} \chi^{2i}\psi^{2j}
          \right] + \nonumber \\
   & & \chi^3 (1 - \chi^2)
          \left[
             \, \Omega_1 - (1 - \chi^2) \sum_{i=0}^{\infty} D_i \chi^{2i}
          \right] . \label{eq:Fab2xy}
\end{eqnarray}
$C_{ij}$ and $D_i$ are derived from $c_{ij}^*$ and $d_i^*$ as given by Chan \&
O'Neill (\cite{kn:CO}).  The other parameters are given by exact formul\ae\
developed by O'Neill \& Laubscher (\cite{kn:OL}), who provide the numeric
values of their parameters in tables and software listings.  Both disagree
with their formul\ae, but the software listings do contain the actual numeric
parameters still in use for the COBE Project (Immanuel Freedman, private
communication, 1993).  They are
\begin{displaymath}
\begin{array}{lcl}
   \gamma^* & = & \SP 1.37484847732   \\
   M        & = & \SP 0.004869491981  \\
   \Gamma   & = &    -0.13161671474   \\
   \Omega_1 & = &    -0.159596235474  \\
   C_{00}   & = & \SP 0.141189631152  \\
   C_{10}   & = & \SP 0.0809701286525 \\
   C_{01}   & = &    -0.281528535557  \\
   C_{20}   & = &    -0.178251207466  \\
   C_{11}   & = & \SP 0.15384112876   \\
   C_{02}   & = & \SP 0.106959469314  \\
   D_0      & = & \SP 0.0759196200467 \\
   D_1      & = &    -0.0217762490699
\end{array}
\end{displaymath}

\begin{figure}
   \centerline{\putfig{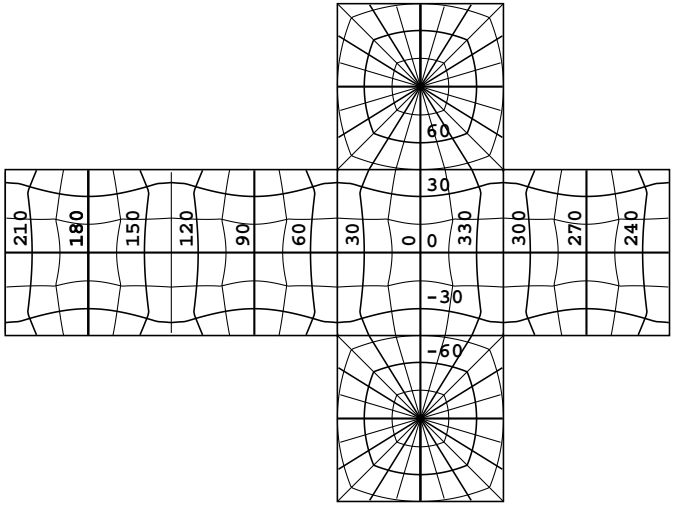}{200pt}}
   \caption[]{Quadrilateralized spherical cube projection (\keyv{QSC}); no
      limits.}
   \label{fig:QSC}
\end{figure}

Chan \& O'Neill (\cite{kn:CO}) actually defined the projection via the inverse
equations.  Their formulation may be rewritten in a more convenient form which
is now the current usage in the COBE Project (Immanuel Freedman, private
communication, 1993):
\begin{eqnarray}
   \chi & = & f(x-\phi\sub{c}, y-\theta\sub{c}) \, , \\
   \psi & = & f(y-\theta\sub{c}, x-\phi\sub{c}) \, ,
\end{eqnarray}
where
\begin{eqnarray}
   \lefteqn{f(x-\phi\sub{c},y-\theta\sub{c}) =} \nonumber \\*
       &  & \hskip 40pt X + X \left(1-X^2\right) \sum_{j=0}^N
              \sum_{i=0}^{N-j} P_{ij} X^{2i} Y^{2j} \, , \label{eq:fxy2ab}
\end{eqnarray}
and
\begin{eqnarray}
   X & = & (x-\phi\sub{c})  /45\degr \nonumber \, , \\
   Y & = & (y-\theta\sub{c})/45\degr \nonumber \, .
\end{eqnarray}
For COBE, $N = 6$ and the best-fit parameters have been taken to be
\begin{displaymath}
\begin{array}{@{\hskip 15pt}lcr@{\hskip 20pt}lcr}
   P_{00} &=& -0.27292696 & P_{04} &=&  0.93412077 \\
   P_{10} &=& -0.07629969 & P_{50} &=&  0.25795794 \\
   P_{01} &=& -0.02819452 & P_{41} &=&  1.71547508 \\
   P_{20} &=& -0.22797056 & P_{32} &=&  0.98938102 \\
   P_{11} &=& -0.01471565 & P_{23} &=& -0.93678576 \\
   P_{02} &=&  0.27058160 & P_{14} &=& -1.41601920 \\
   P_{30} &=&  0.54852384 & P_{05} &=& -0.63915306 \\
   P_{21} &=&  0.48051509 & P_{60} &=&  0.02584375 \\
   P_{12} &=& -0.56800938 & P_{51} &=& -0.53022337 \\
   P_{03} &=& -0.60441560 & P_{42} &=& -0.83180469 \\
   P_{40} &=& -0.62930065 & P_{33} &=&  0.08693841 \\
   P_{31} &=& -1.74114454 & P_{24} &=&  0.33887446 \\
   P_{22} &=&  0.30803317 & P_{15} &=&  0.52032238 \\
   P_{13} &=&  1.50880086 & P_{06} &=&  0.14381585
\end{array}
\end{displaymath}

\noindent
Given the face number, $\chi$, and $\psi$, the native coordinates
$(\phi,\theta)$ may be computed as for the tangential spherical cube
projection.

Equations~(\ref{eq:Fab2xy}) and (\ref{eq:fxy2ab}), the forward and reverse
projection equations used by COBE, are not exact inverses.  Each set could of
course be inverted to any required degree of precision via iterative methods
(in that case Eq.~(\ref{eq:fxy2ab}) should be taken to define the projection).
However, the aim here is to describe the projection in use within the COBE
project.  One may evaluate the closure error in transforming
$(x-\phi\sub{c},y-\theta\sub{c})$ to $(\chi,\psi)$ with
Eq.~(\ref{eq:fxy2ab}) and then transforming back to
$(x-\phi\sub{c},y-\theta\sub{c})$ with Eq.~(\ref{eq:Fab2xy}), i.e.\ 
\begin{eqnarray*}
   E_{ij}^2 & = & \left(F(f(x_i,y_j),f(y_j,x_i)) - x_i\right)^2\\
           & & \mbox{}+\left(F(f(y_j,x_i),f(x_i,y_j)) - y_j\right)^2 \, .
\end{eqnarray*}
The COBE parameterization produces an average error of 4.7 arcsec over the
full field.  The root mean square and peak errors are 6.6 and 24 arcsec,
respectively.  In the central parts of the image ($|X|, |Y| \leq 0.8$), the
average and root mean square errors are 5.9 and 8.0 arcsec, larger than for
the full field.

Measures of equal-area conformance obtained for Eq.~(\ref{eq:fxy2ab}) show
that the rms deviation is 1.06\%\ over the full face and 0.6\%\ over the inner
64\%\ of the area of each face.  The maximum deviation is $+13.7$\%\ and
$-4.1$\%\ at the edges of the face and only $\pm1.3$\%\ within the inner
64\%\ of the face.


\subsubsection{\keyv{QSC}: quadrilateralized spherical cube}
\label{sec:QSC}

O'Neill \& Laubscher (\cite{kn:OL}) derived an exact expression for an
equal-area transformation from a sphere to the six faces of a cube.  At that
time, their formulation was thought to be computationally intractable, but
today, with modern computers and telescopes of higher angular resolution than
COBE, their formulation has come into use.  Fred Patt (1993, private
communication) has provided us with the inverse of the O'Neill \& Laubscher
formula and their expression in Cartesian coordinates.

O'Neill \& Laubscher's derivation applies only in the quadrant
$-45\degr \leq \phi \leq 45\degr$ and must be reflected into the other
quadrants.  This has the effect of making the projection non-differentiable
along the diagonals as is evident in Fig.~\ref{fig:QSC}.  To compute the
forward projection first identify the face and find $(\xi,\eta,\zeta)$ and
$(\phi\sub{c},\theta\sub{c})$ from Table~\ref{ta:cubeface}.  Then
\begin{equation}
   (x,y) = (\phi\sub{c},\theta\sub{c}) +
              \left\{ \begin{array}{ll}
                        (u,v) & \mbox{if $|\xi| > |\eta|$} \\
                        (v,u) & \mbox{otherwise}
                      \end{array} \right. \, ,
\end{equation}
where
\begin{eqnarray}
      u & = & 45\degr S\, \sqrt{\frac{1-\zeta}{1-1/\sqrt{2+\omega^2}}} \, , \\
      v & = & \frac{u}{15\degr} \left[ \tan^{-1}\left( \omega \right) -
              \sin^{-1} \left( \frac{\omega}{\sqrt{2(1+\omega^2)}} \right)
              \right] \, , \\
 \omega & = & \left\{ \begin{array}{ll}
              \eta / \xi  & \mbox{if $|\xi| > |\eta|$} \\
              \xi  / \eta & \mbox{otherwise}
              \end{array} \right. \, , \nonumber \\
      S & = & \left\{ \begin{array}{ll}
              +1 & \mbox{if $\xi > |\eta|$ or $\eta > |\xi|$} \\
              -1 & \mbox{otherwise}
              \end{array} \right. \, . \nonumber
\end{eqnarray}
To compute the inverse first identify the face from the $(x,y)$ coordinates,
then determine $(u,v)$ via
\begin{equation}
  (u,v) = \left\{ \begin{array}{ll}
                     (x-\phi\sub{c},y-\theta\sub{c}) &
                        \mbox{if $|x-\phi\sub{c}| > |y-\theta\sub{c}|$} \\
                     (y-\theta\sub{c},x-\phi\sub{c}) &
                        \mbox{otherwise}
                  \end{array} \right. \, .
\end{equation}
Then
\begin{equation}
  \zeta = 1 - \left(\frac{u}{45\degr}\right)^2 \left(1 -
              \frac{1}{\sqrt{2+\omega^2}} \right) \, ,
\end{equation}
where
\begin{equation}
 \omega = \frac{\sin(15\degr v/u)}{\cos(15\degr v/u) - 1/\sqrt{2}} \, .
\end{equation}
If $|x-\phi\sub{c}|>|y-\theta\sub{c}|$ then
\begin{eqnarray}
    \xi & = & \sqrt{\frac{1-\zeta^2}{1+\omega^2}} \, , \\
   \eta & = & \xi\omega \, ,
\end{eqnarray}
otherwise
\begin{eqnarray}
   \eta & = & \sqrt{\frac{1-\zeta^2}{1+\omega^2}} \, , \\
    \xi & = & \eta\omega \, .
\end{eqnarray}
Given the face number and $(\xi,\eta,\zeta)$, the native coordinates
$(\phi,\theta)$ may be computed with reference to Table~\ref{ta:cubeface} as
for the tangential spherical cube projection.


\section{Support for the AIPS convention}
\label{sec:previous}

A large number of FITS images have been written using the AIPS coordinate
convention and a substantial body of software exists to interpret it.
Consequently, the AIPS convention has acquired the status of a de facto
standard and FITS interpreters will need to support it indefinitely in order
to obey the maxim ``once FITS always FITS''.  Translations between the old and
new system are therefore required.


\subsection{Interpreting old headers}

In the AIPS convention, \CROTA{i} assigned to the latitude axis was used to
define a bulk rotation of the image plane.  Since this rotation was applied
{\em after} \CDELT{i} the translation to the current formalism follows from
\begin{eqnarray}
   \lefteqn{
      \left(
         \begin{array}{cc}
            \keyw{CDELT1} & 0             \\
            0             & \keyw{CDELT2}
         \end{array}
      \right)
      \left(
         \begin{array}{cc}
            \keyw{PC1\_1} & \keyw{PC1\_2} \\
            \keyw{PC2\_1} & \keyw{PC2\_2}
         \end{array}
      \right) =
   } \label{eq:AIPS2NEW} \\
   & & \hspace{70pt}
   \left(
      \begin{array}{rr}
         \cos\rho &   -\sin\rho \\
         \sin\rho &    \cos\rho
      \end{array}
   \right)
   \left(
      \begin{array}{cc}
         \keyw{CDELT1} & 0             \\
         0             & \keyw{CDELT2}
      \end{array}
   \right) \, , \nonumber
\end{eqnarray}
where we have used subscript $1$ for the longitude axis, $2$ for latitude,
and written $\rho$ for the value of \keyw{CROTA2}.
Equation~(\ref{eq:AIPS2NEW}), which includes the added constraint of
preserving \CDELT{i} in the translation, is readily solved for the elements of
the \PC{i}{ja} matrix
\begin{equation}
   \left(
      \begin{array}{cc}
         \keyw{PC1\_1} & \keyw{PC1\_2} \\
         \keyw{PC2\_1} & \keyw{PC2\_2}
      \end{array}
   \right) =
   \left(
      \begin{array}{rr}
                           \cos\rho & -\lambda \sin\rho \\
         \frac{1}{\lambda} \sin\rho &          \cos\rho
      \end{array}
   \right) \, , \label{eq:CROTA2PC}
\end{equation}
where
\begin{equation}
   \lambda = \keyw{CDELT2} / \keyw{CDELT1} \, .
\end{equation}
Note that Eq.~(\ref{eq:CROTA2PC}) defines a rotation if and only if
$\lambda = \pm 1$, which is often the case.  In fact, the operations of
scaling and rotation are commutative if and only if the scaling is isotropic,
i.e.\ $\lambda = +1$; for $\lambda = -1$ the direction of the rotation is
reversed.  However, whatever the value of $\lambda$, the interpretation of
Eq.~(\ref{eq:AIPS2NEW}) as that of a scale followed by a rotation is
preserved.

The translation for \CD{i}{ja} is simpler, effectively because the \CDELT{i}
have an implied value of unity and the constraint on preserving them in the
translation is dropped:
\begin{eqnarray}
   \left(
      \begin{array}{cc}
         \keyw{CD1\_1} & \keyw{CD1\_2} \\
         \keyw{CD2\_1} & \keyw{CD2\_2}
      \end{array}
   \right) & = &
   \left(
      \begin{array}{rr}
         \keyw{CDELT1}\,\cos\rho & -\keyw{CDELT2}\,\sin\rho \\
         \keyw{CDELT1}\,\sin\rho &  \keyw{CDELT2}\,\cos\rho
      \end{array} \right) . \nonumber \\
   & & \label{eq:CROTA2CD}
\end{eqnarray}

The expressions in Hanisch \& Wells (\cite{kn:HW}) and Geldzahler \&
Schlesinger (\cite{kn:NOSTuser}) yield the same results as
Eq.~(\ref{eq:CROTA2CD}) for the usual left-handed sky coordinates and
right-handed pixel coordinates, but can lead to an incorrect interpretation
(namely, possible sign errors for the off-diagonal elements) for other
configurations of the coordinate systems.  The Hanisch \& Wells draft and the
coordinate portions of Geldzahler \& Schlesinger are superseded by this paper.


\subsubsection{\keyv{SIN}}

The \keyv{SIN} projection defined by Greisen (\cite{kn:G1}) is here
generalized by the addition of projection parameters.  However, these
parameters assume default values which reduce to the simple orthographic
projection of the AIPS convention.  Therefore no translation is required.


\subsubsection{\keyv{NCP}}
\label{sec:NCP}

The ``north-celestial-pole'' projection defined by Greisen (\cite{kn:G1}) is
a special case of the new generalized \keyv{SIN} projection.  The old header
cards
\begin{eqnarray*}
   \keyw{CTYPE1} & = & \keyv{`RA---NCP'} \, , \\
   \keyw{CTYPE2} & = & \keyv{`DEC--NCP'} \, ,
\end{eqnarray*}
should be translated to the current formalism as
\begin{eqnarray*}
   \keyw{CTYPE1} & = & \keyv{`RA---SIN'}      \, , \\
   \keyw{CTYPE2} & = & \keyv{`DEC--SIN'}      \, , \\
   \keyw{PV2\_1} & = & 0             \, , \\
   \keyw{PV2\_2} & = & \cot\delta_0  \, .
\end{eqnarray*}


\subsubsection{\keyv{TAN}, \keyv{ARC} and \keyv{STG}}

The \keyv{TAN}, \keyv{ARC} and \keyv{STG} projections defined by Greisen
(\cite{kn:G1}, \cite{kn:G2}) are directly equivalent to those defined here and
no translation is required.


\subsubsection{\keyv{AIT}, \keyv{GLS} and \keyv{MER}}
\label{sec:AITGLSMER}

Special care is required in interpreting the \keyv{AIT} (Hammer-Aitoff),
\keyv{GLS} (Sanson-Flamsteed), and \keyv{MER} (Mercator) projections in the
AIPS convention as defined by Greisen (\cite{kn:G2}).  As explained in
Sect.~\ref{sec:oblique}, the AIPS convention cannot represent oblique
celestial coordinate graticules such as the one shown in Fig.~\ref{fig:Grids}.
\CRVAL{i} for these projections in AIPS does not correspond to the celestial
coordinates $(\alpha_0,\delta_0)$ of the reference point, as understood in
this formalism, unless they are both zero in which case no translation is
required.

A translation into the new formalism exists for non-zero \CRVAL{i} but only if
\CROTA{i} is zero.  It consists of setting \CRVAL{i} to zero and adjusting
\CRPIX{j} and \CDELT{i} accordingly in the AIPS header whereupon the above
situation is obtained.  The corrections to \CRPIX{j} are obtained by
computing the pixel coordinates of $(\alpha,\delta) = (0,0)$ within the AIPS
convention.  For \keyv{AIT} and \keyv{MER} (but not \keyv{GLS}), \CDELT{i}
must also be corrected for the scaling factors $f_\alpha$ and $f_\delta$
incorporated into the AIPS projection equations.

Of the three projections only \keyv{GLS} is known to have been used with
non-zero \CRVAL{i}.  Consequently we have renamed it as \keyv{SFL} as a
warning that translation is required.


\subsection{Supporting old interpreters}

As mentioned in Sect.~\ref{sec:previous}, FITS interpreters will need to
recognize the AIPS convention virtually forever.  It stands to reason,
therefore, that if modern FITS-writers wish to assist older FITS interpreters
they may continue to write older style headers, assuming of course that it is
possible to express the coordinate system in the AIPS convention.

Modern FITS-writers must not attempt to help older interpreters by including
\CROTA{i} together with the new keyword values (assuming the combination of
\CDELT{i} and \PC{i}{ja} matrix, or \CD{i}{ja} matrix, is amenable to such
translation).  We make this requirement primarily to minimize confusion.

Assuming that a header has been developed using the present formalism the
following test may be applied to determine whether the combination of
\CDELT{ia} and \PC{i}{ja} matrix represents a scale followed by a rotation as
in Eq.~(\ref{eq:AIPS2NEW}).  Firstly write
\begin{eqnarray}
   \lefteqn{
      \left(
         \begin{array}{cc}
            \keyw{CD1\_1} & \keyw{CD1\_2} \\
            \keyw{CD2\_1} & \keyw{CD2\_2}
         \end{array}
      \right) =
   } \nonumber \\
   & & \hskip 30pt
   \left(
      \begin{array}{cc}
         \keyw{CDELT1} & 0             \\
         0             & \keyw{CDELT2}
      \end{array}
   \right)
   \left(
      \begin{array}{cc}
         \keyw{PC1\_1} & \keyw{PC1\_2} \\
         \keyw{PC2\_1} & \keyw{PC2\_2}
      \end{array}
   \right) \, ,
   \label{eq:CDji}
\end{eqnarray}
where $1$ is the longitude coordinate and $2$ the latitude coordinate, then
evaluate $\rho\sub{a}$ and $\rho\sub{b}$ as
\begin{eqnarray}
   \rho\sub{a} & = &
      \left\{
         \begin{array}{ll}
            \arg\,(\SP \keyw{CD1\_1}, \SP \keyw{CD2\_1})
               & \mbox{if $\keyw{CD2\_1} > 0$} \\
            0
               & \mbox{if $\keyw{CD2\_1} = 0$} \\
            \arg\,(  - \keyw{CD1\_1},   - \keyw{CD2\_1})
               & \mbox{if $\keyw{CD2\_1} < 0$}
         \end{array}
      \right. \, , \nonumber \\
   \rho\sub{b} & = &
      \left\{
         \begin{array}{ll}
            \arg\,(  - \keyw{CD2\_2}, \SP \keyw{CD1\_2})
               & \mbox{if $\keyw{CD1\_2} > 0$} \\
            0
               & \mbox{if $\keyw{CD1\_2} = 0$} \\
            \arg\,(\SP \keyw{CD2\_2},   - \keyw{CD1\_2})
               & \mbox{if $\keyw{CD1\_2} < 0$}
         \end{array}
      \right. \, .
\end{eqnarray}
If $\rho\sub{a} = \rho\sub{b}$ to within reasonable precision (Geldzahler \&
Schlesinger, \cite{kn:NOSTuser}), then compute
\begin{equation}
   \rho = (\rho\sub{a} + \rho\sub{b}) / 2
\end{equation}
as the best estimate of the rotation angle, the older keywords are
\begin{eqnarray}
  \keyw{CDELT1} & = & \keyw{CD1\_1} / \cos\rho \, , \nonumber \\
  \keyw{CDELT2} & = & \keyw{CD2\_2} / \cos\rho \, , \label{eq:AIPSCD} \\
  \keyw{CROTA2} & = & \rho                     \, . \nonumber
\end{eqnarray}
Note that the translated values of \CDELT{i} in Eqs.~(\ref{eq:AIPSCD}) may
differ from the starting values in Eq.~(\ref{eq:CDji}).

Solutions for \keyw{CROTA2} come in pairs separated by $180\degr$.  The above
formul\ae\ give the solution which falls in the half-open interval
$[0,180\degr)$.  The other solution is obtained by subtracting $180\degr$ from
\keyw{CROTA2} and negating \keyw{CDELT1} and \keyw{CDELT2}.  While each
solution is equally valid, if one makes $\keyw{CDELT1} < 0$ and
$\keyw{CDELT2} > 0$ then it would normally be the one chosen.

Of course, the projection must be one of those supported by the AIPS
convention, which only recognizes \keyv{SIN}, \keyv{NCP}, \keyv{TAN},
\keyv{ARC}, \keyv{STG}, \keyv{AIT}, \keyv{GLS} and \keyv{MER}\@.  Of these, we
strongly recommend that the AIPS version of \keyv{AIT}, \keyv{GLS}, and
\keyv{MER} not be written because of the problems described in
Sect.~\ref{sec:AITGLSMER}.  It is interesting to note that a translation does
exist for the slant orthographic (\keyv{SIN}) projection defined in
Sect.~\ref{sec:SIN} to the simple orthographic projection of AIPS\@.  However,
we advise against such translation because of the likelihood of creating
confusion and so we do not define it here.  The exception is where the
\keyv{SIN} projection may be translated as \keyv{NCP} as defined in
Sect.~\ref{sec:NCP}.


\section{Discussion}
\label{sec:discussion}


\subsection{Oblique projections}
\label{sec:oblique}

The term {\em oblique projection} is often used when a projection's native
coordinate system does not coincide with the coordinate system of interest.
Texts on terrestrial cartography often separately name and derive projection
equations for particular oblique projections.  Thus the Cassini, Transverse
Mercator, and Bartholomew nordic projections are nothing more than the plate
carr\'{e}e, Mercator, and Mollweide projections in disguise.

The view of obliqueness as being a property of a projection arose mainly
because of the computational difficulty of producing oblique graticules in the
days before electronic computers.  The particular aspects chosen were those
for which geometrical construction was possible or for which the mathematical
formulation had a simple form, and this tended to entrench particular oblique
projections as separate entities.  In addition, terrestrial longitude and
latitude are so closely tied to the visual representation of the Earth's
surface that it is shown predominantly in the usual north-south orientation.
The only other natural terrestrial coordinate system is that defined by the
magnetic pole but the difference between it and the geographic graticule is
sufficiently small that it is usually treated as a local correction to the
magnetic bearing.  The special view of obliqueness was probably also
reinforced by traditional methods of map making using sextant and chronometer,
which were based on geographic longitude and latitude.

The situation in astronomical cartography is quite different.  The celestial
sphere has a variety of natural coordinate systems -- equatorial, ecliptic,
galactic, etc.\ -- and oblique and non-oblique graticules are often plotted
together on the same map.  It wouldn't make sense to describe such a
projection as being simultaneously oblique and non-oblique; clearly it is the
particular coordinate {\em graticule} which may be oblique, not the
projection.  Visually, an area of the sky may be seen in different
orientations depending on whether it's rising, transiting, or setting, or
whether seen from the northern or southern hemisphere.  Moreover, oblique
graticules arise as a normal feature of observations in optical, infrared,
radio, and other branches of astronomy, just as they do now in terrestrial
mapping based on aerial and satellite photography.  The center of the field of
view, wherever it may happen to be, typically corresponds to the native pole
of a zenithal projection.  Thus the aim of this paper has been to provide a
formalism whereby obliquity may be handled in a general way.

\begin{figure*}
   \if\dofig F
      \vspace{620pt}
   \else
      \centerline{\putfig{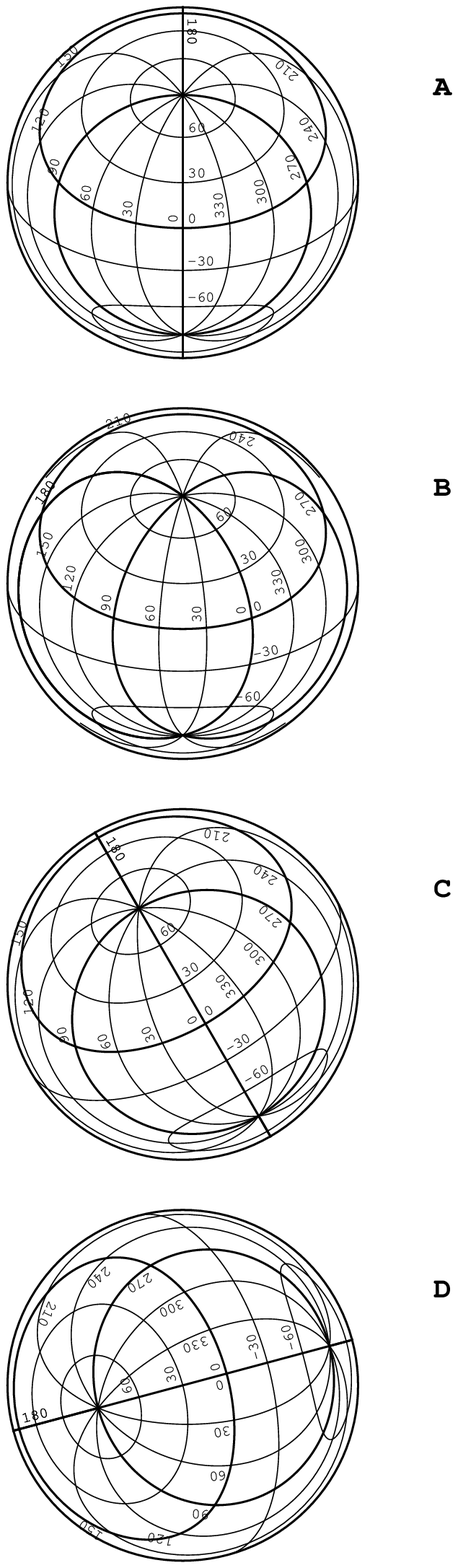}{620pt}
                  \putfig{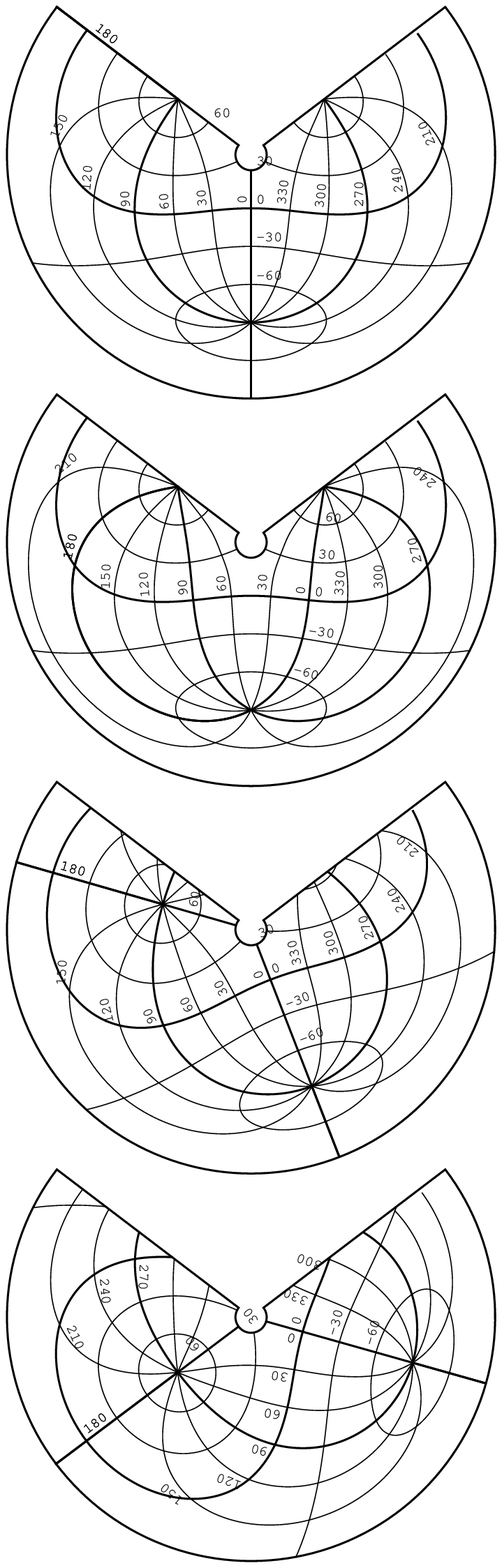}{620pt}}
   \fi
   \vspace{10pt}
   \caption[]{Oblique $(\alpha,\delta)$ graticules plotted for the zenithal
      equal area (\keyv{ZEA}) and conic equidistant (\keyv{COD}) projections
      with parameters
      (A) $\alpha\sub{p} =   0$,
          $\delta\sub{p} =  30\degr$,
          $\phi\sub{p}   = 180\degr$;
      (B) $\alpha\sub{p} =  45\degr$,
          $\delta\sub{p} =  30\degr$,
          $\phi\sub{p}   = 180\degr$;
      (C) $\alpha\sub{p} =  0$,
          $\delta\sub{p} =  30\degr$,
          $\phi\sub{p}   = 150\degr$;
      (D) $\alpha\sub{p} =  0$,
          $\delta\sub{p} =  30\degr$,
          $\phi\sub{p}   =  75\degr$.}
   \label{fig:ZEAex}
\end{figure*}

\begin{figure*}[!ht]
   \if\dofig F
      \vspace{620pt}
   \else
      \centerline{\putfig{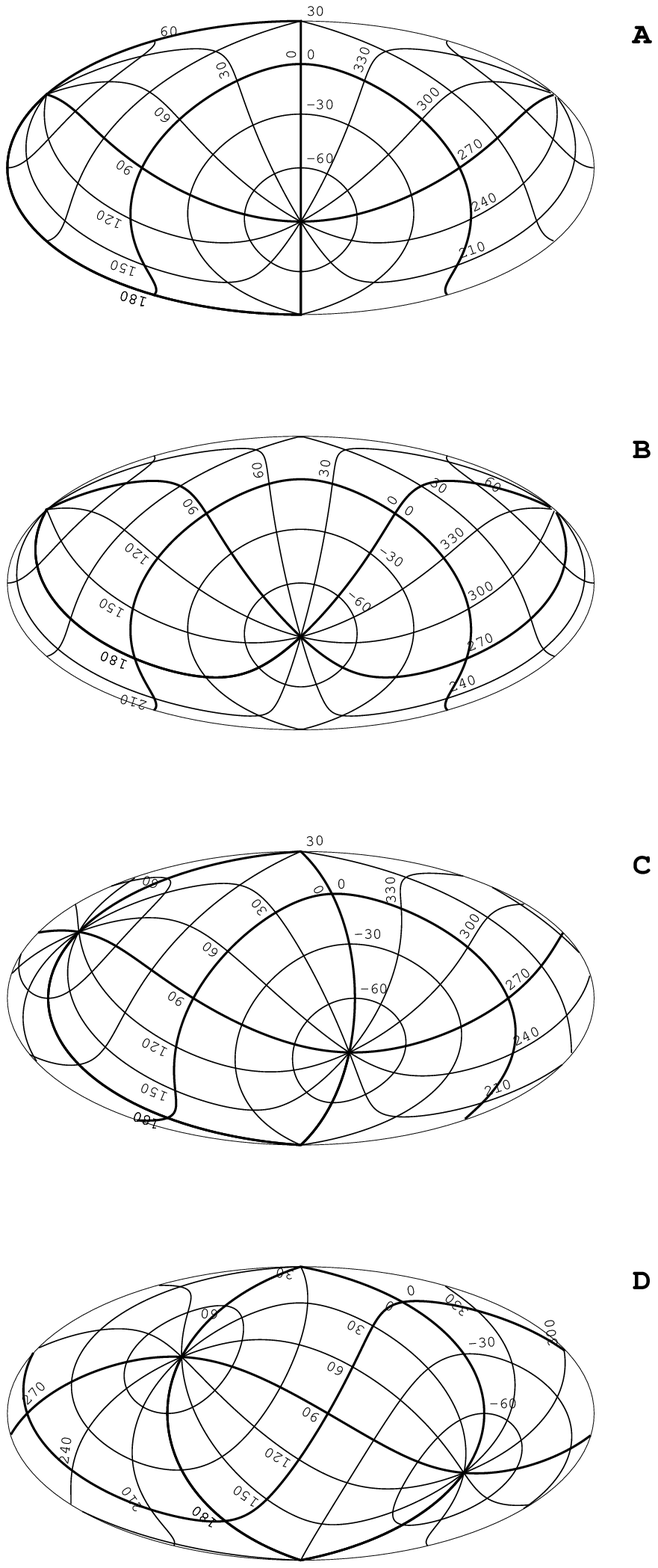}{620pt}
                  \putfig{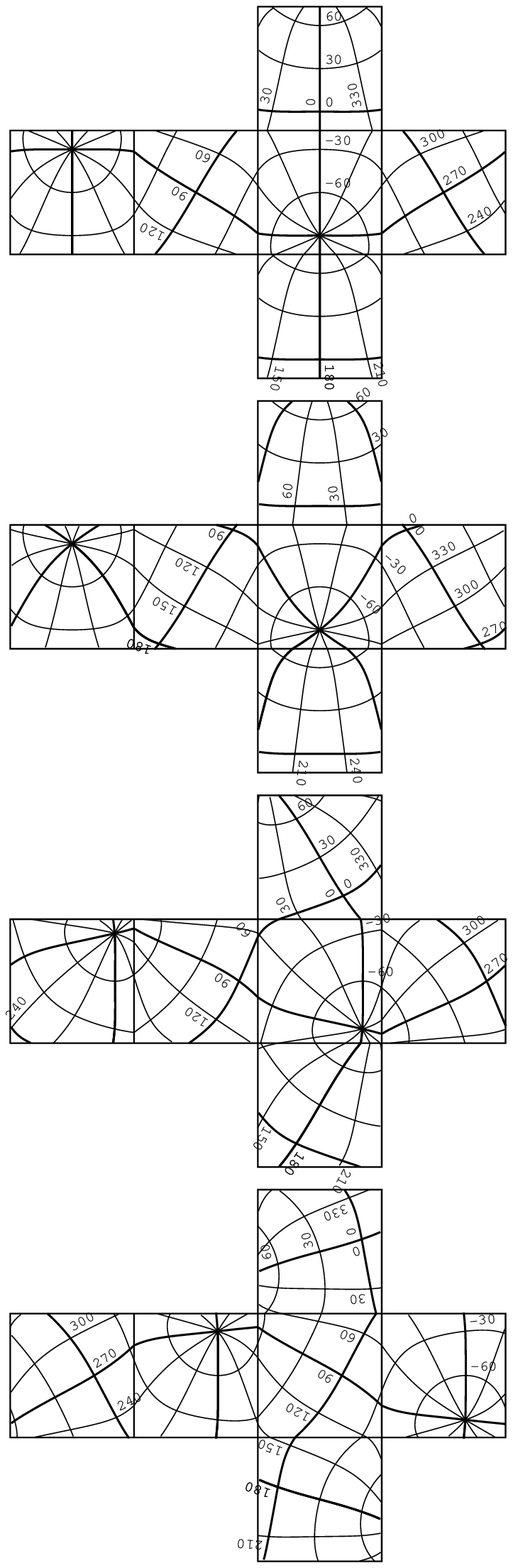}{620pt}}
   \fi
   \vspace{10pt}
   \caption[]{Oblique ($\alpha,\delta$) graticules plotted for the
      Hammer-Aitoff (\keyv{AIT}) and COBE quadrilateralized spherical cube
      (\keyv{CSC}) projections using the same parameters as
      Fig.~\ref{fig:ZEAex}.}
   \label{fig:AITex}
   \vspace{10pt}
\end{figure*}

Figures~\ref{fig:ZEAex} and \ref{fig:AITex} illustrate the same four oblique
celestial graticules for the zenithal equal area, conic equidistant,
Hammer-Aitoff, and COBE quadrilateralized spherical cube projections (at
variable $(x,y)$ scale).  For the sake of intercomparison, these graticules
are defined in terms of the celestial coordinates of the native pole,
$(\alpha\sub{p},\delta\sub{p})$, together with $\phi\sub{p}$ by setting
$(\phi_0,\theta_0) = (0, 90\degr)$ as described in Sect.~\ref{sec:userspec}.

The first graticule, A, when compared to the non-oblique native coordinate
graticules presented earlier for each projection, illustrates the effect of
changing $\delta\sub{p}$ (and hence $\delta_0$).  Comparison of graticules A
and B shows that changing $\alpha\sub{p}$ (and hence $\alpha_0$) results in a
simple change in origin of longitude.  Graticules A, C, and D show the more
interesting effect of varying $\phi\sub{p}$ (conveyed by \LONPOLE{a}).  For
the zenithal projections the result is indistinguishable from a bulk rotation
of the image plane, but this is not the case for any other class of
projection.  This explains why the role of \LONPOLE{a} was covered by
\CROTA{i} in the AIPS convention for the zenithal projections introduced by
Greisen (\cite{kn:G1}), but why this does not work for any other class of
projection.

\subsection{Choice of projection}
\label{sec:choice}

The projected meridians and parallels in Figs.~\ref{fig:ZEAex} and
\ref{fig:AITex} also serve to illustrate the distortions introduced by the
various projections.  In particular, the quad-cube projection, while doing a
good job within each face, is very distorted over the sphere as a whole,
especially where the faces join, so much so that it is difficult even to trace
the path of some of the meridians and parallels.  However, as indicated
previously, this projection is designed for efficient numerical computation
rather than visual representation of the sphere.

This leads us to the question of the choice of projection for a particular
application.  In some cases the projection is the natural result of the
geometry of the observation and there is no choice.  For example, maps
produced by rotational synthesis radio telescopes are orthographic
(\keyv{SIN}) projections with native pole at the phase center of the
observation.  Photographic plates produced by Schmidt telescopes are best
described by the zenithal equidistant (\keyv{ARC}) projection, while the field
of view of other optical telescopes is closer to a gnomonic (\keyv{TAN})
projection.  On the other hand, the great circle scanning technique of the
Sloan Digital Sky Survey produces a cylindrical projection.  Similarly a map
of the surface of the Moon as it appears from Earth requires a zenithal
perspective (\keyv{AZP}) projection with $\mu \approx -220$.  The same is true
for spacecraft generated images of distant moons and planets.  All-sky cameras
should be well served by a \keyv{ZPN} projection with empirically determined
parameters.

Sometimes observational data will be regridded onto a projection chosen for a
particular purpose.  Equal area projections are often used since they preserve
surface density and allow integration, whether numerical or visual, to be
performed with reasonable accuracy by summing pixel values.  The Hammer-Aitoff
(\keyv{AIT}) projection is probably the most widely used for all-sky maps.
However, the Sanson-Flamsteed (\keyv{SFL}) may be preferred as being easier to
take measurements from.  The humble plate carr\'{e}e (\keyv{CAR}) excels in
this regard and may be considered adequate, say, for mapping a few degrees on
either side of the galactic plane.  In general terms, zenithal projections are
good for mapping the region in the vicinity of a point, often a pole;
cylindrical projections are good for the neighborhood of a great circle,
usually an equator; and the conics are suitable for small circles such as
parallels of latitude.  A conic projection might be a good choice for mapping
a hemisphere; distortion at most points is reduced compared to a zenithal
projection although at the expense of the break between native longitudes
$\pm180\degr$.  However, in some cases this break might be considered an
extreme distortion which outweighs other reductions in distortion and,
depending on the application, may mandate the use of a zenithal projection.
Cartographers typically favor conics and polyconics for ``Australia-sized''
regions of the sphere and at this they excel.  Oblique forms of the zenithal
equal-area projection are also commonly used.  Celestial applications might
include large scale maps of the Magellanic clouds.  Tables 3.1 and 4.1 of
Snyder (\cite{kn:Sny}) summarize actual usage in a variety of 19th and 20th
century atlases.

As mentioned in Sect.~\ref{sec:projections}, some projections are not scaled
true at the reference point.  In most cases this is deliberate, usually
because the projection is designed to minimize distortion over a wide area;
the scale will be true at some other place on the projection, typically along
a parallel of latitude.  If the projection is required to have
$(\alpha,\delta)\approx(\alpha_0,\delta_0)+(x,y)$ then a number of projections
will serve provided also that $\phi\sub{p}$ takes its default value.  The
following are always scaled true at the reference point: \keyv{SZP},
\keyv{TAN}, \keyv{SIN}, \keyv{STG}, \keyv{ARC}, \keyv{ZEA}, \keyv{CAR},
\keyv{MER}, \keyv{BON}, \keyv{PCO}, \keyv{SFL}, and \keyv{AIT}\@.  The
following are scaled true at the reference point for the particular conditions
indicated: \keyv{AZP} ($\gamma = 0$), \keyv{ZPN} ($P_0=0, P_1=1$), \keyv{AIR}
($\theta_b=90\degr$), \keyv{CYP} ($\lambda=1$), \keyv{CEA} ($\lambda=1$),
\keyv{COP} ($\eta=0$), \keyv{COD} ($\eta=0$), \keyv{COE} ($\eta=0$), and
\keyv{COO} ($\eta=0$).  The following are never scaled true at the reference
point: \keyv{PAR} (but is scaled true in $x$), \keyv{MOL}, \keyv{TSC},
\keyv{CSC}, and \keyv{QSC}; the latter three are equiscaled at the reference
point.

Of course \CDELT{ia} may be used to control scaling of $(x,y)$ with respect to
$(p_1,p_2)$.  Thus, for example, the plate carr\'{e}e projection (\keyv{CAR})
also serves as the more general equirectangular projection.

\begin{table*}[!t]
   \caption[]{FITS header for example 1.}
   \begin{tabular}{l}
      \hline
      \noalign{\smallskip}
      \verb+123456789 123456789 123456789 123456789 123456789 123456789 123456789 123456789 + \\
      \noalign{\smallskip}
      \hline
      \noalign{\smallskip}
      \verb+NAXIS   =                    4 / 4-dimensional cube+ \\
      \verb+NAXIS1  =                  512 / x axis (fastest)+ \\
      \verb+NAXIS2  =                  512 / y axis (2nd fastest)+ \\
      \verb+NAXIS3  =                  196 / z axis (planes)+ \\
      \verb+NAXIS4  =                    1 / Dummy to give a coordinate+ \\
      \verb+CRPIX1  =                  256 / Pixel coordinate of reference point+ \\
      \verb+CDELT1  =               -0.003 / 10.8 arcsec per pixel+ \\
      \verb+CTYPE1  = 'RA---TAN'           / Gnomonic projection+ \\
      \verb+CRVAL1  =                45.83 / RA at reference point+ \\
      \verb+CUNIT1  = 'deg     '           / Angles are degrees always+ \\
      \verb+CRPIX2  =                  257 / Pixel coordinate of reference point+ \\
      \verb+CDELT2  =                0.003 / 10.8 arcsec per pixel+ \\
      \verb+CTYPE2  = 'DEC--TAN'           / Gnomonic projection+ \\
      \verb+CRVAL2  =                63.57 / Dec at reference point+ \\
      \verb+CUNIT2  = 'deg     '           / Angles are degrees always+ \\
      \verb+CRPIX3  =                    1 / Pixel coordinate of reference point+ \\
      \verb+CDELT3  =               7128.3 / Velocity increment+ \\
      \verb+CTYPE3  = 'VELOCITY'           / Each plane at a velocity+ \\
      \verb+CRVAL3  =             500000.0 / Velocity in m/s+ \\
      \verb+CUNIT3  = 'm/s     '           / metres per second+ \\
      \verb+CRPIX4  =                    1 / Pixel coordinate of reference point+ \\
      \verb+CDELT4  =                    1 / Required here.+ \\
      \verb+CTYPE4  = 'STOKES  '           / Polarization+ \\
      \verb+CRVAL4  =                    1 / Unpolarized+ \\
      \verb+CUNIT4  = '        '           / Conventional unitless = I pol+ \\
      \verb+LONPOLE =                  180 / Native longitude of celestial pole+ \\
      \verb+RADESYS = 'FK5     '           / Mean IAU 1984 equatorial coordinates+ \\
      \verb+EQUINOX =               2000.0 / Equator and equinox of J2000.0+ \\
      \noalign{\smallskip}
      \hline
   \end{tabular}
   \label{ta:interp1}
\end{table*}


\subsection{Header interpretation examples}
\label{sec:interp}

We now consider three examples chosen to illustrate how a FITS reader would
interpret a celestial coordinate header.

Example 1 is a simple header whose interpretation is quite straightforward.

Example 2 is more complicated; it also serves to illustrate the WCS header
cards for 1) image arrays in binary tables, and 2) pixel lists.

Example 3 highlights a subtle problem introduced into a header by the FITS
writer and considers how this should be corrected.  As such, it introduces
the generally more difficult task of composing WCS headers, considered in
greater detail in Sect.~\ref{sec:construct}.


\subsubsection{Header interpretation example 1}
\label{sec:interp1}

\begin{figure*}
   \centerline{\putfig{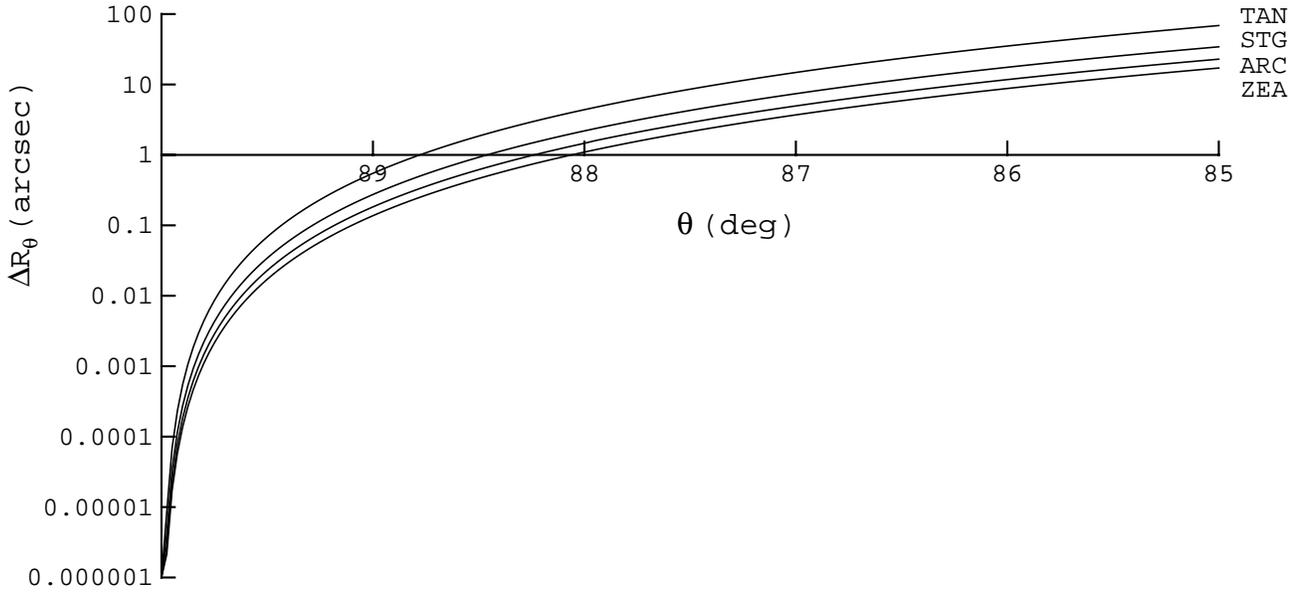}{230pt}}
   \caption[]{$\Rt(projection) - \Rt(\keyv{SIN})$ in arcsec plotted versus
      $\theta$ for various projections.}
   \label{fig:ZenDif}
\end{figure*}

Consider as the first example the relatively simple optical image whose header
is given in Table~\ref{ta:interp1}.  The \keyw{NAXIS} and \keyi{NAXIS}{j}
keywords indicate that we have a four-dimensional image consisting of 512
columns, 512 rows, 196 planes, and one polarization.  The degenerate
\keyv{STOKES} axis is used simply to convey a coordinate value which applies
to every pixel in the image.  The \CRPIX{j} keywords tell us that the
reference point is at pixel coordinate $(256,257,1,1)$.  The \PC{i}{ja}
keywords default to the unit matrix which indicates that no bulk rotation or
shear is applied to the pixel coordinates.  Intermediate world coordinates may
thus be computed from Eq.~(\ref{eq:px}) as
\begin{eqnarray*}
   \left(
      \begin{array}{c}
         x \\
         y \\
         z \\
         s
      \end{array}
   \right) & = &
   \left(
      \begin{array}{cccc}
         -0.003 &   0   &    0   & 0 \\
            0   & 0.003 &    0   & 0 \\
            0   &   0   & 7128.3 & 0 \\
            0   &   0   &    0   & 1
      \end{array}
   \right)
   \left(
      \begin{array}{ccl}
         p_1 & - & 256 \\
         p_2 & - & 257 \\
         p_3 & - & 1 \\
         p_4 & - & 1
      \end{array}
   \right) \, .
\end{eqnarray*}

Since `\keyv{VELOCITY}' and `\keyv{STOKES}' are linear axis types, the
velocity and Stokes value of each point are found simply by adding the
coordinate value at the reference point to the relative coordinate.  Thus,
\begin{eqnarray*}
   {\rm Velocity} & = & 500000.0 + 7128.3\,(p_3 - 1)\ {\rm m\,s}^{-1} \, , \\
   {\rm Stokes}   & = & 1\ {\rm (I\ polarization)} \, .
\end{eqnarray*}

\begin{table}
   \caption[]{Sample calculations for example 1.}
   \begin{tabular}{clrrr}
      \hline
      \noalign{\smallskip}
      parameter & units & SE corner & NE corner & NW corner \\
      \noalign{\smallskip}
      \hline
      \noalign{\smallskip}
      $(p_1,p_2)$ & pixel
                  &         (1, 2) &        (1, 512) &      (511, 512) \\
      $(p_3,p_4)$ & pixel
                  &         (1, 1) &          (1, 1) &        (196, 1) \\
      $x$         & deg
                  &  $0\fdg765000$ &   $0\fdg765000$ &  $-0\fdg765000$ \\
      $y$         & deg
                  & $-0\fdg765000$ &   $0\fdg765000$ &   $0\fdg765000$ \\
      $\phi$      & deg
                  & $45\fdg000000$ & $135\fdg000000$ & $225\fdg000000$ \\
      $\theta$    & deg
                  & $88\fdg918255$ &  $88\fdg918255$ &  $88\fdg918255$ \\
      $\alpha$    & deg
                  & $47\fdg503264$ &  $47\fdg595581$ &  $44\fdg064419$ \\
      $\delta$    & deg
                  & $62\fdg795111$ &  $64\fdg324332$ &  $64\fdg324332$ \\
      Velocity    & ms$^{-1}$
                  &      500000.00 &       500000.00 &      1890018.50 \\
      Stokes      &
                  & $1.0 \equiv$ I &  $1.0 \equiv$ I &  $1.0 \equiv$ I \\
      \noalign{\smallskip}
      \hline
   \end{tabular}
   \label{ta:working1}
\end{table}

The \keyw{CTYPE1} and \keyw{CTYPE2} keywords denote a \keyv{TAN} (gnomonic)
projection for which $(x,y)$ are projection plane coordinates for the zenithal
perspective projection with the source of the projection at the center of the
sphere.  Thus the native longitude and latitude are given by
\begin{eqnarray*}
    \phi & = & \arg\,(-y,x) \, , \\
  \theta & = & \tan^{-1} \left( \rd\frac{1}{\sqrt{x^2+y^2}} \right) \, ,
\end{eqnarray*}
which, on substitution, become
\begin{eqnarray*}
    \phi & = & \arg\,(p_2-257, p_1-256) + 180\degr \, , \\
  \theta & = & \tan^{-1} \left( \frac{19098\fdg5932}{\sqrt{
                        (p_1-256)^2+(p_2-257)^2}} \right) \, ,
\end{eqnarray*}
where we have used Eqs.~(\ref{eq:azrt}), (\ref{eq:azphi}), and
(\ref{eq:TANtheta}).

The celestial coordinate system is equatorial since the \CTYPE{ia} begin with
\keyv{RA} and \keyv{DEC} and the \RADESYS{a} and \EQUINOX{a} cards denote that
these are in the IAU 1984 system.  Zenithal projections have
$(\phi_0,\theta_0) = (0,90\degr)$ for which the \CRVAL{i} give equatorial
coordinates $\alpha\sub{p}=45\fdg83$ in right ascension and
$\delta\sub{p}=63\fdg57$ in declination.  The equatorial north pole has a
longitude of $180\degr$ in the native coordinate system from the \LONPOLE{a}
keyword.  \LATPOLE{a} is never required for zenithal projections and was not
given.  Thus, Eqs.~(\ref{eq:nat2std2}) for the right ascension and declination
become
\begin{eqnarray*}
\begin{array}{l}
   \sin\delta = \sin\theta \sin(63\fdg57) -
                       \cos\theta\cos\phi\cos(63\fdg57) \, , \\
   \cos\delta\sin(\alpha-45\fdg83) = \cos\theta\sin\phi \, , \\
   \cos\delta\cos(\alpha-45\fdg83) = \sin\theta\cos(63\fdg57) \\
      \hphantom{\cos\delta\cos(\alpha-45\fdg83) = } \mbox{} +
         \cos\theta\cos\phi\sin(63\fdg57) \, .
\end{array}
\end{eqnarray*}
Sample calculations for points on the diagonal near the three corners of the
image are shown in Table~\ref{ta:working1}.

If we define the projection non-linearity as the departure of $\theta$ from
$r = \sqrt{x^2 + y^2}$, then in this $1\fdg5 \times 1\fdg5$ image it amounts
to $\sim 0\farcs5$ at the corners.  However, in a $6\degr \times 6\degr$ image
it quickly escalates to $\sim 0\farcm5$, sixty times larger.  Comparison of
$\alpha$ for the southeast and northeast corners indicates the significant
effect of grid convergence even in this moderate sized image.  The two differ
by $22\fs16$, most of which is attributable to the $\cos\delta$ term in
converting longitude offsets to true angular distances.  The effect of grid
convergence is small for $(\alpha_0,\delta_0)$ near the equator and large near
the poles.

Figure~\ref{fig:ZenDif} investigates the effect of projective non-linearities
for moderate field sizes for the commonly used zenithal projections.  It shows
the difference in $\Rt$ between the various projections and the \keyv{SIN}
projection as a function of native latitude $\theta$.  The \keyv{SIN}
projection is used as reference since it always has $\Rt$ less than the other
zenithal projections.  The difference for all projections exceeds 1~arcsec for
values of $\theta$ less than $88\degr$ and the difference for the \keyv{TAN}
projection exceeds one milliarcsec only 440 arcsec from the native pole.  Such
nonlinearities would be significant in optical, VLBI, and other high
resolution observations.


\subsubsection{Header interpretation example 2}
\label{sec:interp2}

While the previous header was a realistic example it overlooked many of the
concepts introduced in this paper.  Consider now the header given in
Table~\ref{ta:interp2}.  Although contrived to illustrate as much as possible
in one example, this header could conceivably arise as a ``tile'' from a conic
equal area projection of a region of the southern galactic hemisphere centered
at galactic coordinates $(\ell,b) = (90\degr,-25\degr)$.  The tile size of
$2048 \times 2048$ pixels is approximately $10\degr \times 10\degr$, and this
particular tile is situated immediately to the north of the central tile.

\begin{table*}
   \caption[]{Second example FITS header (blank lines have been inserted for
      clarity).}
   \begin{tabular}{l}
      \hline
      \noalign{\smallskip}
      \verb+123456789 123456789 123456789 123456789 123456789 123456789 123456789 123456789 + \\
      \noalign{\smallskip}
      \hline
      \noalign{\smallskip}
      \verb+NAXIS   =                    2 / 2-dimensional image+ \\
      \verb+NAXIS1  =                 2048 / x axis (fastest)+ \\
      \verb+NAXIS2  =                 2048 / y axis (2nd fastest)+ \\
      \verb+MJD-OBS =        44258.7845612 / MJD at start of observation+ \\
      \\
      \verb+CRPIX1  =               1024.5 / Pixel coordinate of reference point+ \\
      \verb+CRPIX2  =              -1023.5 / Pixel coordinate of reference point+ \\
      \verb+PC1_1   =                    1 / Transformation matrix element+ \\
      \verb+PC1_2   =               -0.004 / Transformation matrix element+ \\
      \verb+PC2_1   =               -0.002 / Transformation matrix element+ \\
      \verb+PC2_2   =                    1 / Transformation matrix element+ \\
      \verb+CDELT1  =               -0.005 / x-scale+ \\
      \verb+CDELT2  =                0.005 / y-scale+ \\
      \verb+CTYPE1  = 'GLON-COE'           / Conic equal area projection+ \\
      \verb+CTYPE2  = 'GLAT-COE'           / Conic equal area projection+ \\
      \verb+PV2_1   =                -25.0 / Conic mid-latitude+ \\
      \verb+CRVAL1  =                 90.0 / Galactic longitude at reference point+ \\
      \verb+CRVAL2  =                -25.0 / Galactic latitude at reference point+ \\
      \\
      \verb+CRPIX1A =               1024.5 / Pixel coordinate of reference point + \\
      \verb+CRPIX2A =              -1023.5 / Pixel coordinate of reference point + \\
      \verb+PC1_1A  =                    1 / Transformation matrix element+ \\
      \verb+PC1_2A  =               -0.004 / Transformation matrix element+ \\
      \verb+PC2_1A  =               -0.002 / Transformation matrix element+ \\
      \verb+PC2_2A  =                    1 / Transformation matrix element+ \\
      \verb+CDELT1A =               -0.005 / x-scale+ \\
      \verb+CDELT2A =                0.005 / y-scale+ \\
      \verb+CTYPE1A = 'ELON-COE'           / Conic equal area projection+ \\
      \verb+CTYPE2A = 'ELAT-COE'           / Conic equal area projection+ \\
      \verb+PV2_1A  =                -25.0 / Conic mid-latitude+ \\
      \verb+CRVAL1A =           -7.0300934 / Ecliptic longitude at reference point+ \\
      \verb+CRVAL2A =           34.8474143 / Ecliptic latitude at reference point+ \\
      \verb+LONPOLEA=            6.3839706 / Native longitude of ecliptic pole+ \\
      \verb+LATPOLEA=           29.8114400 / Ecliptic latitude of native pole+ \\
      \verb+RADESYSA= 'FK5     '           / Mean IAU 1984 ecliptic coordinates+ \\
      \noalign{\smallskip}
      \hline
   \end{tabular}
   \label{ta:interp2}
\end{table*}

The header defines ecliptic coordinates as an alternate coordinate system,
perhaps to help define the distribution of zodiacal light.  This has the same
reference point and transformation matrix as the primary description and the
reference values are translated according to the prescription given in
Sect.~\ref{sec:basics}.  The \keyw{RADESYSA} card indicates that the ecliptic
coordinates are mean coordinates in the IAU 1984 system, though the author of
the header has been sloppy in omitting the \keyw{EQUINOXA} card which
therefore defaults to J2000.0.  Note that, in accord with Paper~I, all
keywords for the alternate description are reproduced, even those which do not
differ from the primary description.

The problem will be to determine the galactic and ecliptic coordinates
corresponding to a field point with pixel coordinates $(1957.2,775.4)$.  We
begin as before by substituting in Eq.~(\ref{eq:px})
\begin{eqnarray*}
   \left(
      \begin{array}{c}
         x \\
         y
      \end{array}
   \right) & = &
   \left(
      \begin{array}{ll}
         -0.005   & 0.00002 \\
         -0.00001 & 0.005   \\
      \end{array}
   \right)
   \left(
      \begin{array}{ccl}
         p_1 & - & 1024.5 \\
         p_2 & + & 1023.5 \\
      \end{array}
   \right) \, .
\end{eqnarray*}

\noindent
Note that this transformation matrix describes a slight skewness and rotation;
the $x$-, and $y$-axes correspond to the following two non-orthogonal lines in
the pixel plane;
\begin{eqnarray*}
   p_2 & = &             0.002 \left( p_1 - 1024.5 \right) - 1023.5 \, , \\*
   p_2 & = & \hphantom{0.} 250 \left( p_1 - 1024.5 \right) - 1023.5 \, .
\end{eqnarray*}
The values of the $(x,y)$ projection plane coordinates for the chosen point
are shown in Table~\ref{ta:working2} together with the remaining calculations
for this example.

\begin{table}
   \caption[]{Calculations for example 2.}
   \begin{tabular}{lrr}
      \hline
      \noalign{\smallskip}
      $(p_1,p_2)$              & $         1957.2$ & $          775.4$ \\
      $(x,y)$                  & $  -4\fdg6275220$ & $   8\fdg9851730$ \\
      $\theta\sub{a},\eta$     & $ -25\fdg0000000$ & $   0\fdg0000000$ \\
      $\theta_1,\theta_2$      & $ -25\fdg0000000$ & $ -25\fdg0000000$ \\
      $\gamma$                 & $     -0.8452365$ \\
      $C,Y_0$                  & $     -0.4226183$ & $-122\fdg8711957$ \\
      $(\phi,\theta)$          & $  -4\fdg7560186$ & $ -15\fdg8973800$ \\
      \noalign{\medskip}
      $(\ell\sub{p},b\sub{p})$ & $ -90\fdg0000000$ & $  90\fdg0000000$ \\
      $(\ell,b)$               & $  85\fdg2439814$ & $ -15\fdg8973800$ \\
      \noalign{\medskip}
      $(\lambda\sub{p},\beta\sub{p})$
                               & $-179\fdg9767827$ & $  29\fdg8114400$ \\
      $(\lambda,\beta)$        & $ -14\fdg7066741$ & $  43\fdg0457292$ \\
      \noalign{\smallskip}
      \hline
   \end{tabular}
   \label{ta:working2}
\end{table}

The next step is to compute the native longitude and latitude.  Like all
standard conics, the conic equal area projection has two parameters,
$\theta\sub{a}$ and $\eta$, given by the keywords \PVi{1} and \PVi{2} attached
to latitude coordinate~{\it i}.  In this example \keyw{PV2\_1} is given but
the author has again been sloppy in omitting \keyw{PV2\_2}, which thus
defaults to 0.  Explicit inclusion of this keyword in the header would obviate
any suspicion that it had been accidentally omitted.  The native coordinates,
$(\phi,\theta)$ are obtained from $(x,y)$ using Eqs.~(\ref{eq:ConPhi}) and
(\ref{eq:COEtheta}) via the intermediaries $C$, $Y_0$, and $\Rt$ given by
Eqs.~(\ref{eq:COEC}), (\ref{eq:COEy0}), and (\ref{eq:ConRt}).  These in turn
are expressed in terms of $\gamma$, $\theta_1$, and $\theta_2$ given by
Eqs.~(\ref{eq:COEgamma}), (\ref{eq:ConTheta1}), and (\ref{eq:ConTheta2}).  The
calculations are shown in Table~\ref{ta:working2}.

At this stage we have deprojected the field point.  In this example the
alternate coordinate description defines the same projection as the primary
description.  Indeed, it may seem odd that the formalism even admits the
possibility that they may differ.  However, this is a realistic possibility,
for example in defining multiple optical plate solutions based on the
\keyv{TAN} projection.  It now remains to transform the native spherical
coordinates into galactic coordinates, $(\ell,b)$, and ecliptic coordinates,
$(\lambda,\beta)$.  To do this we need to apply Eqs.~(\ref{eq:nat2std}) and in
order to do that we need $(\ell\sub{p},b\sub{p})$ and
$(\lambda\sub{p},\beta\sub{p})$.

Looking first at galactic coordinates, we have $(\ell_0,b_0) =
(90\degr,-25\degr)$ and $\phi\sub{p} = 0$.  This conic projection has
$(\phi_0,\theta_0) = (0\degr,-25\degr)$, and because $b_0 = \theta_0$ and
$\phi\sub{p} = 0\degr$, the native and galactic coordinate systems must
coincide to within an offset in longitude.  However, it is not obvious what
$\ell\sub{p}$ should be to produce this offset.  Equation~(\ref{eq:org2pola})
has one valid solution, namely $b\sub{p}=90\degr$.  The special case in point
2 in the usage notes for Eqs.~(\ref{eq:org2polb}) and (\ref{eq:org2polc}) must
therefore be used to obtain $(\ell\sub{p},b\sub{p}) = (-90\degr,90\degr)$.
The galactic coordinates of the field point listed in Table~\ref{ta:working2}
are then readily obtained by application of Eqs.~(\ref{eq:nat2std}).

The header says that the ecliptic coordinates of the reference point are
$(\lambda_0,\beta_0) = (-7\fdg0300934,34\fdg8474143)$ and the native longitude
of the ecliptic pole is $\phi\sub{p} = 6\fdg3839706$.  It also specifies
\keyw{LATPOLEA} as $29\fdg8114400$.  In this case Eq.~(\ref{eq:org2pola}) has
two valid solutions, $\beta\sub{p} = -25\fdg1367794 \pm 54\fdg9482194$, and
the one closest in value to \keyw{LATPOLEA} (in fact equal to it) is chosen.
If \keyw{LATPOLEA} had been omitted from the header its default value of
$+90\degr$ would have selected the northerly solution anyway, but of course it
is good practice to make the choice clear.  The value of $\lambda\sub{p}$ may
be obtained by a straightforward application of Eqs.~(\ref{eq:org2polb}) and
(\ref{eq:org2polc}), and the ecliptic coordinates of the field point computed
via Eqs.~(\ref{eq:nat2std}) are listed in Table~\ref{ta:working2} as the final
step of the calculation.  The reader may verify the calculation by
transforming the computed galactic coordinates of the field point to mean
ecliptic coordinates.


\subsubsection{Binary table representations of example 2}

Table~\ref{ta:bintable} shows the FITS header for a set of images stored in
binary table image array column format.  The images are similar to that
described in header interpretation example 2, Sect.~\ref{sec:interp2}, with
primary image header illustrated in Table~\ref{ta:interp2}.

In this example the images are stored as 2-D arrays in column 5 of the table
and each row of the table contains a $2048 \times 2048$ pixel image of a
different region on the sky.  This might represent a set of smaller images
extracted from a single larger image.  In this case all coordinate system
parameters except for the reference pixel coordinate are the same for each
image and are given as header keywords.  The reference pixel coordinate for
the primary and secondary description are given in columns 1 to 4 of the
binary table.

Table~\ref{ta:pixlist} shows the header for the same image given in pixel list
format.  There are 10000 rows in this table, each one listing the pixel
coordinates (\keyv{XPOS}, \keyv{YPOS}) of every detected ``event'' or photon
in the image.  For illustration purposes, this table also contains an optional
`\keyv{DATA\_QUALITY}' column that could be used to flag the reliability or
statistical significance of each event.  A real image may be constructed from
this virtual image as the 2-dimensional histogram of the number of events that
occur within each pixel.  The additional \keyi{TLMIN}{n} and \keyi{TLMAX}{n}
keywords shown here are used to specify the minimum and maximum legal values
in \keyv{XPOS} and \keyv{YPOS} columns and are useful for determining the
range of each axis when constructing the image histogram.

\begin{table*}
   \caption[]{Example binary table image array header (blank lines have been
      inserted for clarity).}
   \begin{tabular}{l}
      \hline
      \noalign{\smallskip}
      \verb+123456789 123456789 123456789 123456789 123456789 123456789 123456789 123456789 + \\
      \noalign{\smallskip}
      \hline
      \noalign{\smallskip}
      \verb+XTENSION= 'BINTABLE'           / Binary table extension+ \\
      \verb+BITPIX  =                    8 / 8-bit bytes+ \\
      \verb+NAXIS   =                    2 / 2-dimensional binary table+ \\
      \verb+NAXIS1  =             16777232 / Width of table in bytes+ \\
      \verb+NAXIS2  =                    4 / Number of rows in table+ \\
      \verb+PCOUNT  =                    0 / Size of special data area+ \\
      \verb+GCOUNT  =                    1 / One data group (required keyword)+ \\
      \\
      \verb+TFIELDS =                    5 / number of fields in each row+ \\
      \verb+TTYPE1  = '1CRP5   '           / Axis 1: reference pixel coordinate+ \\
      \verb+TFORM1  = '1J      '           / Data format of column 1: I*4 integer+ \\
      \verb+TTYPE2  = '2CRP5   '           / Axis 2: reference pixel coordinate+ \\
      \verb+TFORM2  = '1J      '           / Data format of column 2: I*4 integer+ \\
      \verb+TTYPE3  = '1CRP5A  '           / Axis 1A: reference pixel coordinate+ \\
      \verb+TFORM3  = '1J      '           / Data format of column 3: I*4 integer+ \\
      \verb+TTYPE4  = '2CRP5A  '           / Axis 2A: reference pixel coordinate+ \\
      \verb+TFORM4  = '1J      '           / Data format of column 4: I*4 integer+ \\
      \verb+TTYPE5  = 'Image   '           / 2-D image array+ \\
      \verb+TFORM5  = '4194304J'           / Data format of column 5: I*4 vector+ \\
      \verb+TDIM5   = '(2048,2048)'        / Dimension of column 5 array+ \\
      \verb+MJDOB5  =        44258.7845612 / MJD at start of observation+ \\
      \\
      \verb+COMMENT  The following keywords define the primary coordinate description+ \\
      \verb+COMMENT  of the images contained in Column 5 of the table.+ \\
      \\
      \verb+11PC5   =                    1 / Transformation matrix element+ \\
      \verb+12PC5   =               -0.004 / Transformation matrix element+ \\
      \verb+21PC5   =               -0.002 / Transformation matrix element+ \\
      \verb+22PC5   =                    1 / Transformation matrix element+ \\
      \verb+1CDE5   =               -0.005 / Axis 1: scale+ \\
      \verb+2CDE5   =                0.005 / Axis 2: scale+ \\
      \verb+1CTY5   = 'GLON-COE'           / Axis 1: conic equal area projection+ \\
      \verb+2CTY5   = 'GLAT-COE'           / Axis 2: conic equal area projection+ \\
      \verb+2PV5_1  =                -25.0 / Conic mid-latitude+ \\
      \verb+1CRV5   =                 90.0 / Axis 1: galactic longitude at reference point+ \\
      \verb+2CRV5   =                -25.0 / Axis 2: galactic latitude at reference point+ \\
      \\
      \verb+COMMENT  The following keywords define the secondary coordinate description+ \\
      \verb+COMMENT  of the images contained in Column 5 of the table.+ \\
      \\
      \verb+11PC5A  =                    1 / Transformation matrix element+ \\
      \verb+12PC5A  =               -0.004 / Transformation matrix element+ \\
      \verb+21PC5A  =               -0.002 / Transformation matrix element+ \\
      \verb+22PC5A  =                    1 / Transformation matrix element+ \\
      \verb+1CDE5A  =               -0.005 / Axis 1A: scale+ \\
      \verb+2CDE5A  =                0.005 / Axis 2A: scale+ \\
      \verb+1CTY5A  = 'ELON-COE'           / Axis 1A: conic equal area projection+ \\
      \verb+2CTY5A  = 'ELAT-COE'           / Axis 2A: conic equal area projection+ \\
      \verb+2PV5_1A =                -25.0 / Conic mid-latitude+ \\
      \verb+1CRV5A  =           -7.0300934 / Axis 1A: ecliptic longitude at reference point+ \\
      \verb+2CRV5A  =           34.8474143 / Axis 2A: ecliptic latitude at reference point+ \\
      \verb+LONP5A  =            6.3839706 / Native longitude of ecliptic pole+ \\
      \verb+LATP5A  =           29.8114400 / Ecliptic latitude of native pole+ \\
      \verb+RADE5A  = 'FK5     '           / Mean IAU 1984 ecliptic coordinates+ \\
      \verb+EQUI5A  =               2000.0 / Coordinate epoch+ \\
      \verb+END+ \\
      \noalign{\smallskip}
      \hline
   \end{tabular}
   \label{ta:bintable}
\end{table*}

\begin{table*}[!ht]
   \caption[]{Example pixel list header (blank lines have been inserted for
              clarity).}
   \begin{tabular}{l}
      \hline
      \noalign{\smallskip}
      \verb+123456789 123456789 123456789 123456789 123456789 123456789 123456789 123456789 + \\
      \noalign{\smallskip}
      \hline
      \noalign{\smallskip}
      \verb+XTENSION= 'BINTABLE'           / Binary table extension+ \\
      \verb+BITPIX  =                    8 / 8-bit bytes+ \\
      \verb+NAXIS   =                    2 / 2-dimensional binary table+ \\
      \verb+NAXIS1  =                    5 / Width of table in bytes+ \\
      \verb+NAXIS2  =                10000 / Number of rows in table+ \\
      \verb+PCOUNT  =                    0 / Size of special data area+ \\
      \verb+GCOUNT  =                    1 / One data group (required keyword)+ \\
      \\
      \verb+TFIELDS =                    3 / Number of fields in each row+ \\
      \verb+TTYPE1  = 'DATA_QUALITY'       / Quality flag value of the photon+ \\
      \verb+TFORM1  = '1B      '           / Data format of the field: 1-byte integer+ \\
      \verb+TTYPE2  = 'XPOS    '           / Axis 1: pixel coordinate of the photon+ \\
      \verb+TFORM2  = '1I      '           / Data format of column 2: I*2 integer+ \\
      \verb+TLMIN2  =                    1 / Lower limit of axis in column 2+ \\
      \verb+TLMAX2  =                 2048 / Upper limit of axis in column 2+ \\
      \verb+TTYPE3  = 'YPOS    '           / Axis 2: pixel coordinate of the photon+ \\
      \verb+TFORM3  = '1I      '           / Data format of column 3: I*2 integer+ \\
      \verb+TLMIN3  =                    1 / Lower limit of axis in column 3+ \\
      \verb+TLMAX3  =                 2048 / Upper limit of axis in column 3+ \\
      \verb+MJDOB3  =        44258.7845612 / MJD at start of observation+ \\
      \\
      \verb+COMMENT  The following keywords define the primary coordinate description.+ \\
      \\
      \verb+TCRP2   =               1024.5 / Axis 1: reference pixel coordinate+ \\
      \verb+TCRP3   =              -1023.5 / Axis 2: reference point pixel coordinate+ \\
      \verb+TPC2_2  =                    1 / Transformation matrix element+ \\
      \verb+TPC2_3  =               -0.004 / Transformation matrix element+ \\
      \verb+TPC3_2  =               -0.002 / Transformation matrix element+ \\
      \verb+TPC3_3  =                    1 / Transformation matrix element+ \\
      \verb+TCDE2   =               -0.005 / Axis 1: scale+ \\
      \verb+TCDE3   =                0.005 / Axis 2: scale+ \\
      \verb+TCTY2   = 'GLON-COE'           / Axis 1: conic equal area projection+ \\
      \verb+TCTY3   = 'GLAT-COE'           / Axis 2: conic equal area projection+ \\
      \verb+TPV3_1  =                -25.0 / Conic mid-latitude+ \\
      \verb+TCRV2   =                 90.0 / Axis 1: galactic longitude at reference point+ \\
      \verb+TCRV3   =                -25.0 / Axis 2: galactic latitude at reference point+ \\
      \\
      \verb+COMMENT  The following keywords define the secondary coordinate description.+ \\
      \\
      \verb+TCRP2A  =               1024.5 / Axis 1A: reference pixel coordinate+ \\
      \verb+TCRP3A  =              -1023.5 / Axis 2A: reference point pixel coordinate+ \\
      \verb+TP2_2A  =                    1 / Transformation matrix element+ \\
      \verb+TP2_3A  =               -0.004 / Transformation matrix element+ \\
      \verb+TP3_2A  =               -0.002 / Transformation matrix element+ \\
      \verb+TP3_3A  =                    1 / Transformation matrix element+ \\
      \verb+TCDE2A  =               -0.005 / Axis 1A: scale+ \\
      \verb+TCDE3A  =                0.005 / Axis 2A: scale+ \\
      \verb+TCTY2A  = 'ELON-COE'           / Axis 1A: conic equal area projection+ \\
      \verb+TCTY3A  = 'ELAT-COE'           / Axis 2A: conic equal area projection+ \\
      \verb+TV3_1A  =                -25.0 / Conic mid-latitude+ \\
      \verb+TCRV2A  =           -7.0300934 / Axis 1A: ecliptic longitude at reference point+ \\
      \verb+TCRV3A  =           34.8474143 / Axis 2A: ecliptic latitude at reference point+ \\
      \verb+LONP3A  =            6.3839706 / Native longitude of ecliptic pole+ \\
      \verb+LATP3A  =           29.8114400 / Ecliptic latitude of native pole+ \\
      \verb+RADE3A  = 'FK5     '           / Mean IAU 1984 ecliptic coordinates+ \\
      \verb+EQUI3A  =               2000.0 / Coordinate epoch+ \\
      \verb+END+ \\
      \noalign{\smallskip}
      \hline
   \end{tabular}
   \label{ta:pixlist}
\end{table*}


\subsubsection{Header interpretation example 3}
\label{sec:interp3}

\begin{table*}
   \caption[]{Third example FITS header.}
   \begin{tabular}{l}
      \hline
      \noalign{\smallskip}
      \verb+123456789 123456789 123456789 123456789 123456789 123456789 123456789 123456789+ \\
      \noalign{\smallskip}
      \hline
      \noalign{\smallskip}
      \verb+NAXIS   =                    2 / 2-dimensional image+ \\
      \verb+NAXIS1  =                  181 / x axis (fastest)+ \\
      \verb+NAXIS2  =                   91 / y axis (2nd fastest)+ \\
      \verb+CRPIX1  =                226.0 / Pixel coordinate of reference point+ \\
      \verb+CRPIX2  =                 46.0 / Pixel coordinate of reference point+ \\
      \verb+CDELT1  =                 -1.0 / x-scale+ \\
      \verb+CDELT2  =                  1.0 / y-scale+ \\
      \verb+CTYPE1  = 'GLON-CAR'           / Plate carree projection+ \\
      \verb+CTYPE2  = 'GLAT-CAR'           / Plate carree projection+ \\
      \verb+CRVAL1  =                 30.0 / Galactic longitude at reference point+ \\
      \verb+CRVAL2  =                 35.0 / Galactic latitude at reference point+ \\
      \noalign{\smallskip}
      \hline
   \end{tabular}
   \label{ta:interp3}
\end{table*}

This example has been adapted from a real-life FITS data file.  The simplicity
of the header shown in Table~\ref{ta:interp3} is deceptive; it is actually
presented as an example of how {\em not} to write a FITS header, although the
latent problem with its interpretation is quite subtle.

Observe that the image spans $180\degr$ in native longitude and $90\degr$ in
native latitude and that the reference pixel lies outside the image.  In fact,
the reference pixel is located so that the native longitude runs from
$45\degr$ to $225\degr$ and hence the image lies partly inside and partly
outside the normal range of native longitude, [$-180\degr$,$180\degr$].

In fact, as might be expected, this makes no difference to the computation of
celestial coordinates.  For example, in computing the celestial coordinates of
pixel $(1,1)$ we readily find from Eqs.~(\ref{eq:CARx}) and (\ref{eq:CARy})
that the native coordinates are $(\phi,\theta) = (225\degr,-45\degr)$.  The
fact that $\phi$ exceeds $180\degr$ becomes irrelevant once
Eqs.~(\ref{eq:nat2std}) are applied since the trigonometric functions do not
distinguish between $\phi = 225\degr$ and $\phi = -135\degr$.  The latter
value is the appropriate one to use if the cylinder of projection is
considered to be ``rolled out'' over multiple cycles.  Consequently the
correct galactic coordinates are obtained.

The problem only surfaces when we come to draw a coordinate grid on the image.
A meridian of longitude, for example, is traced by computing the pixel
coordinates for each of a succession of latitudes along the segment of the
meridian that crosses the image.  As usual, in computing pixel coordinates,
the celestial coordinates are first converted to native coordinates by
applying Eqs.~(\ref{eq:std2nat}), and the native longitude will be returned in
the normal range [$-180\degr$,$180\degr$].  Consequently, in those parts of
the image where $\phi~>~180\degr$ the pixel coordinates computed will
correspond to the point at $\phi~-~360\degr$, i.e.\ in the part of the
principle cycle of the cylindrical projection outside the image.

In principle it is possible to account for this, at least in specific cases,
particularly for the cylindrical projections which are somewhat unusual in
this regard.  In practice, however, it is difficult to devise a general
solution, especially when similar problems may arise for projection types
where it is not desirable to track $\phi$ outside its normal range.  For
example, consider the case where a Hammer-Aitoff projection is used to
represent the whole sky; since its boundary is curved there will be
out-of-bounds areas in the corner of the image.  Normally a grid drawing
routine can detect these by checking whether the inverse projection equations
return a value for $\phi$ outside its normal range.  It may thus determine the
proper boundary of the projection and deal with the discontinuity that arises
when a grid line passes through it.

How then should the header have been written?  Note that the problem exists at
the lowest level of the coordinate description, in the conversion between
$(x,y)$ and $(\phi,\theta)$, and the solution must be found at this level.
The problem arose from a particular property of cylindrical projections in
that they have $x~\propto~\phi$.  We must use this same property, which we
might call ``$\phi$-translation similarity'', to recast the coordinate
description into a more manageable form.  $\phi$-translation similarity simply
means that changing the origin of $\phi$ corresponds to shifting the image in
the $x$-direction.  In other words, we can transfer the reference point of the
projection from its current location to another location along the native
equator without having to regrid the image.  The fact that the \PC{i}{ja}
matrix is unity in this example makes this task a little simpler than
otherwise.

Note first that because the image straddles $\phi = 180\degr$ we can't simply
reset \keyw{CRPIX1} so as to shift the reference point to $\phi = -360\degr$;
the image would then straddle $\phi = -180\degr$, which is no improvement.  In
this example it is convenient and sufficient to shift the reference point to
$(\phi,\theta) = (180\degr,0\degr)$, which corresponds to pixel coordinate
$(p_1,p_2) = (46.0,46.0)$.  Hence we need to reset \keyw{CRPIX1} to $46.0$ and
adjust the keywords which define the celestial coordinate system.  The reader
may readily verify that the galactic coordinates of the new reference point
are $(\ell,b) = (210\degr,-35\degr)$ and whereas the old, implied value of
\keyw{LONPOLE} was $0\degr$ when $\delta_0~>~0$, now that $\delta_0~<~0$ its
new implied value is $180\degr$, and this is correct.  However, we will set it
explicitly anyway.  The keywords to be changed are therefore
\begin{eqnarray*}
   \keyw{CRPIX1}  & = &     46.0 \, , \\
   \keyw{CRVAL1}  & = & 210\degr \, , \\
   \keyw{CRVAL2}  & = & -35\degr \, , \\
   \keyw{LONPOLE} & = & 180\degr \, .
\end{eqnarray*}

What if the \PC{i}{ja} matrix was not unity?  The problem of determining the
pixel coordinates where $(\phi,\theta) = (-180\degr,0\degr)$ would have
presented little extra difficulty, although in general \keyw{CRPIX2} would
also need to be changed.  On reflection it may come as a surprise that
changing \CRPIX{j} like this does not fundamentally alter the linear
transformation.  However it may readily be verified that the only effect of
changing \CRPIX{j} is to change the origin of the $(x,y)$ coordinates.


\subsection{Header construction examples}
\label{sec:construct}

This section considers the more difficult problem faced by FITS writers; that
of constructing world coordinate system headers.

Example 1 is contrived to illustrate the general methods used in constructing
a spherical coordinate representation.  Paying homage to Claudius~Ptolemy, it
actually constructs a terrestrial coordinate grid for the Mediterranean region
as seen from space.

Example 2 constructs headers for the infra-red dust maps produced by Schlegel
et al.\ (\cite{kn:SFD}) who regridded data from the COBE/DIRBE and IRAS/ISSA
surveys onto two zenithal equal area projections.

Example 3 considers the case of long-slit optical spectroscopy.  It is
concerned in particular with solving the problem of producing three world
coordinate elements for a data array of only two dimensions.

Example 4 constructs a dual coordinate description for an image of the moon
and considers the problem of producing consistent scales for each.  It also
suggests an extension to deal with the rings of the planet Saturn.


\subsubsection{Header construction example 1}
\label{sec:construct1}

\begin{figure}
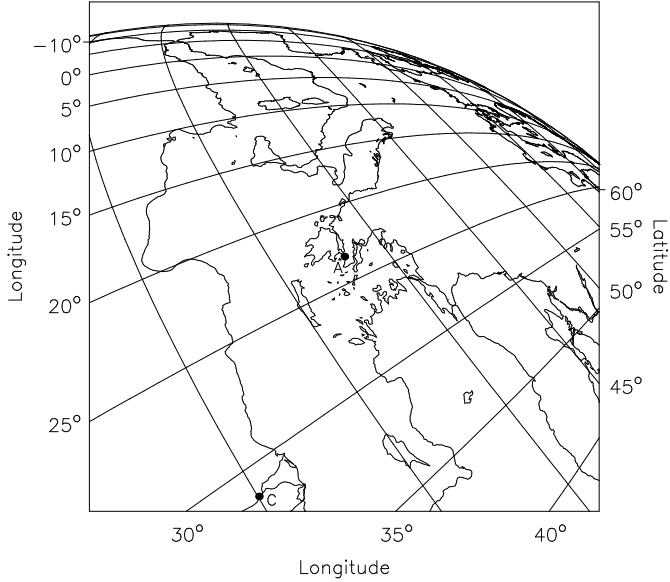

   \centerline{\putfig{AZPhoto}{220pt}}
   \caption[]{The Earth from 2230\,km above Cairo looking directly towards
     Athens.}
   \label{fig:AZPhoto}
\end{figure}

Our first example of FITS header construction concerns satellite photography
of the Earth.  Figure~\ref{fig:AZPhoto} shows a part of the world which
probably would have been recognizable to Claudius~Ptolemy.

A satellite 2230\,km immediately above Cairo aims its digital camera directly
towards Athens and adjusts the orientation and focal length to include Cairo
in the field near the bottom edge of the frame.  The $2048 \times 2048$ pixel
CCD detector array employed by the camera is centered on its optical axis so
Athens is at pixel coordinate $(1024.5,1024.5)$.  Cairo -- at the satellite's
nadir -- is later found to be at $(681.67,60.12)$.  The task is to overlay a
terrestrial coordinate grid on this image.

\begin{figure}
   \vspace{10pt}
   \centerline{\putfig{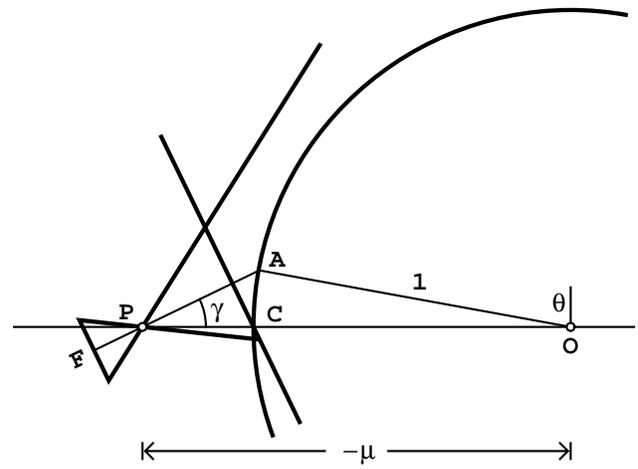}{175pt}}
   \caption[]{The geometry of Fig.~\ref{fig:AZPhoto}.  Cairo at $\mathbf{C}$
     is at the reference point of the projection with Athens at $\mathbf{A}$.
     The camera with aperture at $\mathbf{P}$ and focal plane $\mathbf{F}$ is
     not drawn to scale, though the rest of the diagram is.  Note that
     $\mu < 0$ so $-\mu$, as marked, is a positive value.}
   \label{fig:AZPgeom}
\end{figure}

For the sake of simplicity we will assume pin-hole camera optics.
Figure~\ref{fig:AZPgeom} identifies the geometry as that of a tilted,
near-sided zenithal perspective projection with the nadir at the reference
point.  The point of projection $\mathbf{P}$ corresponds to the pin-hole of
the camera and the tilted plane of projection is parallel to the camera's
focal plane $\mathbf{F}$.  The image in the focal plane is simply scaled (and
inverted) with respect to that on the plane of projection.  Clearly a tilted
\keyv{AZP} projection is a better match to the geometry than \keyv{SZP} in
this instance.

The satellite altitude of 2230\,km is 0.350 Earth radii and since the
projection is near-sided we may immediately write $\mu = -1.350$.

Determination of $\gamma$ requires knowledge of the coordinates of Cairo
$(31\fdg15\,\mathrm{E}, 30\fdg03\,\mathrm{N})$ and Athens
$(23\fdg44\,\mathrm{E}, 38\fdg00\,\mathrm{N})$.  From spherical trigonometry
we may deduce that the angular separation between the two cities is
$10\fdg2072$ and also that Athens is on a bearing $36\fdg6252$ west of north
from Cairo.

Now since Cairo is at the native pole of the projection the angular separation
between the two cities is just their difference in native latitude,
$90\degr - \theta\sub{A}$.  Using the sine rule in triangle $\mathbf{PAO}$ in
Fig.~\ref{fig:AZPgeom} we obtain
\begin{equation}
   90\degr - \theta = -\gamma - \sin^{-1}(\mu \sin\gamma) \, .
\end{equation}
This equation, which takes account of the fact that $\mu$ is negative, may be
solved iteratively to obtain $\gamma = 25\fdg8458$.

The obliqueness of this view of the Earth, occasioned by the fact that the
image is not oriented along a meridian, is handled partly via \LONPOLE{a} and
partly as a bulk image rotation via \PC{i}{ja}.  Note that these are not
complementary; they produce distinct effects since the tilted \keyw{AZP}
projection does not have point symmetry.  Figure~\ref{fig:AZPgeom} shows the
situation -- the generating sphere with Cairo at the native pole must be
oriented so that Athens is at native longitude $180\degr$.  The native
longitude of the terrestrial pole (substituting for the celestial pole in this
example) is offset from this by the bearing of Athens from Cairo,
i.e.\ $\phi\sub{p} = 180\degr - 36\fdg6252 = 143\fdg3748$.

Since Athens is at native longitude $180\degr$ its $x$-coordinate, like
Cairo's, must be zero.  The inequality of the corresponding pixel coordinates
must have arisen by the satellite rotating the camera about its optical axis
thereby rotating the CCD detector in the focal plane.  The angle of this bulk
rotation is readily deduced from the pixel coordinates of Athens and Cairo,
$\arg(j\sub{A}-j\sub{C}, i\sub{A}-i\sub{C}) = 19\fdg570$.  This rotation is to
be applied via \PC{i}{ja}.  The direction of rotation is completely determined
by the requirement that Athens' $x$-coordinate be zero, regardless of the sign
of \CDELT{ia}, and the resulting \PC{i}{ja} rotation matrix is shown below.

The reference pixel coordinates, \CRPIX{ja}, were measured from the image and
the reference coordinates for Cairo, \CRVAL{ia}, were obtained from an atlas,
so the last remaining unknowns are the scales, \CDELT{ia}.  From previous
calculations we know that Athens is at $(\phi,\theta) = (180\degr,
79\fdg7928)$ and we may apply Eqs.~(\ref{eq:AZPx}) and (\ref{eq:AZPy}) to
obtain $(x,y) = (0,8\fdg7424)$.  The distance in pixel coordinates between the
two cities is readily found to be 1023.5 so the $y$-scale must be
$8\fdg7424 / 1023.5 = 0\fdg008542$ per pixel.

The $x$-scale cannot be determined like this since both cities have the same
$x$-coordinate.  However, the $x$-, and $y$-scales must be equal because the
focal plane is parallel to the plane of projection and ray-tracing through the
pin-hole therefore results in an isotropic change of scale.  Do not confuse
this with the statement made in Sect.~\ref{sec:AZP} that with $\gamma \neq 0$
the projection is not scaled true at the reference point; this refers to the
differential scale between $(x,y)$ and $(\phi,\theta)$.

Though the image in the focal plane is inverted through the pin-hole, thus
indicating a negative scale, we can assume that the camera compensated by
reading out the CCD array in reverse order.  Thus for the image of
Fig.~\ref{fig:AZPhoto} we make both of the \CDELT{ia} positive in order to
have east to the right as befits a sphere seen from the outside.  Putting this
all together we have
\begin{eqnarray*}
   \keyw{  NAXIS} & = &                  2 \, , \\
   \keyw{NAXIS1}  & = &               2048 \, , \\
   \keyw{NAXIS2}  & = &               2048 \, , \\
   \keyw{CRPIX1}  & = &             681.67 \, , \\
   \keyw{CRPIX2}  & = &  \hphantom{6}60.12 \, , \\
   \keyw{PC1\_1}  & = & \hphantom{-}0.9422 \, , \\
   \keyw{PC1\_2}  & = &            -0.3350 \, , \\
   \keyw{PC2\_1}  & = & \hphantom{-}0.3350 \, , \\
   \keyw{PC2\_2}  & = & \hphantom{-}0.9422 \, , \\
   \keyw{CDELT1}  & = &        0\fdg008542 \, , \\
   \keyw{CDELT2}  & = &        0\fdg008542 \, , \\
   \keyw{CTYPE1}  & = &  \keyv{`TLON-AZP'} \, , \\
   \keyw{CTYPE2}  & = &  \keyv{`TLAT-AZP'} \, , \\
   \keyw{PV2\_1}  & = &             -1.350 \, , \\
   \keyw{PV2\_2}  & = & \hphantom{1}25\fdg8458 \, , \\
   \keyw{CRVAL1}  & = & \hphantom{1}31\fdg15   \, , \\
   \keyw{CRVAL2}  & = & \hphantom{1}30\fdg03   \, , \\
   \keyw{LONPOLE} & = &            143\fdg3748 \, , \\
   \keyw{WCSNAME} & = & \keyv{`Terrestrial coordinates'} \, .
\end{eqnarray*}

This example was simplified by the fact that the nadir was known and was
included in the field of view.  This probably would not be the case in a more
realistic example where there may also be some uncertainty in the target
coordinates and the value of $\mu$.  In such cases the mapping parameters
would have to be determined by least squares via the identification of an
adequate number of surface features with known coordinates.


\subsubsection{Header construction example 2}
\label{sec:construct2}

The following example comes from the $4096 \times 4096$ pixel infrared dust
maps produced by Schlegel, Finkbeiner, \& Davis (SFD, \cite{kn:SFD}).  The
authors chose to regrid data from the COBE/DIRBE and IRAS/ISSA maps onto two
zenithal equal area (\keyv{ZEA}) projections centered on the galactic poles.
The projection formula given in their Appendix C expressed in terms of
standard 1-relative FITS pixel coordinates $(p_1,p_2)$ is
\begin{eqnarray*}
   p_1 - 1 & = & \hphantom{-n}2048\sqrt{1 - n\sin b}\,\cos\ell + 2047.5 \, , \\
   p_2 - 1 & = &           -n 2048\sqrt{1 - n\sin b}\,\sin\ell + 2047.5 \, ,
\end{eqnarray*}
where $n = +1$ for the north Galactic pole and $n = -1$ for the south Galactic
pole maps.  Now for \keyv{ZEA} from Eqs.~(\ref{eq:px}), (\ref{eq:azx}),
(\ref{eq:azy}), and (\ref{eq:ZEARt})
\begin{eqnarray*}
   \keyw{CDELT1}\, (p_1 - \keyw{CRPIX1}) & = &
      \SP\sqrt{2} \rd\sqrt{1-\sin\theta}\sin\phi \, , \\
   \keyw{CDELT2}\, (p_2 - \keyw{CRPIX2}) & = &
        -\sqrt{2} \rd\sqrt{1-\sin\theta}\cos\phi \, .
\end{eqnarray*}
If we take the north Galactic pole case first, the SFD equations may be
rewritten as
\begin{eqnarray*}
   p_1 - 2048.5 & = & -2048 \sqrt{1 - \sin b}\,\sin(\ell-90\degr) \, , \\
   p_2 - 2048.5 & = & -2048 \sqrt{1 - \sin b}\,\cos(\ell-90\degr) \, .
\end{eqnarray*}
By inspection of the two sets of equations we must have
\begin{eqnarray*}
   \keyw{  NAXIS} & = & 2 \, , \\
   \keyw{NAXIS1}  & = & 4096 \, , \\
   \keyw{NAXIS2}  & = & 4096 \, , \\
   \keyw{CRPIX1}  & = & 2048.5 \, , \\
   \keyw{CRPIX2}  & = & 2048.5 \, , \\
   \keyw{CDELT1}  & = &    -180\degr \sqrt{2} / (2048\pi) \, , \\
   \keyw{CDELT2}  & = & \SP 180\degr \sqrt{2} / (2048\pi) \, , \\
   \keyw{CTYPE1}  & = & \keyv{`GLON-ZEA'} \, , \\
   \keyw{CTYPE2}  & = & \keyv{`GLAT-ZEA'} \, , \\
             \ell & = & \phi + 90\degr \, , \\
                b & = & \theta \, .
\end{eqnarray*}
Now, writing $\ell$ and $b$ in place of $\alpha$ and $\delta$ in
Eqs.~(\ref{eq:nat2std}), we have to determine $\ell\sub{p}$, $b\sub{p}$, and
$\phi\sub{p}$ to give $(\ell,b)$ in terms of $(\phi,\theta)$.  This is easy
because we know $b\sub{p}=90\degr$ and the simple special case
Eqs.~(\ref{eq:nat2std90}) apply, so
\begin{eqnarray*}
   \ell = \phi + (\ell\sub{p} - \phi\sub{p} - 180\degr) \, .
\end{eqnarray*}
We have a degree of freedom here since only $\ell\sub{p}-\phi\sub{p}$ is
determined.  It's best to let $\phi\sub{p}$ take its default value of $0\degr$
for zenithal projections with the celestial pole at the native pole (it's
$180\degr$ otherwise), so we must have
\begin{eqnarray*}
   \keyw{CRVAL1}  & = &             270\degr \, , \\
   \keyw{CRVAL2}  & = & \hphantom{-} 90\degr \, , \\
   \keyw{LONPOLE} & = & \hphantom{-9} 0\degr \, .
\end{eqnarray*}
Although \keyw{LONPOLE} assumes its default value here it would of course be
wise to write it explicitly into the header.  The procedure for the south
Galactic pole case is similar.  The SFD equations may be rewritten
\begin{eqnarray*}
   p_1 - 2048.5 & = & 2048 \sqrt{1 - \sin(-b)}\,\sin(90\degr-\ell) \, , \\
   p_2 - 2048.5 & = & 2048 \sqrt{1 - \sin(-b)}\,\cos(90\degr-\ell) \, ,
\end{eqnarray*}
whence
\begin{eqnarray*}
   \keyw{ CRPIX1} & = & 2048.5 \, , \\
   \keyw{CRPIX2}  & = & 2048.5 \, , \\
   \keyw{CDELT1}  & = & \SP 180\degr \sqrt{2} / (2048\pi) \, , \\
   \keyw{CDELT2}  & = &    -180\degr \sqrt{2} / (2048\pi) \, , \\
             \ell & = & 90\degr - \phi \, , \\
                b & = & -\theta \, .
\end{eqnarray*}
Equations~(\ref{eq:nat2stdm90}) apply for $b\sub{p}=-90\degr$ so
\begin{eqnarray*}
   \ell & = & (\ell\sub{p} + \phi\sub{p}) - \phi \, .
\end{eqnarray*}
Again we let $\phi\sub{p}$ take its default value of $180\degr$, so
\begin{eqnarray*}
   \keyw{CRVAL1}  & = & 270\degr \, , \\
   \keyw{CRVAL2}  & = & -90\degr \, , \\
   \keyw{LONPOLE} & = & 180\degr \, .
\end{eqnarray*}

It's generally easier to interpret coordinate headers than to construct them
so it's essential after formulating the header to test it at a few points and
make sure that it works as expected.  Note that this sort of translation
exercise wouldn't be necessary if the formalism of this paper was used right
from the start, i.e.\ in the regridding operation used to produce the maps.


\subsubsection{Header construction example 3}
\label{sec:construct3}

Consider now the coordinate description for the two-dimensional image formed
by a long slit spectrograph.  We assume that the wavelength axis of length
1024 and dispersion $\Delta\lambda$ nm/pixel corresponds to the $p_1$ pixel
coordinate, and the 2048 pixel spatial axis corresponds to $p_2$.  The slit is
centered on equatorial coordinates $(\alpha_0,\delta_0)$ and oriented at
position angle $\rho$ measured such that when $\rho = 0$ the first spectrum is
northwards.  We will assume that the telescope and spectrograph optics are
such that the distance along the slit is in proportion to true angular
distance on the sky with a separation between pixels of $\sigma$ arcsec.
We do not consider curvature of the slit here, that is a distortion of the
sort to be handled by the methods of Paper~IV\@.

Equiscaling along the length of the slit together with the one-dimensional
nature of the spatial geometry indicate that any projection could be used that
is equiscaled along a great circle projected as a straight line.  Many
projections satisfy this criterion.  However, in practice the zenithal
equidistant (\keyv{ARC}) projection is the most convenient choice.

The one-dimensional spatial geometry may at first seem problematical, with
each point along the slit having a different $(\alpha,\delta)$.  However, this
is easily handled by introducing a third, degenerate axis and introducing a
rotation.  The rotation in position angle may be handled via the linear
transformation matrix.  Moreover, since we've opted for a zenithal projection,
bearing in mind the discussion of Sect.~\ref{sec:oblique}, the rotation can
also be handled via $\phi\sub{p}$ (i.e.\ \keyw{LONPOLE}).  We will demonstrate
both methods.

Use of $\phi\sub{p}$ is perhaps more straightforward, and the header may be
written without further ado:
\begin{eqnarray*}
   \keyw{  NAXIS}  & = &                 3 \, , \\
   \keyw{NAXIS1}  & = &              1024 \, , \\
   \keyw{NAXIS2}  & = &              2048 \, , \\
   \keyw{NAXIS3}  & = &                 1 \, , \\
   \keyw{CRPIX1}  & = &                 1 \, , \\
   \keyw{CRPIX2}  & = &            1024.5 \, , \\
   \keyw{CRPIX3}  & = &                 1 \, , \\
   \keyw{CDELT1}  & = &     \Delta\lambda \, , \\
   \keyw{CDELT2}  & = &      -\sigma/3600 \, , \\
   \keyw{CDELT3}  & = &                 1 \, , \\
   \keyw{CTYPE1}  & = & \keyv{`WAVELEN'}  \, , \\
   \keyw{CTYPE2}  & = & \keyv{`RA---ARC'} \, , \\
   \keyw{CTYPE3}  & = & \keyv{`DEC--ARC'} \, , \\
   \keyw{CUNIT1}  & = &       \keyv{`nm'} \, , \\
   \keyw{CRVAL1}  & = &         \lambda_0 \, , \\
   \keyw{CRVAL2}  & = &          \alpha_0 \, , \\
   \keyw{CRVAL3}  & = &          \delta_0 \, , \\
   \keyw{LONPOLE} & = &    90\degr + \rho \, .
\end{eqnarray*}
\keyw{LONPOLE} is the only card which requires explanation.  Since no rotation
is introduced by the linear transformation, the slit, which coincides with the
$p_2$-axis, maps directly to the $x$-axis.  However, because \keyw{CDELT2} is
negative, the two axes run in opposite directions and, given that when
$\rho = 0$ the $p_2$-axis runs from north to south, $x$ must run from south to
north.  Referring to the left side of Fig.~\ref{fig:Native} for zenithal
projections, we see that when $\rho = 0$, the north celestial pole must be at
$\phi = 90\degr$.  Since position angle increases in an anticlockwise
direction (i.e.\ north through east) in the plane of the sky, the celestial
pole must rotate clockwise as $\rho$ increases.  Thus, as $\rho$ increases so
must $\phi\sub{p}$, hence $\phi\sub{p} = 90\degr + \rho$.

To verify this representation we will test it with $(\alpha_0,\delta_0) =
(150\degr,-35\degr)$, $\rho = 30\degr$, and $\sigma = 2"$ for pixel coordinate
$(1,1,1)$.  The corresponding $(s,x,y)$ coordinates are
$(0,0\fdg5686111,0\degr)$.  The wavelength is therefore $\lambda_0$.  From
Eqs.~(\ref{eq:azrt}), (\ref{eq:azphi}), and (\ref{eq:ARCRt}) the native
longitude and latitude are $(\phi,\theta) = (90\degr,89\fdg4313889)$, $\phi$
is always $\pm 90\degr$, and the value of $\theta$ corresponds to an offset of
$2047"$ from the center of the slit as it should.  From
Eqs.~(\ref{eq:nat2std}) the right ascension and declination are
$(\alpha,\delta) = (150\fdg3450039,-34\fdg5070794)$, which are in the correct
quadrant.

Most optical telescopes are better described by the \keyv{TAN} projection.  In
this case the only difference in the header would be
\begin{eqnarray*}
   \keyw{CTYPE2} & = & \keyv{`RA---TAN'} \, , \\
   \keyw{CTYPE3} & = & \keyv{`DEC--TAN'} \, .
\end{eqnarray*}
Verifying it with the same values as before yields the same $(s,x,y)$ and thus
the same wavelength.  From Eqs.~(\ref{eq:azrt}), (\ref{eq:azphi}), and
(\ref{eq:TANRt}) we get $(\phi,\theta) = (90\degr,89\fdg4314076)$, the offset
of 2046\farcs933 differing by only 67~mas.  The right ascension and
declination are $(\alpha,\delta) = (150\fdg3449926,-34\fdg5070956)$.

As an alternative, the original header could also have been written with the
\CTYPE{i} interchanged, so that the slit, still coincident with the
$p_2$-axis, maps directly to the $y$-axis.  The header would be as above but
with the following changes:
\begin{eqnarray*}
   \keyw{CTYPE2}  & = & \keyv{`DEC--ARC'} \, , \\
   \keyw{CTYPE3}  & = & \keyv{`RA---ARC'} \, , \\
   \keyw{CRVAL2}  & = &          \delta_0 \, , \\
   \keyw{CRVAL3}  & = &          \alpha_0 \, , \\
   \keyw{LONPOLE} & = &   180\degr + \rho \, .
\end{eqnarray*}
\keyw{LONPOLE} differs because the slit now runs along the $y$-axis rather
than the $x$-axis; the negative sign on \keyw{CDELT2} is still needed to make
$y$ run counter to $p_2$.

There are several practical ways to rewrite the header using the \PC{i}{j} or
\CD{i}{j} matrices.  Looking first at the \CD{i}{j} form we have
\begin{eqnarray*}
   \left(
      \begin{array}{c}
         s \\ x \\ y
      \end{array}
   \right)
   & = &
   \left(
      \begin{array}{ccc}
         \keyw{CD1\_1} & \keyw{CD1\_2} & \keyw{CD1\_3} \\
         \keyw{CD2\_1} & \keyw{CD2\_2} & \keyw{CD2\_3} \\
         \keyw{CD3\_1} & \keyw{CD3\_2} & \keyw{CD3\_3}
      \end{array}
   \right)
   \left(
      \begin{array}{c}
         p_1 - \keyw{CRPIX1} \\
         p_2 - \keyw{CRPIX2} \\
         p_3 - \keyw{CRPIX3}
      \end{array}
   \right) .
\end{eqnarray*}
Since the third axis is degenerate, with $\keyw{CRPIX3} = 1$, we have
$p_3 - \keyw{CRPIX3} = 0$, so the values of
\keyw{CD1\_3}, \keyw{CD2\_3}, and \keyw{CD3\_3} are irrelevant.  Moreover,
\keyw{CD1\_2}, \keyw{CD2\_1}, and \keyw{CD3\_1} are all zero, hence
\begin{eqnarray}
   s & = & \keyw{CD1\_1} \  (p_1 - \keyw{CRPIX1}) \, , \nonumber       \\
   x & = & \keyw{CD2\_2} \  (p_2 - \keyw{CRPIX2}) \, , \label{eq:slit} \\
   y & = & \keyw{CD3\_2} \  (p_2 - \keyw{CRPIX2}) \, . \nonumber
\end{eqnarray}
Effectively this provides $(x,y)$ coordinates along the slit via the
parametric equations of a line, where $p_2 - \keyw{CRPIX2}$ is the distance
parameter.

As before, the main problem is to determine the correct angle of rotation.
For zenithal projections $\phi\sub{p} = 180\degr$ by default, and referring to
the left side of Fig.~\ref{fig:Native} we see that this corresponds to the
$y$-axis.  Thus when $\rho = 0$ we require a rotation of $90\degr$ to
transform the $p_2$-axis onto the $y$-axis.  For $\rho > 0$ we need to rotate
further so the angle of rotation is $90\degr + \rho$.

Therefore, \CDELT{i} and \keyw{LONPOLE} in the original header must be
replaced with
\begin{eqnarray*}
   \keyw{CD1\_1} & = &                                \Delta\lambda \, , \\
   \keyw{CD2\_2} & = & \hphantom{-}\cos(90\degr+\rho) (\sigma/3600) \, , \\
   \keyw{CD3\_2} & = &           - \sin(90\degr+\rho) (\sigma/3600) \, , \\
   \keyw{CD2\_3} & = &                                            1 \, , \\
   \keyw{CD3\_3} & = &                                            1 \, .
\end{eqnarray*}
The negative sign on \keyw{CD3\_2} is associated with the rotation, not the
scale.  Note that, although the value of \keyw{CD3\_3} is irrelevant, it must
be set non-zero otherwise the zero defaults for \CD{i}{j} would make the third
column zero, thereby producing a singular matrix.  Likewise, in this example
and similarly below, we set \keyw{CD2\_3} non-zero to prevent the second row
of the matrix becoming zero for values of $\rho$ such that
$\cos(90\degr+\rho) = 0$.

The \PC{i}{j} matrix formulation is similar but allows more flexibility in
the way the scale is handled:

\noindent
\begin{enumerate}
\item
   Since \CDELT{i} defaults to unity, omitting it allows \PC{i}{j} to
   duplicate the functionality of \CD{i}{j}.  However, since \PC{i}{j}
   defaults to the unit matrix rather than zero, there is no need to set
   \keyw{PC3\_3} specifically, hence
   \begin{eqnarray*}
      \keyw{PC1\_1} & = &                                \Delta\lambda \, , \\
      \keyw{PC2\_2} & = & \hphantom{-}\cos(90\degr+\rho) (\sigma/3600) \, , \\
      \keyw{PC3\_2} & = &           - \sin(90\degr+\rho) (\sigma/3600) \, , \\
      \keyw{PC2\_3} & = &                                            1 \, .
   \end{eqnarray*}

\item
   Equation~(\ref{eq:slit}), as a simple scaling relation, suggests the use of
   \CDELT{i}.  However, \keyw{PC3\_2} must be non-zero since $x$ and $y$ both
   depend on $p_2$.  Setting it to unity and allowing the remaining \PC{i}{j}
   to default to the unit matrix we have
   \begin{eqnarray*}
      \keyw{PC3\_2} & = &                                            1 \, , \\
      \keyw{CDELT1} & = &                                \Delta\lambda \, , \\
      \keyw{CDELT2} & = & \hphantom{-}\cos(90\degr+\rho) (\sigma/3600) \, , \\
      \keyw{CDELT3} & = &           - \sin(90\degr+\rho) (\sigma/3600) \, .
   \end{eqnarray*}

\item
   The ``orthodox'' method is to encode the rotation and scale separately in
   \PC{i}{j} and \CDELT{i}:
   \begin{eqnarray*}
      \keyw{PC2\_2} & = & \hphantom{-}\cos(90\degr+\rho) \, , \\
      \keyw{PC3\_2} & = &           - \sin(90\degr+\rho) \, , \\
      \keyw{PC2\_3} & = &                              1 \, , \\
      \keyw{CDELT1} & = &                  \Delta\lambda \, , \\
      \keyw{CDELT2} & = &                    \sigma/3600 \, , \\
      \keyw{CDELT3} & = &                    \sigma/3600 \, .
   \end{eqnarray*}

   \noindent
   Arguably this is more amenable to human interpretation, especially if
   thoughtful comments are added.
\end{enumerate}

\noindent
The above six methods should all be regarded as equally legitimate.  In fact,
there are infinitely many ways to encode this header, though most would
disguise the essential simplicity of the geometry.


\subsubsection{Header construction example 4}
\label{sec:construct4}

Thompson (\cite{kn:Tho}) has applied the methods of this paper to the
definition of solar coordinates for a variety of coordinate systems.  As the
final header construction example we will consider a specific coordinate
description for another solar system body, the Moon.

A short exposure plate taken at Sydney Observatory at 1:00\,am on 15 February
1957 AEST (GMT +1000) of the full Moon near lunar perigee has been digitized
and converted to a $4096 \times 4096$ pixel image in FITS format.  The scale
is $1\arcsec$ per pixel with the moon centered in the image, and it is desired
to construct a dual coordinate description.  The first system is geocentric
apparent equatorial coordinates in a gnomonic projection.  The second is
selenographic coordinates in a zenithal perspective projection attached to the
surface of the Moon.

The ephemeris gives the geocentric apparent right ascension of the Moon at the
time as $09^{\rm h}41^{\rm m}13\fs1$ and declination as
$+08\degr34\arcmin26\arcsec$.  Diurnal parallax would have caused the Moon to
appear slightly offset from this position as seen from Sydney Observatory, but
to a good approximation the coordinate systems may be defined with the center
of the Moon at the stated geocentric coordinates.  The image was digitized in
the normal orientation with north up and east to the left so the header for
the primary description is straightforward:
\begin{eqnarray*}
   \keyw{  NAXIS} & = &                 2 \, , \\
   \keyw{NAXIS1}  & = &              4096 \, , \\
   \keyw{NAXIS2}  & = &              4096 \, , \\
   \keyw{MJD-OBS} & = &         35883.625 \, , \\
   \keyw{CRPIX1}  & = &            2048.5 \, , \\
   \keyw{CRPIX2}  & = &            2048.5 \, , \\
   \keyw{CDELT1}  & = &        -0.0002778 \, , \\
   \keyw{CDELT2}  & = &     \SP 0.0002778 \, , \\
   \keyw{CTYPE1}  & = & \keyv{`RA---TAN'} \, , \\
   \keyw{CTYPE2}  & = & \keyv{`DEC--TAN'} \, , \\
   \keyw{CRVAL1}  & = &         145.30458 \, , \\
   \keyw{CRVAL2}  & = &\hphantom{14}8.57386 \, , \\
   \keyw{LONPOLE} & = &             180.0 \, , \\
   \keyw{RADESYS} & = &    \keyv{`GAPPT'} \, .
\end{eqnarray*}
If any rotation had been introduced it could have been corrected for in the
linear transformation matrix.

The first step in constructing the secondary description is to determine the
distance of the observer from the Moon.  The ephemeris gives the equatorial
horizontal parallax as $61\arcmin29\farcs3$, corresponding to a distance
between the centers of the Earth and Moon of 55.91 Earth radii.  However, the
observer may be closer or further away by up to one Earth radius depending on
location and time of day and this 2\% effect is deemed worthy of correction.
At Sydney Observatory (longitude $10^{\rm h}04^{\rm m}49\fs2$, latitude
$-33\degr51\arcmin41\arcsec$) the Greenwich apparent sidereal time was
$00^{\rm h}51^{\rm m}43\fs6$.  Thus the apparent right ascension and
declination of the zenith was $10^{\rm h}56^{\rm m}32\fs8$,
$-33\degr51\arcmin41\arcsec$.  The vector dot product then gives the distance
correction as $0.69$ Earth radii.  Using the ratio of the Earth and Moon radii
of 3.670 the corrected distance of $55.22$ Earth radii indicates that the
distance parameter for the zenithal perspective projection is $\mu = 202.64$
Moon radii.

\begin{figure}
   \centerline{\putfig{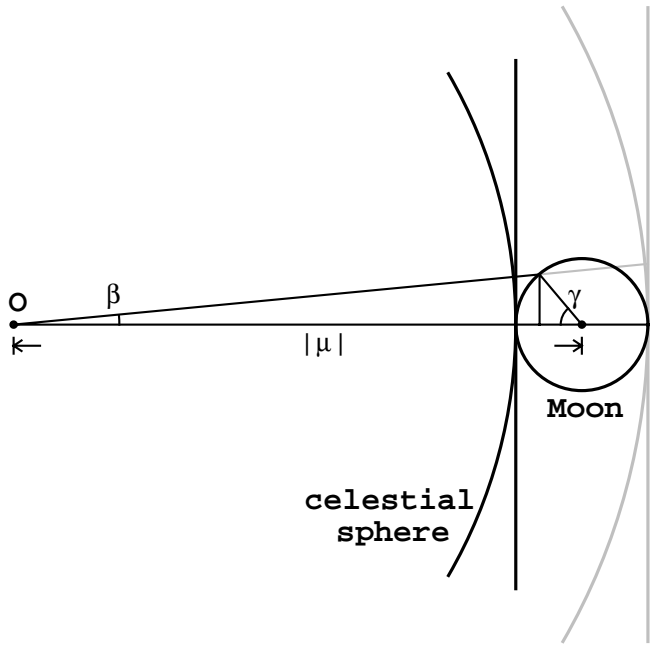}{250pt}}
   \caption[]{Geometry of header construction example 4 (not to scale).  The
      observer is at $O$, a distance of $\vert\mu\vert$ Moon radii from the
      center of the Moon.  The Moon is projected as a near-sided zenithal
      perspective projection ($\mu < -1$), and the celestial sphere as a
      gnomonic projection.  An alternative plane of projection is shown.}
   \label{fig:Moon}
\end{figure}

We now need to determine the correct relative scale.  Figure~\ref{fig:Moon}
shows the geometry of the two projections where we note, by analogy with
Fig.~\ref{fig:Zenithal}, that $\mu < -1$.  From the diagram we have
\begin{equation}
   \beta = \tan^{-1} \left(
                        \frac{\sin\gamma}{\vert\mu\vert - \cos\gamma}
                     \right) \, .
\end{equation}
The figure shows that this equation is not influenced by the choice of the
plane of projection.  Evaluating the derivative ${\rm d}\beta/{\rm d}\gamma$
at $\gamma = 0$ we find the relative scale is $1/(\vert\mu\vert - 1)$.

The ephemeris records that the selenographic coordinates of the Earth at the
time were $(\ell,b) = (+0\fdg26,+6\fdg45)$ and the position angle of the
Moon's axis was $C = 19\fdg03$.  However, since the Earth subtends an angle of
over $2\degr$ in the lunar sky, topocentric optical libration -- the
correction for the observatory's location -- is significant.  Application of
the correction formul\ae\ derived by Atkinson (\cite{kn:Atk}) gives the
selenographic coordinates of Sydney Observatory as $(\ell',b') =
(-0\fdg23,5\fdg89)$ and $C' = 18\fdg94$.  Since the image was taken in the
normal orientation and we have a zenithal projection it is convenient to
account for the position angle by setting $\phi\sub{p} = 180\degr - C'$.
Adopting keyword values of \keyv{SELN} and \keyv{SELT} for selenographic
coordinates we may write
\begin{eqnarray*}
   \keyw{CRPIX1S} & = &            2048.5 \, , \\
   \keyw{CRPIX2S} & = &            2048.5 \, , \\
   \keyw{CDELT1S} & = &         0.0562896 \, , \\
   \keyw{CDELT2S} & = &         0.0562896 \, , \\
   \keyw{CTYPE1S} & = & \keyv{`SELN-AZP'} \, , \\
   \keyw{CTYPE2S} & = & \keyv{`SELT-AZP'} \, , \\
   \keyw{PV2\_1S} & = &            202.64 \, , \\
   \keyw{CRVAL1S} & = &             -0.23 \, , \\
   \keyw{CRVAL2S} & = &              5.89 \, , \\
   \keyw{LONPOLES}& = &            161.06 \, , \\
   \keyw{WCSNAMES}& = & \keyv{`SELENOGRAPHIC COORDINATES'} \, .
\end{eqnarray*}
We have used the letter `\keyv{S}' to denote the alternate coordinate system
simply to demonstrate that there is no requirement to start with `\keyv{A}'.
A \WCSNAME{a} keyword, defined in Paper~I, is used to identify the coordinate
system.  Note that selenographic coordinates are right-handed so that
selenographic longitude increases towards the west on the celestial sphere as
seen from the Earth and this is handled by setting \keyw{CDELT1S}
{\em positive}.

As an extension of this example, a FITS header with three coordinate systems
might be constructed for an image of Saturn; a celestial grid for the
background, a saturnographic system for the surface of the planet, and a third
system for its rings.  The rings might be described by a zenithal equidistant
(\keyv{ARC}) projection with associated linear transformation matrix set to
match the oblique viewing angle.


\subsection{Realization}
\label{sec:realization}

Calabretta (\cite{kn:C2}) has written, and made available under a GNU license,
a package of routines, {\sc wcslib}, which implements all projections and
coordinate conversions defined here.  It contains independent C and Fortran
libraries.

The Fortran library includes a routine, {\sc pgsbox}, which is based on
{\sc pgplot} and uses {\sc wcslib} to draw general curvilinear coordinate
axes.  It was used to generate Fig.~\ref{fig:AZPhoto}.


\section{Summary}
\label{sec:summary}

We have developed a flexible method for associating celestial coordinates with
a FITS image and implemented it for all projections likely to be of use in
astronomy.  It should be a relatively simple matter to add new projections
should the need ever arise.

The FITS-header keywords defined in this paper are summarized in
Table~\ref{ta:keyword} with the 3-letter projection codes listed in
Table~\ref{ta:projname}.  The column labeled $\theta_0$ gives the native
latitude of the reference point in degrees ($\phi_0 = 0$ always) and the
parameters associated with each projection are listed in the nomenclature of
Sect.~\ref{sec:projections}.

\begin{table*}
   \caption[]{Summary of new coordinate keywords introduced in this paper; see
              also Table~\ref{ta:bintdef} for alternate types used in binary
              tables.}
   \protect\begin{tabular}{llllll}
      \hline
      \noalign{\smallskip}
      Keyword & Type & Sect. & Use & Status & Comments \\
      \noalign{\smallskip}
      \hline
      \noalign{\smallskip}
      \LONPOLE{a}    & floating
                     & \ref{sec:refpoint}
                     & coordinate rotation
                     & new
                     & Longitude in the native coordinate system of the \\
                     & & & & & celestial system's north pole. \\
                     & & & & & Default: $0\degr$ if $\delta_0 \ge \theta_0$,
                               $180\degr$ otherwise. \\
      \LATPOLE{a}    & floating
                     & \ref{sec:nonpolar}
                     & coordinate rotation
                     & new
                     & Latitude in the native coordinate system of the \\
                     & & & & & celestial system's north pole, or
                               equivalently, the \\
                     & & & & & latitude in the celestial coordinate
                               system of the \\
                     & & & & & native system's north pole. \\
                     & & & & & Default: $90\degr$, unless
                               $(\theta_0, \delta_0, \phi\sub{p} - \phi_0) =
                               (0,0, \pm 90\degr)$ \\
                     & & & & & in which case it has no default. \\
      \RADESYS{a}    & character
                     & \ref{sec:equatorial}
                     & frame of reference
                     & new
                     & Reference frame of equatorial and ecliptic \\
                     & & & & & coordinates; recognized values are given in
                               Table~\ref{ta:radesys}. \\
                     & & & & & Default: `\keyv{FK4}' if \EQUINOX{a}
                               $< 1984.0$, `\keyv{FK5}' if \\
                     & & & & & $\ge 1984.0$, or \keyv{ICRS} if \EQUINOX{a} is
                               not given. \\
      \EQUINOX{a}    & floating
                     & \ref{sec:equatorial}
                     & coordinate epoch
                     & new
                     & Epoch of the mean equator and equinox in years; \\
                     & & & & & Besselian if \RADESYS{a} is \keyv{FK4} or
                               \keyv{FK4-NO-E}, \\
                     & & & & & Julian if \keyv{FK5}; not applicable to
                               \keyv{ICRS} or \keyv{GAPPT}\@. \\
                     & & & & & Default: \keyw{EPOCH} if given, else 1950.0 if
                               \RADESYS{a} is \keyv{FK4} \\
                     & & & & & or \keyv{FK4-NO-E}, or 2000.0 if \keyv{FK5}. \\
      \keyw{EPOCH}   & floating
                     & \ref{sec:equatorial}
                     & coordinate epoch
                     & deprecated
                     & Replaced by \EQUINOX{a}. \\
      \keyw{MJD-OBS} & floating
                     & \ref{sec:equatorial}
                     & time of observation
                     & new
                     & Modified Julian Date (JD - 2400000.5) of
                       observation. \\
                     & & & & & Default: \keyw{DATE-OBS} if given, else
                               no default. \\
      \noalign{\smallskip}
      \hline
   \end{tabular}
   \label{ta:keyword}
\end{table*}


\begin{acknowledgements}
The authors would particularly like to thank Patrick~Wallace (U.K. Starlink)
for his helpful comments which included the text for some of the paragraphs of
Sect.~\ref{sec:equatorial} and the suggestion of the zenithal polynomial
projection.  Also, Steve~Allen (UCO/Lick Observatory) who provided valuable
feedback and suggestions over the long period of this work's development.
We are grateful for the assistance provided by Bob~Hanisch (Space Telescope
Science Institute) in bringing the manuscript to its final form.  We also
thank William~Pence, Arnold~Rots, and Lorella~Angelini (NASA Goddard Space
Flight Center) for contributing the text of Sect.~\ref{sec:bintab}.

The authors thank the following for comments and suggestions: Rick~Balsano,
David~Berry, Emmanuel~Bertin, Jeff~Bloch, Peter~Bunclark, Julian~Daniels,
Lindsey~Davis, Rick~Ebert, Douglas Finkbeiner, Immanuel~Freedman,
Brian~Glendenning, Bill~Gray, Dave~Green, Mike~Kesteven, Doug~Mink,
Jonathan~McDowell, Fred~Patt, Tim~Pearson, William~Thompson, Doug~Tody,
Stephen~Walton, and Don~Wells.

Many others contributed indirectly via requests for clarification of
particular points leading potentially to improvements in the wording or
formalism.

Mark Calabretta would also like to thank the following for feedback on
{\sc wcslib}: Klaus~Banse, David~Barnes, David~Berry, Wim~Brouw,
Lindsey~Davis, Rick~Ebert, Ken~Ebisawa, Brian~Glendenning, Neil~Killeen,
Thomas~A.~McGlynn, Doug~Mink, Clive~Page, Ray~Plante, Keith~A.~Scollick,
and Peter~Teuben.

Coastline data for Fig.~\ref{fig:AZPhoto} are from the ``CIA World DataBank
II'' map database, which is in the public domain.  Dave~Pape's plain-text
version was obtained from {\tt http://www.evl.uic.edu/pape/data/WDB/} and this
site also provides references to the original, highly compressed binary
encodings.

The Australia Telescope is funded by the Commonwealth of Australia for
operation as a National Facility managed by CSIRO\@.

The National Radio Astronomy Observatory is a facility of the (U.S.) National
Science Foundation operated under cooperative agreement by Associated
Universities, Inc.
\end{acknowledgements}

\begin{table*}
   \caption[]{Summary of projection codes, default values of $\phi_0$ and
      $\theta_0$, full name and required parameters.}
   \protect\begin{tabular}{llclccccc}
      \hline
      \noalign{\smallskip}
      FITS & & &
           & \multicolumn{5}{l}{Projection parameters associated with
                               {\em latitude}$^\dag$ axis $i$} \\
      Code & $\phi_0$
           & $\theta_0$
           & Projection name
           & \PVi{0} & \PVi{1} & \PVi{2} & \PVi{3} & \PV{i}{ma} \\
      \noalign{\smallskip}
      \hline
      \noalign{\smallskip}
      \keyv{AZP} & $0\degr$
                 & $90\degr$
                 & Zenithal perspective
                 & & $\mu$ & $\gamma$ \\
      \keyv{SZP} & $0\degr$
                 & $90\degr$
                 & Slant zenithal perspective
                 & & $\mu$ & $\phi\sub{c}$ & $\theta\sub{c}$ \\
      \keyv{TAN} & $0\degr$
                 & $90\degr$
                 & Gnomonic \\
      \keyv{STG} & $0\degr$
                 & $90\degr$
                 & Stereographic \\
      \keyv{SIN} & $0\degr$
                 & $90\degr$
                 & Slant orthographic
                 & & $\xi$ & $\eta$ \\
      \keyv{ARC} & $0\degr$
                 & $90\degr$
                 & Zenithal equidistant \\
      \keyv{ZPN} & $0\degr$
                 & $90\degr$
                 & Zenithal polynomial
                 & $P_0$ & $P_1$ & $P_2$ & $P_3$
                 & $P_m$ for $m = 0,\ldots 29$ \\
      \keyv{ZEA} & $0\degr$
                 & $90\degr$
                 & Zenithal equal-area \\
      \keyv{AIR} & $0\degr$
                 & $90\degr$
                 & Airy
                 & & $\theta\sub{b}$ & \\
      \noalign{\medskip}
      \keyv{CYP} & $0\degr$
                 & $0\degr$
                 & Cylindrical perspective
                 & & $\mu$ & $\lambda$ \\
      \keyv{CEA} & $0\degr$
                 & $0\degr$
                 & Cylindrical equal area
                 & & $\lambda$ & \\
      \keyv{CAR} & $0\degr$
                 & $0\degr$
                 & Plate carr\'{e}e \\
      \keyv{MER} & $0\degr$
                 & $0\degr$
                 & Mercator \\
      \noalign{\medskip}
      \keyv{SFL} & $0\degr$
                 & $0\degr$
                 & Sanson-Flamsteed \\
      \keyv{PAR} & $0\degr$
                 & $0\degr$
                 & Parabolic \\
      \keyv{MOL} & $0\degr$
                 & $0\degr$
                 & Mollweide \\
      \keyv{AIT} & $0\degr$
                 & $0\degr$
                 & Hammer-Aitoff \\
      \noalign{\medskip}
      \keyv{COP} & $0\degr$
                 & $\theta\sub{a}$
                 & Conic perspective
                 & & $\theta\sub{a}$ & $\eta$ \\
      \keyv{COE} & $0\degr$
                 & $\theta\sub{a}$
                 & Conic equal-area
                 & & $\theta\sub{a}$ & $\eta$ \\
      \keyv{COD} & $0\degr$
                 & $\theta\sub{a}$
                 & Conic equidistant
                 & & $\theta\sub{a}$ & $\eta$ \\
      \keyv{COO} & $0\degr$
                 & $\theta\sub{a}$
                 & Conic orthomorphic
                 & & $\theta\sub{a}$ & $\eta$ \\
      \noalign{\medskip}
      \keyv{BON} & $0\degr$
                 & $0\degr$
                 & Bonne's equal area
                 & & $\theta_1$ \\
      \keyv{PCO} & $0\degr$
                 & $0\degr$
                 & Polyconic \\
      \noalign{\medskip}
      \keyv{TSC} & $0\degr$
                 & $0\degr$
                 & Tangential Spherical Cube \\
      \keyv{CSC} & $0\degr$
                 & $0\degr$
                 & COBE Quadrilateralized Spherical Cube \\
      \keyv{QSC} & $0\degr$
                 & $0\degr$
                 & Quadrilateralized Spherical Cube \\
      \noalign{\smallskip}
      \hline
   \end{tabular}

   $^\dag$ Parameters \PVi{0}, \PVi{1}, and \PVi{2} associated with
   {\em longitude} axis $i$ implement user-specified values of
   $(\phi_0,\theta_0)$ as discussed in Sect.~\ref{sec:userspec}, and \PVi{3},
   and \PVi{4} encapsulate the values of \LONPOLE{a} and \LATPOLE{a}
   respectively, as discussed in Sect.~\ref{sec:encapsulation}.
   \label{ta:projname}
\end{table*}


\appendix
\section{Mathematical methods}
\label{apx:methods}


\subsection{Coordinate rotation with matrices}

The coordinate rotations represented in Eqs.~(\ref{eq:nat2std}) or
(\ref{eq:std2nat}) may be represented by a matrix multiplication of a vector
of direction cosines.  The matrix and its inverse (which is simply the
transpose) may be precomputed and applied repetitively to a variety of
coordinates, improving performance.  Thus, we have
\begin{equation}
   \left( \begin{array}{c} l \\ m \\ n \end{array} \right) =
   \left( \begin{array}{ccc}
         r_{11} & r_{12} & r_{13} \\
         r_{21} & r_{22} & r_{23} \\
         r_{31} & r_{32} & r_{33}
   \end{array} \right)
   \left( \begin{array}{c} l^{\prime} \\ m^{\prime} \\ n^{\prime}
       \end{array} \right) \, ,
\end{equation}
where
\begin{eqnarray*}
   (l',m',n') & = & (\cos\delta\cos\alpha,\cos\delta\sin\alpha, \sin\delta) \, , \\
   (l,m,n)    & = & (\cos\theta\cos\phi,\cos\theta\sin\phi,\sin\theta) \, , \\
   r_{11}     & = & -\sin\alpha\sub{p}\sin\phi\sub{p} -
                     \cos\alpha\sub{p}\cos\phi\sub{p}\sin\delta\sub{p} \, , \\
   r_{12}     & = & \cos\alpha\sub{p}\sin\phi\sub{p} -
                    \sin\alpha\sub{p}\cos\phi\sub{p}\sin\delta\sub{p} \, , \\
   r_{13}     & = & \cos\phi\sub{p} \cos\delta\sub{p} \, , \\
   r_{21}     & = & \sin\alpha\sub{p}\cos\phi\sub{p} -
                    \cos\alpha\sub{p}\sin\phi\sub{p}\sin\delta\sub{p} \, , \\
   r_{22}     & = & -\cos\alpha\sub{p}\cos\phi\sub{p} -
                     \sin\alpha\sub{p}\sin\phi\sub{p}\sin\delta\sub{p} \, , \\
   r_{23}     & = & \sin\phi\sub{p} \cos\delta\sub{p} \, , \\
   r_{31}     & = & \cos\alpha\sub{p} \cos\delta\sub{p} \, , \\
   r_{32}     & = & \sin\alpha\sub{p} \cos\delta\sub{p} \, , \\
   r_{33}     & = & \sin\delta\sub{p} \, .
\end{eqnarray*}
The inverse equation is
\begin{equation}
   \left( \begin{array}{c} l^{\prime} \\ m^{\prime} \\ n^{\prime}
       \end{array} \right) =
   \left( \begin{array}{ccc}
         r_{11} & r_{21} & r_{31} \\
         r_{12} & r_{22} & r_{32} \\
         r_{13} & r_{23} & r_{33}
   \end{array} \right)
   \left( \begin{array}{c} l \\ m \\ n \end{array} \right) \, .
\end{equation}


\subsection{Iterative solution}

Iterative methods are required for the inversion of several of the projections
described in this paper.  One, Mollweide's, even requires solution of a
transcendental equation for the forward equations.  However, these do not give
rise to any particular difficulties.

On the other hand, it sometimes happens that one pixel and one celestial
coordinate element is known and it is required to find the others; this
typically arises when plotting graticules on image displays.  Although
analytical solutions exist for a few special cases, iterative methods must be
used in the general case.  If, say, $p_2$ and $\alpha$ are known, one would
compute pixel coordinate as a function of $\delta$ and determine the value
which gave $p_2$.  The unknown pixel coordinate elements would be obtained in
the process.

This prescription glosses over many complications, however.  All bounded
projections may give rise to discontinuities in the graph of $p_2$ versus
$\delta$ (to continue the above example), for example where the meridian
through $\alpha$ crosses the $\phi = \pm180\degr$ boundary in cylindrical,
conic and other projections.  Even worse, if the meridian traverses a pole
represented as a finite line segment then $p_2$ may become multivalued at a
particular value of $\delta$.  The derivative $\partial p_2/\partial\delta$
will also usually be discontinuous at the point of discontinuity, and it
should be remembered that some projections such as the quad-cubes may also
have discontinuous derivatives at points within their boundaries.

We will not attempt to resolve these difficulties here but simply note that
{\sc wcslib} (Calabretta, \cite{kn:C2}) implements a solution.

\begin{table*}
   \caption[]{Projection aliases.}
   \protect\begin{tabular*}{510pt}{ll|ll}
      \hline
      \noalign{\smallskip}
      Name                   & Alias for \hphantom{\hskip 90pt} &
      Name                   & Alias for \\
      \noalign{\smallskip}
      \hline
      \noalign{\smallskip}

      Gnomonic               & \keyv{AZP} with $\mu = 0$ &
      Miller                 & \keyv{CAR} with $x$ scaled by $2/\pi$ \\
         \Ma = Central & &
      Equirectangular        & \keyv{CAR} with unequal scaling  \\
      Near-sided perspective & \keyv{AZP} with $\mu = 1.35$  &
      Cartesian              & \keyv{CAR} \\
      Clarke's (first)       & \keyv{AZP} with $\mu = 1.35$  &
         \Ma = Equidistant \\
      Clarke's (second)      & \keyv{AZP} with $\mu = 1.65$  &
            \Mb cylindrical \\
      James'                 & \keyv{AZP} with $\mu = 1.367$ &
      Cassini                & \keyv{CAR} transverse case \\
      La Hire's              & \keyv{AZP} with $\mu = 1.71$  &
      Transverse Mercator    & \keyv{MER} transverse case \\
      Approximate equidistant & &
         \Ma = Transverse cylindrical \\
         \Ma zenithal        & \keyv{AZP} with $\mu = 1.7519$ &
            \Mb orthomorphic \\
      Approximate equal area  & &
      Sinusoidal             & \keyv{SFL} \\
         \Ma zenithal        & \keyv{AZP} with $\mu = 2.4142$ &
         \Ma = Global sinusoid (\keyv{GLS}) \\
      Postel                 & \keyv{ARC} &
         \Ma = Mercator equal-area \\
         \Ma = Equidistant & &
         \Ma = Mercator-Sanson \\
         \Ma = Globular & &
         \Ma = Sanson's \\
      Lambert azimuthal & &
      Craster                & \keyv{PAR} \\
         \Mb equivalent      & \keyv{ZEA} &
      Bartholomew's atlantis & \keyv{MOL} oblique case \\
         \Ma = Lambert azimuthal & &
      Mollweide's homolographic & \keyv{MOL} \\
             \Mb equal area & &
         \Ma = Homolographic \\
         \Ma = Lambert polar & &
         \Ma = Homalographic \\
             \Mb azimuthal & &
         \Ma = Babinet \\
         \Ma = Lorgna & &
         \Ma = Elliptical \\
      Gall's cylindrical     & \keyv{CYP} with $\mu = 1,
                               \lambda = \sqrt{2}/2$ &
      Hammer equal area      & \keyv{AIT} \\
      Cylindrical equal area & \keyv{CYP} with $\mu = \infty$ &
         \Ma = Aitoff \\
      Simple cylindrical     & \keyv{CYP} with $\mu = 0, \lambda = 1$ &
         \Ma = Aitov \\
         \Ma = Central cylindrical & &
      Bartholomew's nordic   & \keyv{AIT} oblique case \\
         \Ma = Cylindrical central & &
      One-standard conic     & Conic with $\theta_1 = \theta_2 $\\
             \Mb perspective & &
         \Ma = Tangent conic \\
         \Ma = Gall's stereographic & &
      Two-standard conic     & Conic with $\theta_1 \neq \theta_2 $\\
      Lambert's cylindrical  & \keyv{CYP} with $\mu = \infty, \lambda = 1$ &
         \Ma = Secant conic \\
         \Ma = Lambert's equal area & &
      Murdoch conic          & similar to \keyv{COD} \\
      Behrmann equal area    & \keyv{CEA} with $\lambda = 3/4$ &
      Alber's                & \keyv{COE} \\
      Gall's orthographic    & \keyv{CEA} with $\lambda = 1/2$ &
         \Ma = Alber's equal area \\
         \Ma = Approximate Peter's & &
      Lambert equal area     & \keyv{COE} with $\theta_2 = 90\degr$ \\
      Lambert's equal area   & \keyv{CEA} with $\lambda = 1$ &
      Lambert conformal conic& for spherical Earth = \keyv{COO} \\
      & &
      Werner's               & \keyv{BON} with $\theta_1 = 90\degr$\\
      \noalign{\smallskip}
      \hline
   \end{tabular*}
   \label{ta:aliases}
\end{table*}


\section{The slant orthographic projection}
\label{apx:SYN}

The slant orthographic or generalized \keyv{SIN} projection derives from the
basic interferometer equation (e.g.\ Thompson et al., \cite{kn:TMS}).  The
phase term in the Fourier exponent is
\begin{eqnarray}
    \wp = \mathbf{(e - e_0) \cdot B} \, ,
\end{eqnarray}
where $\mathbf{e_0}$ and $\mathbf{e}$ are unit vectors pointing towards the
field center and a point in the field, $\mathbf{B}$ is the baseline vector,
and we measure the phase $\wp$ in rotations so that we don't need to carry
factors of $2\pi$.  We can write
\begin{equation}
   \wp = p_u u + p_v v + p_w w \, ,
   \label{eq:phase2}
\end{equation}
where $(u,v,w)$ are components of the baseline vector in a right-handed
coordinate system with the $w$-axis pointing from the geocenter towards the
source and the $u$-axis lying in the equatorial plane, and
\begin{equation}
   \begin{array}{rccl}
      p_u & = &   & \cos \theta \sin \phi \, , \\
      p_v & = & - & \cos \theta \cos \phi \, , \\
      p_w & = &   & \sin \theta - 1       \, ,
      \label{eq:puvw}
   \end{array}
\end{equation}
are the coordinates of $\mathbf{(e - e_0)}$, where $(\phi,\theta)$ are the
longitude and latitude of $\mathbf{e}$ in the spherical coordinate system with
the pole towards $\mathbf{e_0}$ and origin of longitude towards negative $v$,
as required by Fig.~\ref{fig:Native}.  Now, for a planar array we may write
\begin{equation}
   n_u u + n_v v + n_w w = 0
\end{equation}
where $(n_u,n_v,n_w)$ are the direction cosines of the normal to the plane.
Using this to eliminate $w$ from Eq.~(\ref{eq:phase2}) we have
\begin{equation}
   \wp = [p_u - \frac{n_u}{n_w} p_w] u + [p_v - \frac{n_v}{n_w} p_w] v
   \label{eq:phase3}
\end{equation}
Being the Fourier conjugate variables, the quantities in brackets become the
Cartesian coordinates, in radians, in the plane of the synthesized map.
Writing
\begin{equation}
   \begin{array}{rcl}
      \xi  & = & n_u / n_w \, , \\
      \eta & = & n_v / n_w \, ,
   \end{array}
\end{equation}
Eqs.~(\ref{eq:SYNx}) and (\ref{eq:SYNy}) are then readily derived from
Eqs.~(\ref{eq:puvw}) and (\ref{eq:phase3}).  For 12-hour synthesis by an
east-west interferometer the baselines all lie in the Earth's equatorial plane
whence $(\xi,\eta) = (0, \cot\delta_0)$, where $\delta_0$ is the declination
of the field center.  For a ``snapshot'' observation by an array such as the
VLA, Cornwell \& Perley (\cite{kn:CP}) give $(\xi,\eta) = (-\tan Z\sin\chi,
\tan Z\cos\chi)$, where $Z$ is the zenith angle and $\chi$ is the parallactic
angle of the field center at the time of the observation.

In synthesizing a map a phase shift may be applied to the visibility data in
order to translate the field center.  If the shift applied is
\begin{equation}
   \Delta\wp = q_u u + q_v v + q_w w
\end{equation}
where $(q_u,q_v,q_w)$ is constant then equation (\ref{eq:phase2}) becomes
\begin{equation}
   \wp = (p_u - q_u) u + (p_v - q_v) v + (p_w - q_w) w \, ,
\end{equation}
whence equation (\ref{eq:phase3}) becomes
\begin{equation}
   \begin{array}{rcl}
      \wp & = & [(p_u - q_u) - \xi (p_w - q_w)] u \, + \\
          &   & [(p_v - q_v) - \eta(p_w - q_w)] v \, .
   \end{array}
\end{equation}
Equations~(\ref{eq:SYNx}) and (\ref{eq:SYNy}) become
\begin{eqnarray}
   x & = & \SP \rd \left[\, \cos\theta \sin\phi +
                   \xi  \, (\sin\theta - 1)\, \right]  -
               \rd \left[q_u - \xi  q_w \right] \, ,
               \nonumber \\
   y & = &   - \rd \left[\, \cos\theta \cos\phi -
                   \eta \, (\sin\theta - 1)\, \right]  -
               \rd \left[q_v - \eta q_w \right] \, ,
               \nonumber \\
\end{eqnarray}
from which on comparison with Eqs.~(\ref{eq:SYNx}) and (\ref{eq:SYNy}) we see
that the field center is shifted by
\begin{equation}
   (\Delta x, \Delta y) = \rd (q_u - \xi q_w, \, q_v - \eta q_w) \, .
\end{equation}
The shift is applied to the coordinate reference pixel.


\section{Projection aliases}
\label{apx:aliases}

Table~\ref{ta:aliases} provides a list of aliases which have been used by
cartographers for special cases of the projections described in
Sect.~\ref{sec:projections}.


\end{document}